\documentclass[pdflatex,sn-mathphys-num]{sn-jnl}

\usepackage{tabularx}
\usepackage{ragged2e}
\usepackage{graphicx}%
\usepackage{subcaption}
\usepackage{multirow}%
\usepackage{amsmath,amssymb,amsfonts}%
\usepackage{amsthm}%
\usepackage{mathrsfs}%
\usepackage[title]{appendix}%
\usepackage{xcolor}%
\usepackage{textcomp}%
\usepackage{manyfoot}%
\usepackage{booktabs}%
\usepackage{algorithm}%
\usepackage{algorithmicx}%
\usepackage{algpseudocode}%
\usepackage{listings}%
\usepackage[most]{tcolorbox}
\usepackage{soul}
\usepackage{array}
\usepackage{longtable}
\usepackage{placeins}
\usepackage{lineno}
\newcommand{\projecturl}{https://inyourownwords.github.io}
\newcommand{\codeurl}{https://github.com/jennyw23/InYourOwnWords}

\begin{document}
\title[In your own words: computationally identifying interpretable themes in free-text survey data]{In your own words: computationally identifying interpretable themes from free-text survey data}

\author[1]{\fnm{Jenny S.} \sur{Wang}}\email{jewang@hbs.edu}
\author[2]{\fnm{Aliya} \sur{Saperstein}}\email{asaper@stanford.edu}
\author*[3]{\fnm{Emma} \sur{Pierson}}\email{emmapierson@berkeley.edu}

\affil[1]{\orgname{Technology and Operations Management Unit, Harvard Business School}, \orgaddress{\city{Boston}, \state{Massachusetts}, \country{USA}}}
\affil[2]{\orgdiv{Department of Sociology}, \orgname{Stanford University}, \orgaddress{\city{Palo Alto}, \state{CA}, \country{USA}}}
\affil[3]{\orgdiv{Department of Electrical Engineering and Computer Sciences}, \orgname{University of California, Berkeley}, \orgaddress{\city{Berkeley}, \state{CA}, \country{USA}}}
\affil[*]{\textbf{Corresponding author}: emmapierson@berkeley.edu}

\abstract{Free-text survey responses can provide nuance often missed by structured questions, but remain difficult to statistically analyze. To address this, we introduce \textsc{In Your Own Words}, a computational framework for exploratory analyses of free-text survey data that identifies structured, interpretable themes in free-text responses, facilitating systematic analysis. To illustrate the benefits of this approach, we apply it to a new dataset of free-text descriptions of race, gender, and sexual orientation from 1,004 U.S. participants. The themes our approach produces on this dataset are more coherent and interpretable than those produced by past computational methods. The themes have three practical applications in survey research. First, they can \emph{suggest structured questions} to add to future surveys by surfacing salient constructs---such as belonging and identity fluidity---that existing surveys do not capture. Second, the themes \emph{reveal heterogeneity within standardized categories}, explaining additional variation in health, well-being, and identity importance. Third, the themes \emph{illuminate systematic discordance between self-identified and perceived identities}, highlighting mechanisms of misrecognition that existing measures do not reflect. More broadly, our framework can be deployed in a wide range of survey settings to identify interpretable themes from free text, complementing existing qualitative methods.}

\maketitle
\section{Introduction}\label{sec:intro}
Free-text data provide valuable insights for survey research. When individuals describe their lives in their own words, they often provide personalized and context-sensitive details that standardized categories cannot capture \cite{singer2017some, pao2025case}. For example, open-ended questions about identity can afford greater visibility to smaller or less-recognized populations \cite{wong2025other, pao2025case}; similarly, open-ended healthcare questions reveal actionable information that can improve patient experiences \cite{riiskjaer2012value, o2004any}. 

But a major challenge in using free-text data is that it is much more difficult to statistically analyze than structured data~\cite{singer2017some, prewitt2005racial, aspinall2012answer,magliozzi2016scaling}. Even the most basic quantitative questions---\emph{What are the major groups in the population under study? How large are they? Who is a member of each?}---cannot be directly answered from raw text data. To answer them, researchers must typically rely on labor-intensive hand-coding, where trained analysts iteratively develop and apply codes to capture the meaning in responses \cite{glaser2017discovery, venkatesh2013bridging, beresford2022coding}. This process precludes the use of free-text data in resource-constrained settings or in very large datasets. 

Addressing this, we introduce \textsc{In Your Own Words}, a computational framework for exploratory analyses of survey data that complements existing qualitative approaches. Our framework learns interpretable, human-readable themes from free-text responses by leveraging the powerful abilities of sparse autoencoders (SAEs) and large language models (LLMs), inspired by recent work showing that SAEs can successfully identify interpretable themes in text corpora~\cite{hypothesaes,movva2025s,jiangsun2025interp_embed}. Our approach does not require pre-specifying themes or manually annotating text; instead, it identifies recurring themes in how respondents describe themselves and annotates each free-text response with whether it contains each theme. Each response is then represented as a vector of the free-text themes it contains, preserving the richness and nuance of open-ended free text while converting the data into a structured form that can be easily used in downstream statistical analysis. Our work builds on a long tradition of classical and modern methods for computational text analysis~\cite{blei2003latent, grootendorst2022bertopic, pham-etal-2024-topicgpt, zhong2025hicode}; we show that our approach outperforms several methods from this literature on the data we collect.

We illustrate the benefits of the \textsc{In Your Own Words} framework by applying it to an important setting in which free-text responses confer particular benefits: the study of how people describe their identities. Identity categories for race, gender, and sexual orientation are widely used in analyses of inequality and other social phenomena; they also inform important policy decisions ranging from resource allocation to legal recognition \cite{ennis2024examining, garland2018legislating, morgan2022equal, weideman2025research}. 
But younger generations increasingly identify outside traditional standardized categories \cite{pao2025case}, raising questions about whether existing standardized items capture the aspects of identity that matter most to respondents.
To study this, we design and administer a survey to over 1,000 U.S.-based participants, collecting both categorical and free-text descriptions of race, gender, and sexual orientation, along with measures of health, discrimination, and other social and demographic variables. We release this dataset to other researchers upon request to facilitate subsequent work.\footnote{\href{\projecturl}{\projecturl}}

The free-text themes our method learns from this data---such as ``mentions feelings of not fully belonging or being out of place related to race/ethnicity'' or ``mentions fluidity in gender identity''---surface patterns of identity expression that are not captured by standardized identity categories. For example, many of our free-text themes span multiple standardized categories, revealing shared cultural, linguistic, and developmental dimensions of identity. Similarly, standardized categories account for only a small fraction of variation in the free-text themes (median $R^2$ 0.15 for race, 0.12 for gender, and 0.22 for sexual orientation). We also show that, compared to three widely-used methods for computational text analysis (LDA, BERTopic, and TopicGPT \cite{blei2003latent, grootendorst2022bertopic, pham-etal-2024-topicgpt}), our approach yields more interpretable and coherent identity representations, while alternate methods tend to produce overly broad, overly rare, or category-specific themes. 

We then show that the additional information our free-text themes provide (beyond standardized categories) has three practical applications in survey research. First, the themes can \emph{suggest additional structured questions to add to future surveys}. 
We first validate that our free-text themes can recover structured questions that past work has already recognized as important, including those about nativity, gender expression, or sexual attraction  \cite{bostwick2010dimensions, gottgens2022impact}. 
We then show, importantly, that the free-text themes highlight nuanced dimensions of lived experiences that have \emph{not} traditionally appeared in demographic surveys---for example, uncertainty about identity, lack of belonging to an identity, or fluidity in identity---suggesting structured questions that could be added to future surveys.

Second, we show that free-text themes can reveal \emph{important heterogeneity within standardized identity categories}. A general concern about these categories is that they may fail to capture within-group variation: for example, past work has shown that there is a great deal of heterogeneity in health outcomes within race groups \cite{read2021disaggregating, movva2023coarse}. We show that our free-text themes help describe this heterogeneity, explaining within-group variation in self-reported health, life satisfaction, and identity importance. The free-text themes also trace this variation to specific lived experiences: for example, feelings of not fully belonging to one's race group are associated with worse mental and physical health.

Third, we show that free-text themes can illuminate the gap between \emph{self-described identity} and \emph{perceived identity}. Applying our method to participants’ free-text responses describing how they believe others perceive their identities, we discover distinct mechanisms underlying perceived race, gender, and sexual orientation. Perceived race and gender are described primarily through \emph{appearance-based cues}---such as how people look and dress---whereas perceived sexual orientation is primarily described through \emph{people's relationships}, driven by factors like past romantic partners and heteronormative assumptions. 
These free-text themes also reveal processes that extend beyond what is typically captured in structured identity measures, such as how individuals make sense of and react to others' misperceptions. 
Together, these findings highlight limitations in current measures of perceived identity and suggest opportunities to design more nuanced approaches, particularly for gender and sexual orientation.

Beyond the study of identity, \textsc{In Your Own Words} is a practical framework for exploratory analysis of free-text survey data that complements existing qualitative methods.
The framework is likely to be especially helpful in two survey settings. First, it is useful when researchers suspect that existing standardized questions do not suffice, necessitating analyses of free-text data---for example, if relevant constructs are evolving over time, surveys are being expanded to new populations, or entirely new concepts are being surveyed \cite{gentile2014generational, ennis2024examining, shiller2017narrative}. Second, it provides a scalable alternative in settings where comprehensive qualitative coding is infeasible, such as large-scale surveys or resource-constrained environments \cite{ferrario2022eliciting,dunivin2025scaling}.
By identifying interpretable themes in free-text responses, our framework enables a wider range of survey research to benefit from the richness of open-ended data.

\section{Results}\label{sec:results}

We first describe the new free-text identity dataset we collect (\S \ref{sec:dataset}); we then describe and validate our computational framework (\S \ref{results:computational-framework}) and analyze the themes it identifies (\S \ref{sec:present-themes}). 

\subsection{Dataset}
\label{sec:dataset}

\subsubsection*{Data collection procedure}
We conducted a survey of self-reported identity with 1,004 participants, using the Prolific survey platform \cite{prolific2025} to collect responses from U.S.-based, English-speakers ages 18 years and above. Each participant provided both free-text and categorical descriptions of their race, gender, and sexual orientation. Following prior work~\cite{vaughan2017oversampling, anderssen2017oversampling}, we upsampled minority groups to ensure representation across a diverse set of identities. Specifically, our full sample comprises three separate subsamples, upsampling (1) minority racial identities, (2) minority gender identities, and (3) minority sexual orientation identities. $\S$\ref{sec:methods-sample} provides full details on the sampling procedure. 

Each participant was asked, in three separate prompts, to describe their race, gender identity, and sexual orientation in at least two to three sentences. The purpose of asking for longer responses was to elicit substantive, qualitative responses; when piloting question wordings, we found that questions omitting this yielded brief (e.g., single-word) answers, consistent with prior research on identity data collection \cite{croll2019race,garbarski2025improving}. We also included a short, italicized sub-prompt to encourage more detailed and thoughtful self-descriptions beneath each main question. The full prompt for race was as follows: \textit{In at least 2--3 sentences, how would you describe your \texttt{race and/or ethnicity}? {\small For example, you could discuss specific traditions or customs, cultural practices or norms, languages or dialects spoken in your family, or ways race and/or ethnicity has influenced your life experiences.}} The analogous prompts for gender identity and sexual orientation are described in $\S$\ref{sec:methods-freetext}. Participants also provided a free-text description of how their self-described identity compared to how others perceived them (e.g., for race: \textit{How does your self-identified \texttt{race/ethnicity} compare to how you
believe others perceive your \texttt{race/ethnicity}?}).  Alternate prompt wordings might result in different patterns of responses, an interesting direction for future work; our computational framework is flexible and can be applied regardless of prompt wording.

After answering the free-text identity questions, participants also chose from standardized categories describing their race, gender, and sexual orientation in response to a multiple-choice question. For example, for race, participants selected all that applied from the following choices: ``American Indian or Alaska Native'',
``Asian'',
``Black or African American'',
``Hispanic or Latino'',
``Middle Eastern or North African'',
``Native Hawaiian or Pacific Islander'',
``White'', and ``Some Other Race''.
The question wording and answer choices for the categorical identity questions were informed by federal standards \cite{revesz2024revisions}, research-based recommendations \cite{amaya2020adapting}, and methodological guidance advocating for more granular categories, particularly for sexual orientation \cite{hughes2022guidance, suen2020}. More details on standardized categories are provided in $\S$\ref{sec:methods-categories}. Participants also answered multiple-choice questions indicating whether they believed their free-text descriptions of their race, gender, and sexual orientation conveyed important information not captured by the categorical identity questions. 

After answering the identity-related questions, respondents answered questions about health and social measures, like self-rated mental health and perceived discrimination. Participants also answered questions about other demographic characteristics (e.g., age, religion, and political affiliation). More details on these variables are also provided in $\S$\ref{sec:methods-outcomes}. We have made the survey instrument and data available upon request at \href{\projecturl}{\projecturl} to support future research.

\subsubsection*{Most respondents, especially minorities, believe that their free-text responses add important context}\label{sec:min-write-more}

To motivate our computational approach, we first confirm that most respondents indeed believed that their free-text responses conveyed meaningful details not captured by standardized categories (Figure \ref{fig:minority_say_more}). Specifically, 67\% of respondents believed that their free-text description added important information about their race; 61\% believed this for gender, and 63\% for sexual orientation. Respondents with minority identities had higher odds of holding this view: people with racial minority identities had 2.0$\times$ higher odds than non-minorities ($p < 0.001$); people with gender minority identities had 5.0$\times$
higher odds ($p < 0.001$); and people with sexual orientation minority identities had 2.6$\times$ higher odds ($p < 0.001$). Respondents with minority identities also wrote longer responses than non-minority respondents: people with racial minority identities wrote 15\% more than non-minorities ($p = 0.001$); people with gender minority identities 72\% more ($p < 0.001$); and people with sexual orientation minority identities 66\% more ($p < 0.001$). Together, these results suggest that respondents, particularly respondents with minority identities, felt that their free-text responses conveyed important information that standardized categories missed. Our finding that respondents with minority identities were particularly likely to hold this view is consistent with past work showing that racial and ethnic identity is a more salient aspect of self-concept for Black, Hispanic, and Asian individuals than for White individuals \cite{phinney1990ethnic,porter1993minority,pew2019race}, and that gender and sexual minorities often engage in more frequent and explicit identity reflection due to social misrecognition and marginalization \cite{gulgoz2019similarity,galupo2014conceptualization}. 

Collectively, these findings motivate our central question: is it possible to computationally and interpretably represent the additional information that respondents said their free-text responses conveyed? 

\subsection{Computational approach}
\label{results:computational-framework}

Our goal is to computationally identify interpretable themes in participants' free-text responses---for example, ``mentions speaking or understanding multiple languages or specific non-English languages'' or ``mentions fluidity or fluctuation in gender identity''. These themes should be \emph{automatically learned from the data} (that is, we should not have to pre-specify them), and should accurately and interpretably describe the free-text responses. After learning these themes, we represent each response by whether it expresses each theme, thus converting difficult-to-analyze, unstructured free text into structured data that is natural to use in exploratory downstream statistical analyses. 

\subsubsection*{Identifying themes from free-text responses}
\label{sec:sae_themes}

\begin{figure}
    \centering
    \includegraphics[width=\linewidth]{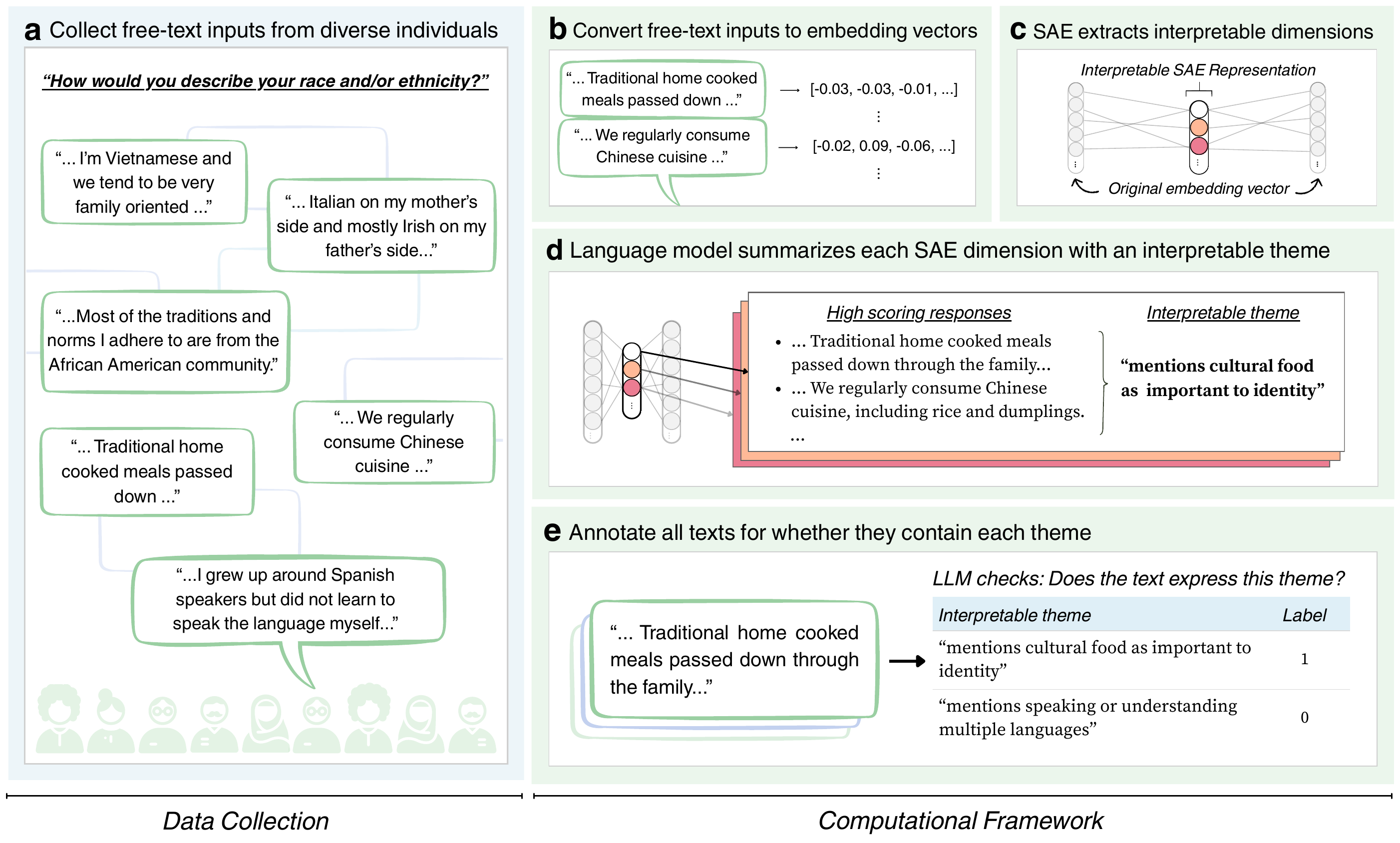}
    \caption{\textbf{Overview of data collection (a) and computational framework (b -- e).} \textbf{(a)}: To create the dataset analyzed via the \textsc{In Your Own Words} framework, we ask participants to describe their race, gender, and sexual orientation in free text. \textbf{(b)}: We convert the free-text responses into embedding vectors that capture their semantic meaning but are not readily interpretable.  \textbf{(c)}: We then use a sparse autoencoder (SAE) to extract more interpretable dimensions from the embeddings; each dimension captures a recurring pattern in how identity is expressed, such as references to cultural heritage, language, or childhood experiences. \textbf{(d)} To produce a text interpretation of each dimension, we prompt a large language model (LLM) to identify the common theme among the free-text responses that score highly along that dimension. \textbf{(e)}: We then use an LLM to annotate each free-text response for whether it contains each theme. Themes described in the illustration are abbreviated for space.}
    \label{fig:framework-overview}
\end{figure}

We summarize our approach, the \textsc{In Your Own Words} framework, in Figure \ref{fig:framework-overview}. First, we convert each free-text response to an \emph{embedding vector}---a  high-dimensional numeric representation capturing the meaning or context of the text (Figure \ref{fig:framework-overview}b) \cite{openai2024embeddings}. For example, the vectors for ``I like apples'' and ``I like oranges'' will be more similar to each other than to the vector for ``Trains are fast''. While such vectors are widely used in computational social science and more broadly \cite{arseniev2024theoretical, mediabiasdetector, petukhova2025text}, their dimensions do not correspond to human-interpretable concepts; rather, a single dimension may simultaneously mix together many concepts \cite{elhage2022toymodelssuperposition}. To extract interpretable dimensions, we apply a sparse autoencoder (SAE) to convert the original embedding vector into a vector whose dimensions are more interpretable (Figure \ref{fig:framework-overview}c). While SAEs have been successfully applied to text analysis tasks in past work \cite{o2024disentangling,hypothesaes,movva2025s}, they have not been previously applied to analyze free-text survey data, motivating the development of our framework.
We then produce a text interpretation of each dimension of the SAE representation by using a large language model (LLM) to identify the common theme among the free-text examples that score highly along that dimension (Figure \ref{fig:framework-overview}d). For example, for one SAE dimension, the free-text examples that score most highly mention ``traditional home-cooked meals passed down through the family'' and ``we regularly consume Chinese cuisine, including rice and dumplings''; the LLM identifies the common pattern in these examples---\emph{mentions cultural food as important to identity}---to produce the theme summarizing this dimension. These themes are the ones we use in our downstream analysis; we use an LLM to annotate all free-text responses for whether they contain each theme (Figure \ref{fig:framework-overview}e). At the end of this process, each free-text response is represented as a binary vector whose entries indicate whether it expresses each theme, creating a structured representation that can be easily used in quantitative analyses. 

As an example of how our method represents participants' free-text responses, consider the following response in our dataset: 

\begin{quote}
    ``I am Mexican-American, I grew up around Spanish speakers but did not learn to speak the language myself. We ate a mixture of Mexican and American food. I am also from the South so that has affected my dialect and food I eat as well."
\end{quote}

\noindent This text activates five of our computationally identified free-text themes: ``mentions specific regions or countries of ancestral origin", ``mentions being Mexican-American or Mexican", ``mentions the languages spoken by themselves or their family", ``mentions speaking Spanish or Spanish as part of family or cultural identity", and ``mentions traditional or cultural food as an important aspect of their identity".

We apply our approach separately to the race, gender, and sexual orientation free-text responses, since each identity axis exhibits different recurring patterns in how people describe themselves. For each set of responses, we identify $M=32$ free-text themes. 
We choose $M$ based on guidelines from prior work \cite{hypothesaes} and our own experiments. Empirically, $M=32$ produces a set of granular, nuanced free-text themes with limited redundancy; increasing $M$ (for example, to $M=64$) uncovers some additional patterns but also yields many duplicated themes.
We remove a small number of free-text themes where the LLM is unable to accurately summarize the SAE dimension, or where the theme captures minor writing details (e.g., “uses single-word or very short descriptions of race/ethnicity") rather than meaningful identity content. This process produces 26 free-text themes for race, 27 for gender, and 28 for sexual orientation (Tables \ref{tab:race_interpretation_fidelity}-\ref{tab:sexual_orientation_interpretation_fidelity}, respectively) which we use in our subsequent analysis. $\S$\ref{sec:methods-framework} provides further details on the computational framework. 

Several conceptual features of this approach are worth highlighting. First, unlike computational methods that classify individuals into a single cluster or group, our approach allows each response to reflect multiple themes at once, acknowledging that identities are often multifaceted and overlapping \cite{crenshaw2013demarginalizing}. Second, our approach does not require the themes to be pre-specified, allowing it to capture unexpected, context-specific nuances rather than relying on predefined hypotheses. Third, the approach is scalable: because it can be performed computationally and does not require hand-coding of the entire dataset, it can be applied even to datasets with millions of free-text responses. Finally, we emphasize that the goal of computationally learning themes from the data is to \emph{complement}, not replace, qualitative coding methods \cite{nelson2020computational}, and specifically to facilitate the exploratory analyses we discuss in greater detail below.

\subsubsection*{Validating the computationally identified free-text themes}

We perform extensive validations to ensure our computationally identified free-text themes faithfully describe the underlying free-text responses; these validations can also be performed when applying our approach to other free-text datasets. We summarize these validations here and provide additional details in $\S$\ref{sec:methods-framework}. We confirm two properties: first, when our framework identifies a theme in a free-text response, the theme actually occurs in that response; second, our framework does not \emph{miss} themes that do occur in free-text responses. These two properties are conceptually analogous to \emph{precision} and \emph{recall}, respectively---standard properties to verify when applying machine learning approaches \cite{powers2011evaluation}.

We perform several validations to assess \emph{precision}---namely, that when our framework identifies a theme in a response, that theme actually occurs. First, we randomly sample 24 of the computationally identified free-text themes (8 themes each for race, gender, and sexual orientation) and compare the LLM annotations of each theme to human annotations for 100 randomly sampled responses. For example, given a theme like ``mentions cultural food as important to identity", both the LLM and human annotators independently evaluate whether each free-text response expresses that theme, allowing us to measure the reliability of the LLM's annotations.
Across the 24 themes, the median inter-rater agreement is $\kappa=0.65$, indicating substantial agreement \cite{landis1977measurement, watkins2000interobserver} and supporting the reliability of the annotations for downstream analysis. 
This validation aligns with prior work demonstrating that LLMs can perform annotation tasks at levels comparable to human coders \cite{asirvatham2026gpt}. Agreement is highest for race ($\kappa=0.77$), followed by sexual orientation ($\kappa=0.67$) and gender ($\kappa=0.54$). We also validate that each computationally identified free-text theme accurately reflects its underlying SAE dimension, verifying that the themes faithfully summarize the patterns learned by the SAE; see $\S$\ref{sec:methods-framework-interp} for full details. Taken together, these validations confirm that when our framework identifies a theme in a free-text response, the theme actually occurs in that response. 

We also assess \emph{recall}---confirming, in other words, that the computationally identified free-text themes do not miss salient motifs in the free-text responses. We perform a manual review of 100 randomly sampled responses per identity axis (race, gender, and sexual orientation). We find that, while our framework does miss some motifs, they generally occur rarely (1-10\% of sampled responses). This is expected: because we constrain our method to identify only $M=32$ free-text themes, it prioritizes frequently-occurring motifs and therefore omits some low-frequency motifs. Manual validation suggests two main reasons motifs are missing. First, some are close variants of content already represented by our free-text themes, but expressed using adjacent labels or a different level of abstraction---for example, Hispanic/Latino identification, a missed motif, partially aligns with an existing free-text theme about speaking Spanish being part of cultural identity. Second, a smaller set of motifs reflect truly distinct content from those we capture with $M=32$, including motifs discussing discrimination (5\%) and racial privilege (5\%) for race; feminism (1\%) and fear or discomfort discussing gender (1\%) in gender; and dating (9\%) in sexual orientation.
Importantly, however, we show that many of these motifs can be captured simply by increasing the number of free-text themes our framework learns (e.g., to $M=64$); we opt against doing this in our main analysis because it also produces more semantically overlapping or split themes, highlighting a tradeoff between coverage and redundancy. 
These findings underscore the value of complementing our computationally identified free-text themes with human-in-the-loop qualitative validation: manual review clarifies how the SAE and LLM represent and interpret responses and can pinpoint motifs that do not surface at an initial choice of $M$. Different analytic use cases may motivate different choices of $M$, or incorporating themes learned at multiple values of $M$; our framework can flexibly accommodate all these choices. We report detailed examples and analyses of these patterns in Appendix $\S$\ref{supp:manual-coverage-val}.

\subsubsection*{Comparison to alternate computational methods}\label{sec:result-baseline-comp}

We compare the \textsc{In Your Own Words} framework to three alternate, widely-used methods for learning themes from free-text data: LDA \cite{blei2003latent}, a classical topic modeling approach that operates over bag-of-words representations; BERTopic \cite{grootendorst2022bertopic}, a modern topic modeling approach that performs unsupervised clustering on text embeddings; and TopicGPT \cite{pham-etal-2024-topicgpt}, a prompt-based topic modeling approach that uses LLMs to generate natural language topics directly from text. For clarity, we refer to the outputs of the baseline methods as ``themes'' rather than ``topics", following the terminology used in the \textsc{In Your Own Words} framework.
To enable a fair comparison, we tune the hyperparameters for each method so that it produces a similar number of themes ($M=32$) to ours (Appendix $\S$\ref{supp:baseline_hyperparams}).

We find that while each baseline method recovers some meaningful themes, none provides a stable or comprehensive representation of identity-related free-text responses. LDA produces many themes that either occur very rarely (e.g., applies to only one respondent) or consist of incoherent word lists, limiting their interpretability for short survey responses. BERTopic reliably recovers major identity category-specific themes but struggles to capture themes shared across multiple standardized categories and excludes a substantial fraction of responses from analysis. TopicGPT produces very different numbers of themes for different identity axes, collapsing most race and gender responses into a small number of broad themes while generating many rare and redundant themes for sexual orientation. In contrast, our framework yields a more consistent and interpretable set of free-text themes that recur across responses, represent the full corpus, and are better suited for systematic exploratory analysis. Detailed comparisons are provided in Appendix $\S$\ref{supp:baseline_results}.

\subsection{Analysis of free-text themes}\label{sec:present-themes}

We first describe the free-text themes themselves and show that many themes within each identity axis span multiple categories. 
We then demonstrate how these themes can facilitate three important tasks in survey research: suggesting new structured questions for future waves; capturing heterogeneity within standardized categories; and illuminating the discordance between self-identified and perceived identity. 

\begin{figure}[tbp]
    \centering
    \includegraphics[width=\linewidth]{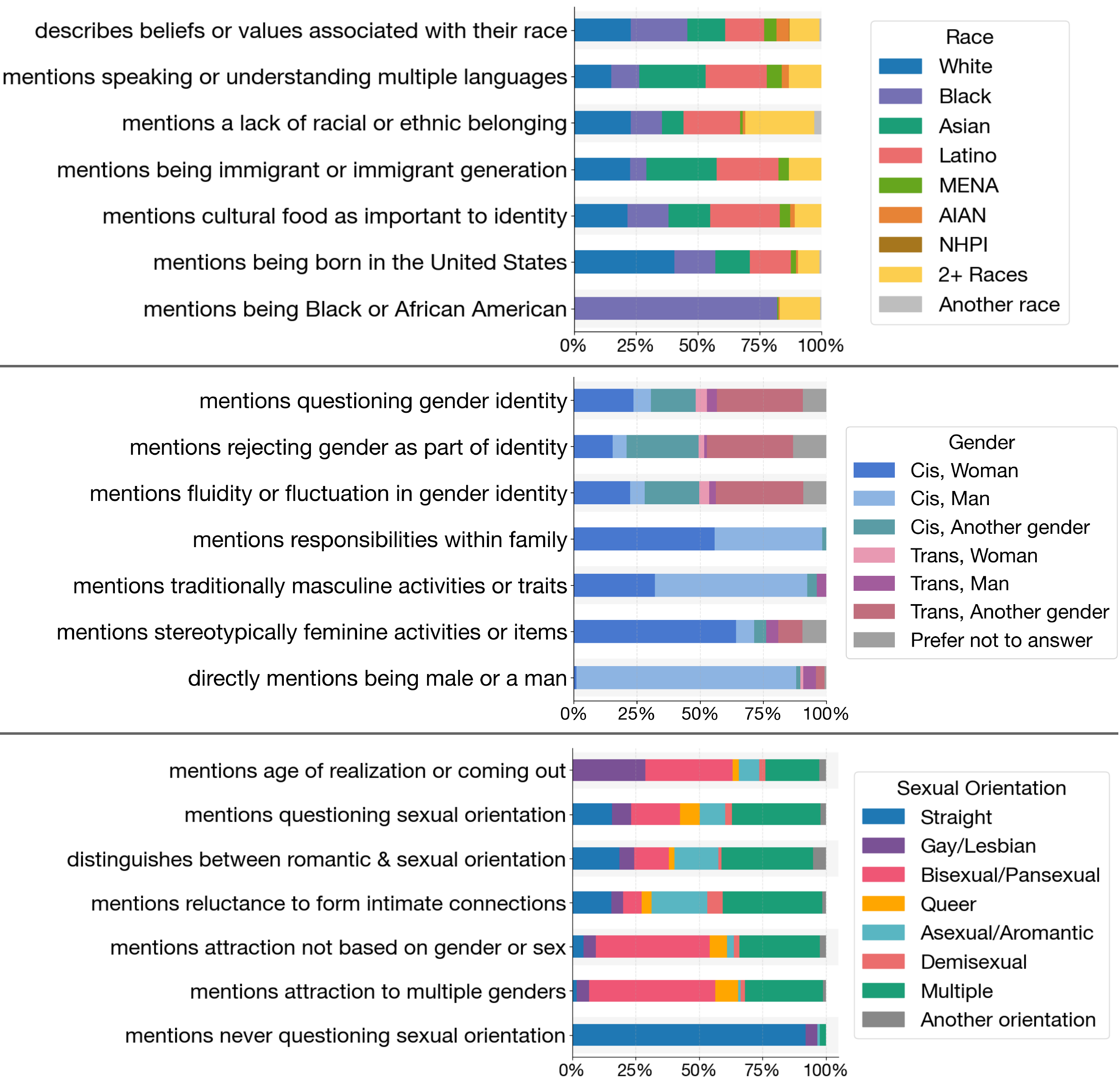}
    \caption{\textbf{Free-text themes produced by our computational framework}. 
    Each row shows one free-text theme, and colored bars indicate the standardized category of respondents whose free-text responses contain the theme. While some themes are predominantly associated with a single category (e.g., ``mentions never questioning sexual orientation'' is mostly expressed by straight respondents) many themes cut across categories within each identity axis. Very rare standardized categories (with fewer than 10 respondents) are excluded. For space, free-text themes are abbreviated, and only a subset of themes are shown; the full text of all themes is provided in Figures \ref{fig:race-themes-barchart}-\ref{fig:sexual-orientation-themes}. Section \ref{supp:category-details} describes the standardized categories in more detail.}
    \label{fig:all-themes}
\end{figure}

Figure \ref{fig:all-themes} presents a selection of the free-text themes for race, gender, and sexual orientation (Figures \ref{fig:race-themes-barchart}-\ref{fig:sexual-orientation-themes} plot the full set of themes). While some themes are expressed predominantly by members of a single standardized category---e.g., most respondents who express the ``mentions being a male in a direct and explicit manner'' theme are cisgender men---many themes cut across standardized categories rather than being expressed primarily by members of one category.
The cross-cutting race themes, for example, capture cultural and social dimensions not represented by standardized categories, such as shared values and traditions (e.g., “describes beliefs or values associated with their race” or “mentions cultural food as important to identity”), language (“mentions speaking or understanding multiple languages”), immigration (“mentions being an immigrant or immigrant generation”), and belonging (“mentions a lack of racial or ethnic belonging”). For gender and sexual orientation, we similarly observe many cross-cutting themes: for example, themes mentioning questioning or struggling with gender identity or sexual orientation and themes mentioning fluidity. We also observe a noticeable divide between the themes expressed by non-minority\footnote{For gender, we define non-minority respondents as cisgender men and cisgender women; for sexual orientation, we define non-minority respondents as straight participants; for race, we define non-minority respondents as White participants.} and minority respondents: for example, non-minority gender respondents were more likely to emphasize gender roles or responsibilities, while minority respondents were more likely to discuss fluidity in identity (both $p<0.001$). 

We assess how much new information these free-text themes provide, relative to standardized categories, by computing the proportion of variance in each theme explained by the categories ($R^2$). An $R^2$ of 1 means that whether a respondent expresses a theme can be perfectly captured by standardized categories, making the theme redundant; an $R^2$ of 0 means the categories do not capture the theme at all. While a few themes are well-explained by standardized categories (e.g., ``mentions being Black or African American” has an $R^2=0.77$), the median $R^2$ values across identity axes are low: 0.15 for race, 0.12 for gender, and 0.22 for sexual orientation, indicating that the themes are not well-captured by standardized categories.
These results remain similar when using alternate metrics designed for binary outcomes (Appendix $\S$\ref{supp:themes_r2_robust}). 

Collectively, these results show that free-text themes provide substantial additional information beyond standardized categories: many themes appear across multiple categories rather than mapping directly onto one, and categories account for only a small share of the variation in the themes. We now show that the additional information the themes provide has several practical applications in survey research. 

\subsubsection*{Free-text themes suggest additional structured questions}
\label{sec:results-structured-qs}

The \textsc{In Your Own Words} framework offers a data-driven approach to iteratively improve survey instruments by suggesting additional structured questions for future waves. This is particularly valuable for complex, multifaceted constructs like identity, where changes in lived experiences often outpace the evolution of standardized questions; for instance, the MENA (Middle Eastern or North African) category was proposed for the U.S. Census only after it became clear that existing categories failed to represent these identities \cite{ennis2024examining}. In exploratory contexts, such as pilot studies, or for recurring polls on dynamic topics where salient themes can shift over time (e.g., elections and voter sentiment) \cite{rothschild2024opportunities}, researchers often do not know \emph{a priori} which structured questions are most relevant. 

To validate that our free-text themes can suggest additional structured questions, we first verify that they can identify structured questions that past work has \emph{already} recognized as important. For race, themes related to nativity, such as ``mentions being first, second, or third generation American", suggest a structured question on nativity or immigration status---a critical and well-studied variable for understanding racialization \cite{kim1999racial, schachter2021intersecting}. For gender, themes like ``mentions enjoyment or participation in activities or items stereotypically associated with femininity" and ``mentions traditionally masculine activities or traits" highlight the distinction between gender identity and gendered traits or expression---concepts that have been operationalized by personality inventories like the BSRI \cite{bem1974measurement} or questions about appearance and mannerisms from surveys like the YRBS \cite{wylie2010socially}. For sexual orientation, several themes emphasize the dimension of \emph{attraction} (e.g., ``mentions attraction to more than two genders or a spectrum of genders"). This distinguishes attraction from sexual identity, validating research that argues sexual orientation is a multidimensional construct where desire, behavior, and identity do not always align \cite{mishel2019intersections}. These examples illustrate that our method can suggest structured questions that past work has indeed recognized as important. We note that none of the examples highlighted above were explicitly suggested by the prompt; rather, they emerged organically in respondents' answers.  

We also show, importantly, that our approach can suggest \emph{new} structured questions by highlighting dimensions of identity not commonly addressed in prior work. For instance, for race, the theme ``mentions feelings of not fully belonging or being out of place related to race/ethnicity" highlights a need for standardized questions that measure social integration and psychological belonging. While these concepts are recognized as key social determinants of health and well-being \cite{walton2011brief}, validated measures that capture these experiences are underdeveloped or still emerging \cite{lee2024ibelong}. For gender and sexual orientation, we find numerous themes describing identity as a dynamic process; specifically relating to fluidity (e.g., ``mentions fluidity or fluctuation in gender identity") and questioning (e.g., ``mentions uncertainty or questioning about their sexual orientation").  
Our findings challenge the common survey practice of treating identity as a static snapshot of current identity status, which may obscure this dynamic process of navigating identity \cite{saperstein2025recognizing}.
These themes suggest several concrete improvements to survey design. Longitudinal surveys can re-ask identity questions over time, distinguishing between current identity, past identity, and whether respondents have ever identified differently \cite{saperstein2025recognizing, pew2015multiracial_appendix}.
The ``questioning'' theme supports recommendations to add ``Questioning" as a formal answer option \cite{hughes2022guidance,russell2009teens}. 
Collectively, these results illustrate that our free-text themes both recover structured questions past work has recognized as important, and suggest new ones. 

\subsubsection*{Free-text themes reveal within-group heterogeneity}
\label{sec:results-explain-outcomes}

A known limitation of standardized categories is their tendency to conceal meaningful within-group heterogeneity \cite{read2021disaggregating, movva2023coarse}: for example, a single coarse race category (e.g., ``Asian") may group individuals with diverse lived experiences. Here, we investigate whether our free-text themes help illuminate this heterogeneity by explaining additional variation in key health and social measures, relative to using standardized categories alone.

Specifically, we test whether adding our free-text themes increases explained variance, as measured by adjusted $R^2$, relative to using standardized categories alone, in six commonly-studied outcomes: life satisfaction, physical health, mental health, income, identity importance (i.e., respondents' self-assessed importance of race, gender, or sexual orientation to their identity), and perceived discrimination \cite{conron2010population,verkuyten1989happiness,michalos2001ethnicity,yip2018ethnic,burrow2010racial}. For each outcome-identity pair, we compare two nested regression models: one \emph{base model} that uses only the standardized identity categories (i.e., race, gender, or sexual orientation) to explain variation in the outcome, and one \emph{full model} that additionally includes the free-text themes for that identity. We use an F-test to evaluate whether the inclusion of the free-text themes produces a statistically significant increase in explained variation, when correcting for the additional parameters added to the model, a standard approach \cite{allen1997testing, greene2018econometric}. $\S$\ref{sec:nestedf} provides further details on this regression analysis. 

Across all three identity axes, we find that free-text themes explain additional variation in several outcomes (after correcting for multiple hypothesis testing using the Benjamini-Hochberg procedure with $FDR=0.05$ \cite{benjamini1995controlling}).\footnote{All $p$-values reported in this section are Benjamini-Hochberg adjusted.} The free-text themes explain additional variation in mental and physical health, particularly for race ($1.5\times$ and $1.9\times$ improvement in adjusted $R^2$ compared to the base model, respectively, $p < 0.05$) and gender ($1.4\times$ and $1.7\times$ improvement, $p < 0.001$). Additionally, gender free-text themes increase adjusted $R^2$ for life satisfaction by $1.5\times$ ($p < 0.001$). 
The largest and most consistent gains emerge for identity importance, where adding free-text themes increases adjusted $R^2$ by 1.8$\times$ for gender ($p < 0.001$), 3.1$\times$ for sexual orientation ($p < 0.001$), and 1.1$\times$ for race ($p < 0.001$). Overall, these results show that free-text themes explain substantial additional variation in consequential health and social outcomes, highlighting important within-group heterogeneity concealed by standardized categories. Table \ref{tab:nested-f-results} reports full results for all outcome-identity pairs. 

Free-text themes can also highlight specific life experiences that account for this heterogeneity. We examine the free-text themes that are most strongly associated with outcomes after controlling for standardized categories (Figures \ref{fig:race-theme-predictiveness}-\ref{fig:sexual-orientation-theme-predictiveness}). For race, several themes correlate with worse mental or physical health, including mentioning feelings of not fully belonging, mentioning being born in the United States, or making explicit reference to parentage (e.g., “my mom is X and my dad is Y”), which multiracial respondents mention especially frequently. 
We also observe substantial within-group heterogeneity in race identity importance. Respondents who express pride, cultural connection, or values (e.g., ``mentions being Mexican-American or Mexican'', ``mentions pride in Black identity'') tend to report higher identity importance; respondents who express detachment from culture (``lacking cultural or ethnic identity or traditions”) tend to report lower identity importance.
Several patterns also warrant deeper exploration. Gender themes reflecting instability or ambiguity---such as ``fluidity and fluctuation” and ``questioning or struggling with identity'' show consistently negative associations with life satisfaction, mental health, physical health, and gender importance. Similarly, sexual orientation themes around questioning, avoiding labels, or hesitating to form relationships show negative associations with mental health and sexual orientation importance. We emphasize that these specific associations are exploratory and non-causal: our sample is small, and future work should investigate whether these associations generalize. However, all these associations are supported by past social science findings. Prior work linking social belonging to better health outcomes suggests that shared culture, social networks, and social capital may provide emotional support and help buffer stress---resources that may be less accessible to individuals who feel out of place \cite{house1988social, berkman1979social, pickett2008people}. Similarly, U.S.-born individuals, particularly in immigrant families, often exhibit worse health outcomes than their foreign-born counterparts, a pattern known as the ``unhealthy assimilation” effect \cite{antecol2006unhealthy, garcia2016converging}. Research on multiracial identity has also linked identity complexity and perceived marginalization to poorer health outcomes \cite{udry2003health, vora2024influence}. 

More broadly, these findings suggest that the \textsc{In Your Own Words} framework can identify nuance concealed within coarse identity groups. By surfacing themes like social belonging and identity fluidity, our approach allows researchers to generate new, testable hypotheses about \emph{how} and \emph{why} this within-group heterogeneity matters for key social and health outcomes. More generally, the framework identifies data-driven concepts that can then be rigorously tested with the development of formal survey measures and domain-specific theory.

\subsubsection*{Free-text themes describe discordance between different measures of identity}
\label{results:perception}

General surveys often omit questions about how people \emph{believe} their own identities are perceived, primarily due to practical constraints on survey length and a lack of established justification for inclusion of these questions \cite{national2022measuring, garbarski2023measurement}. But how people believe they are perceived by others fundamentally shapes their experiences of social recognition, belonging, and inequality \cite{kang2015multiple}. While existing measures (e.g., socially assigned race \cite{jones2008using} and street race \cite{lopez2018s}) can capture \emph{whether} self-described identity differs from perceived identity, they offer little insight into \emph{how and why} they differ. Existing gender-related instruments assess masculine or feminine presentation or perceived gender nonconformity \cite{bem1974measurement, wylie2010socially}, and sexual orientation measures typically focus on identity labels, attraction, or behavior \cite{klein1978klein, national2022measuring}. To illuminate what these measures miss, we apply the \textsc{In Your Own Words} framework to identify themes in free-text responses about how respondents believe their identity is perceived (\S\ref{sec:methods-perceive}). We focus on respondents who reported a difference between their self-described identity and how they believe their identity is perceived (64\% of responses for race, 39\% for gender, and 52\% for sexual orientation), as this subset highlights themes associated with perceived identity discordance \cite{jones2008using}.

Our approach identifies distinct cues through which respondents believe their race, gender, and sexual orientation are perceived by others. Respondents believe their race and gender are predominantly perceived through \emph{appearance-based} cues, driven by personal and physical presentation (as evidenced by themes like “mentions assumptions others make about ethnicity based on appearance or name” and “mentions physical attributes...when describing how others perceive their gender identity”). In contrast, respondents believe their sexual orientation is more often perceived through \emph{relational context}, particularly the gender of one's current or previous partners (e.g., ``mentions being perceived as straight due to being married to a man”)---suggesting that interpersonal and relationship history plays a more prominent role in sexual orientation perception than in race or gender perception.

Beyond identifying the cues people use to perceive identity, our themes also illuminate how respondents react to these perceptions (and misperceptions). For gender, themes surface active resistance to stereotypes (e.g., ``discusses the perception of women as weak and contrasts it with their own or women's actual strength or capabilities''). For sexual orientation, respondents with minority identities frequently described their identities as overlooked due to heteronormative expectations, causing their bisexual, pansexual, or asexual identities to go unrecognized \cite{bartholomay2023doing}. Many respondents also expressed uncertainty about how others perceive them across identity axes, reflecting the ambiguity that often accompanies perceived identity discordance. These experiences---resistance, erasure, and uncertainty---are rarely measured directly, even though they shape how individuals manage their identities, anticipate stigma, and navigate social situations \cite{goffman2009stigma, quinn2013concealable}, and point to specific opportunities for developing more nuanced survey items.
\section{Discussion}
\label{sec:discussion}

Free-text survey responses can provide rich information but are challenging to analyze systematically. In this work, we introduce \textsc{In Your Own Words}, a computational framework that identifies interpretable themes in free-text survey data and represents each survey response as a vector of the themes it contains, aiding downstream statistical analysis.
We apply our approach to a new dataset of free-text self-descriptions of race, gender, and sexual orientation, showing that it produces more interpretable themes than three widely-used computational text analysis methods and surfaces substantively meaningful patterns such as racial and cultural belonging, identity fluidity, and identity questioning. Our free-text themes provide information beyond standardized identity categories: many themes span multiple standardized categories within each identity axis, and categories account for only a small fraction of the explained variation in the themes. 
We further demonstrate that the information our free-text themes provide has practical applications in exploratory survey research: themes suggest new structured questions to add to future surveys, reveal heterogeneity within standardized categories, and illuminate mechanisms underlying discordance between self-identified and perceived identities. 

Like all computational text analysis methods, our framework has limitations. Our free-text themes may not perfectly capture all salient dimensions present in the data. The quality of the themes depends on hyperparameters including (1) the models used to produce the embeddings, interpret the dimensions, and annotate the responses; (2) the prompts used for interpretation and annotation; and (3) the total number of dimensions learned ($M$) and the number of dimensions in each response ($K$). Different settings of these hyperparameters yield different trade-offs between theme coverage (the range of concepts captured), granularity (how specific the themes are), and redundancy (whether multiple themes are very similar). 
To assess and mitigate these trade-offs, we perform many validations both to choose the aforementioned hyperparameters, and to assess the quality of the free-text themes ($\S$\ref{sec:methods-framework-interp}, Tables \ref{tab:race_interpretation_fidelity}-\ref{tab:sexual_orientation_interpretation_fidelity}, and Appendix $\S$\ref{supp:manual-coverage-val}): manual review of each theme's highest-activating responses to verify semantic agreement; interpretation fidelity scores to quantify how well themes recover the responses they summarize; inter-rater agreement with human annotators to assess annotation reliability; and manual coverage checks to flag salient patterns not captured by the free-text themes. Together, these validations indicate that the free-text themes generally align well with underlying patterns in the data, while also highlighting the importance of performing similar checks in new contexts and complementing computational analysis with qualitative review.

Our choice of $M$ illustrates this flexibility in practice. 
With $M=32$ dimensions, the SAE prioritizes frequent patterns, so themes that appear relatively rarely (for example, race-related references to discrimination or privilege) may be absorbed into themes that occur more frequently, rather than being highlighted independently. When finer-grained coverage is needed, increasing the dimensionality (e.g., to $M=64$) surfaces these themes more explicitly, allowing researchers to tune the framework to suit their exploratory goals.
In general, we emphasize that the \textsc{In Your Own Words} framework is a tool to support exploratory analysis:
it is designed to surface interpretable patterns at scale that can guide deeper analysis, hypothesis generation, and survey refinement. 

Another consideration in interpreting our empirical results is that they reflect the data collection context. The free-text responses were collected from a non-representative sample of U.S.-based participants; thus, some computationally identified free-text themes from this study may differ in other cultural or national settings. However, this limitation reflects the data rather than the framework itself; indeed, applying our framework to free-text identity data collected in other contexts represents an interesting direction for future work. Additionally, in our free-text survey prompt ($\S$\ref{sec:methods-freetext}), we suggested identity-related topics participants might discuss, since we found empirically this was useful to encourage rich and detailed responses. It is possible that the suggested topics influenced the free-text responses participants provided. However, the responses surfaced many meaningful free-text themes that were not mentioned by the prompt (e.g., ``mentions being immigrant or immigrant generation''; ``mentions fluidity or fluctuation in gender identity"; ``mentions discomfort or rejection of labels for their sexual orientation"), suggesting that participants articulated genuine and heterogeneous experiences. As in all survey research, prompt design should be viewed as a flexible methodological choice; future work could apply our general framework to free-text identity data collected using different question prompts, which might surface different themes. 

Looking ahead, the \textsc{In Your Own Words} framework offers substantial opportunities to inform the design, iteration, and analysis of surveys. The framework can be applied to analyze responses \emph{after} a survey is completed, and can also support an iterative survey design process in which free-text responses are used to identify relevant features, guide the development of new structured questions, and evaluate whether existing measures adequately capture respondents’ lived experiences. This is particularly useful in settings where key concepts are evolving, theory is underdeveloped, or standardized measures lag behind cultural shifts in terminology or social norms. 
Importantly, we view the framework as a complement to qualitative inquiry rather than a replacement: it can surface interpretable patterns at scale that help prioritize where deeper reading, coding, or interviews are most needed, and can support mixed-method workflows for validating and refining constructs \cite{nelson2020computational}. While we demonstrate the benefits of our approach in the context of identity, the computational method could be applied much more generally to free-text survey questions in many domains (e.g., workplace climate, healthcare, or public opinion). Future work could extend this framework to incorporate free-text responses into theory-building and iterative survey design, enabling surveys that remain interpretable, scalable, and responsive to how people describe their experiences over time.
\renewcommand{\thesubsection}{M.\arabic{subsection}} 
\renewcommand{\thesubsubsection}{M.\arabic{subsection}.\arabic{subsubsection}} 

\section*{Methods}\label{sec:methods}

Our survey was considered exempt from full institutional review board (IRB) review by the Cornell Office of Research Integrity and Assurance (protocol number IRB0148894). Before the survey, all participants provided electronic informed consent. 

\subsection{Sample recruitment}
\label{sec:methods-sample}

Participants were recruited between September 29, 2024 and January 8, 2025 through Prolific, an online survey platform \cite{prolific2025}. We filtered for U.S.-based, English-speaking participants ages 18 years and older. To ensure representation across identity groups, we used Prolific’s quota sampling feature,\footnote{\href{https://researcher-help.prolific.com/en/article/bc8a0e}{https://researcher-help.prolific.com/en/article/bc8a0e}} which allows researchers to set target proportions for individual demographic variables (e.g., race, gender, or sexual orientation). 
Our full sample consists of three separate subsamples: one upsampling minority racial identities (approximately 400 participants); one upsampling minority gender identities (approximately 300 participants); and one upsampling minority sexual orientation identities (approximately 300 participants). We combine these subsamples into a single dataset of 1,004 participants for analysis. Our aim is not to obtain a perfectly representative sample of these subpopulations, but rather to recruit a cohort that captures a diverse set of self-described free-text identities on which to assess our computational approach. Prolific, like other non-probability panels, does not provide a fully representative sample \cite{douglas2023data}. Demographic characteristics of participants are available in Table \ref{tab:demographic_breakdown}.

For race, we oversampled for minorities by recruiting participants who identify as Black/African American (20\% of the sample), Latino/Hispanic (20\%), Asian (21\%, split into equally-sized subgroups from East, South, and Southeast Asia), White (15\%), Middle Eastern (10\%), Native American or Alaskan Native (10\%), and other (5\%). For sexual orientation, participants are recruited evenly across five categories: heterosexual (20\%), homosexual (20\%), bisexual (20\%), asexual (20\%), and other (20\%). 

Our sampling procedure for gender was constrained by the fact that the Prolific platform does not support joint quota sampling across multiple different variables simultaneously. This means we could not simultaneously set quotas for participants' transgender status and their gender identity (man, woman, or non-binary). We therefore drew two samples of 150 participants each: (1) a balanced sample of transgender (50\%) and non-transgender (50\%) participants and (2) an oversample of nonbinary participants (80\%), with the remaining 20\% evenly split between men and women. We used this two-step approach to ensure we oversampled both nonbinary participants and transgender participants. 

Prolific's demographic response options differ slightly from those used in the multiple-choice demographic questions in our survey. For instance, Prolific offers participants five main sexual orientation options, along with a ``Rather not say" option. In contrast, our survey implemented \citeauthor{hughes2022guidance} (\citeyear{hughes2022guidance})'s recommended inclusive sexual orientation, which offers a broader set of options such as Demisexual, Queer, and Sexually fluid. Throughout our analysis, we used the identity categories collected through our own survey instrument which are outlined below in $\S$\ref{sec:survey_instrument}.

\subsection{Survey instrument}
\label{sec:survey_instrument}

We now describe the survey questions. The full survey instrument and response data are available upon request. Details are provided at \href{\projecturl}{\projecturl}.

\subsubsection*{Free-text self-described identity questions}
\label{sec:methods-freetext}

Each participant answered three free-text questions asking them to describe their race, gender, and sexual orientation in their own words. The main prompt for each question read: “In at least 2-3 sentences, how would you describe your \texttt{[identity]}?” 
Each question was followed by a short italicized sub-prompt that provided examples of topics participants might discuss, such as cultural practices, life experiences, or identity-related milestones. We included these sub-prompts to encourage more detailed responses and mitigate the tendency toward brief, low-effort answers in online open-ended formats \cite{aspinall2012answer}, as prior work shows that such prompts can improve response length and quality \cite{smyth2009open}.
For race, the sub-prompt read: “For example, you could discuss specific traditions or customs, cultural practices or norms, languages or dialects spoken in your family, or ways race and/or ethnicity has influenced your life experiences.” The gender identity and sexual orientation questions followed the same format, with tailored examples suggesting that participants might describe milestones, social perceptions, or aspects of daily life related to that dimension of identity.
The sub-prompts were grounded in established frameworks for each identity axis: the race prompt emphasized cultural norms, practices, language, and shared experiences, consistent with frameworks that define racial and ethnic identity through shared behaviors and values \cite{phinney1996we}, while the gender and sexual orientation prompts focused on narrative identity and development, aligning with life-story models of identity formation \cite{mcadams2001psychology}. Table \ref{tab:survey_questions} provides full question wordings. As with any guided prompt, these sub-prompts may have shaped the range of responses we received---for instance, themes related to language, ancestry, and immigration emerged frequently across identity axes. Our computational framework is flexible; researchers who wish to focus on different identity-related themes could apply the same methodology using alternative prompts. 

\begin{table}[t]
\centering
\caption{Race, Gender, and Sexual Orientation Free-Text Questions}
\label{tab:survey_questions}
\renewcommand{\arraystretch}{1.5}{
\begin{tabular}{p{0.9\linewidth}}
\hline
\textbf{Free-Text Question} \\
\hline
{\normalsize In at least 2--3 sentences, how would you describe your \texttt{race}?} 
\emph{For example, you could discuss specific traditions or customs, cultural practices or norms, languages or dialects spoken in your family, or ways race and/or ethnicity has influenced your life experiences.} \\[0.5em]
\hline
{\normalsize In at least 2--3 sentences, how would you describe your \texttt{gender}?} 
\emph{For example, you could discuss ways your gender identity influences your life or significant milestones or moments in your life related to your gender identity.} \\[0.5em]
\hline
{\normalsize In at least 2--3 sentences, how would you describe your \texttt{sexual orientation}?} 
\emph{For example, you could discuss ways your sexual orientation influences your life, significant milestones or moments in your life related to your sexual orientation, or any differences between your romantic and sexual identities.} \\[0.5em]
\hline
\end{tabular}}
\end{table}

\subsubsection*{Free-text perceived identity questions}
\label{sec:methods-perceive}

Each participant also answered three free-text questions across race, gender, and sexual orientation asking them: ``How does your self-identified \texttt{[identity]} compare to how you believe others perceive your \texttt{[identity]}?" This question was immediately followed by a multiple-choice question that asked: ``How similar is your self-identified \texttt{[identity]} to how you believe others perceive your \texttt{[identity]}?" The options presented were: ``Mostly the same", ``Somewhat the same and somewhat different", ``Mostly different", ``Completely different", and ``Unsure". To focus our analysis on instances of perceived identity discordance, we excluded free-text responses from participants who selected ``Mostly the same" \cite{jones2008using}.

\subsubsection*{Multiple-choice identity questions}
\label{sec:methods-categories}

In addition to the free-text identity questions, each participant selected from detailed multiple-choice options describing their race, gender, and sexual orientation. The demographic breakdown of participants is shown in Table \ref{tab:demographic_breakdown}. 
For race, each participant could select multiple categories from the following set: ``American Indian or Alaska Native'', ``Asian'', ``Black or African American'', ``Hispanic or Latino'', ``Middle Eastern or North African'', ``Native Hawaiian or Pacific Islander'', ``White'', and ``Some Other Race''. For gender, each participant first selected their current gender identity (``Man'', ``Woman'', or ``Some other way''), then also indicated whether they identify as transgender by selecting ``Yes'', ``No'', or ``Prefer not to answer''.
For sexual orientation, each participant could select one or more of the following: ``Asexual or aromantic'', ``Bisexual'', ``Demisexual'', ``Gay'', ``Lesbian'', ``Pansexual'', ``Queer'', ``Questioning'', ``Sexually fluid'', ``Straight or heterosexual'', ``Other sexual identity or orientation (please specify)'', and ``Prefer not to answer''.
In Figures \ref{fig:all-themes} and \ref{fig:race-themes-barchart}-\ref{fig:sexual-orientation-themes}, we visualize the concordance between standardized identity categories and free-text themes. To construct these visualizations, we create mutually exclusive identity categories from the original raw responses, which allow participants to select multiple categories; the details of this procedure are described in Appendix $\S$\ref{sec:mutually-exclusive-cat}. Aside from these figures and one supplementary analysis (Appendix $\S$\ref{supp:minority_share_more}), all other analyses use the original multiple-choice data without regrouping them into mutually exclusive categories.

\subsubsection*{Self-reported health and social outcomes}\label{sec:methods-outcomes}

After completing the free-text identity and perceived identity questions, participants answered questions about six health and social outcomes. These measures are used in the analysis described above of whether free-text identity themes explain additional variance relative to using only standardized identity categories. All question wordings were derived from prior surveys \cite{kff2023_consumer_experiences_health_insurance,kilpatrick1960self,gonzalez2024kff,pew2025_discrimination_questionnaire,williams1997racial}. 
\medskip

\noindent \emph{Physical and mental health.} 
Participants rated their physical health and mental health/emotional well-being using single-item self-assessments on 5-point scales (``Poor", ``Fair", ``Good", ``Very Good", and ``Excellent"). Such self-rated health items are standard in public health research and public opinion polling \cite{kff2023_consumer_experiences_health_insurance}. 
\medskip

\noindent \emph{Life satisfaction.} 
To assess life satisfaction, we used the Cantril Self-Anchoring Ladder, a 0-10 scale in which participants indicate where they currently stand between the `worst possible life' (0) and the `best possible life' (10) \cite{kilpatrick1960self}. This scale is a widely used measure of global well-being in large-scale population surveys (e.g., Gallup World Poll) \cite{tortora2010gallup}.
\medskip

\noindent \emph{Identity importance.}
To assess the importance of participants' different identities, we prompted participants with \emph{How important are the following aspects to how you think about yourself?} Responses are measured on a 5-point Likert scale ranging from ``Not at all important" (1) to ``Extremely important" (5) \cite{gonzalez2024kff}. 
\medskip

\noindent \emph{Perceived discrimination (group-level).}
Respondents rated how much discrimination they believed exists in society today against several race, gender, and sexual orientation groups (e.g., Hispanic people, transgender people, or women) using a four-point scale from ``None at all" to ``A lot". These items follow formats used in national surveys such as the Pew Research Center’s discrimination and attitudes modules \cite{pew2025_discrimination_questionnaire}.
\medskip

\noindent \emph{Everyday discrimination (individual-level).}
To measure perceived discrimination, we used a version of the Everyday Discrimination Scale \cite{williams1997racial}. Respondents reported how frequently they experienced discrimination in their daily lives; specifically, the questions focused on events such as being treated with less respect, receiving poorer service, or being insulted, using a 6-point frequency scale ranging from `Never' (1) to `Almost everyday' (6). Participants then selected the perceived reasons for these negative experiences (e.g., race, gender, age, religion, sexual orientation, skin tone).
\medskip

\subsection*{Additional demographic variables} 

Finally, respondents answered an additional set of multiple-choice questions capturing demographic and socioeconomic characteristics including age, religion, political affiliation, income, education, employment status, nativity, years lived in the United States, and citizenship status. Survey items follow standard wording from major national polls, including Gallup and the Pew Research Center. 

\subsection{Computational framework}
\label{sec:methods-framework}

Here, we provide additional details about the computational framework summarized in $\S$\ref{results:computational-framework}.

\subsubsection*{SAE training}
\label{sec:methods-framework-sae}

We train SAEs on semantic embeddings of the free-text responses \cite{gao2024scalingevaluatingsparseautoencoders}. As described in \S \ref{results:computational-framework}, SAEs convert the original high-dimensional embedding vector, whose components are not interpretable, into a lower-dimensional vector whose components capture interpretable themes (Figure \ref{fig:framework-overview}, $\S$\ref{sec:sae_themes}): for example, one dimension of the SAE vector might correspond to responses describing experiences of immigration. 

We first encode each free-text response using OpenAI’s \texttt{text-embedding-3-large} model \cite{openai2024embeddings}, producing a 3,072-dimensional vector representation. We then train a SAE on these high-dimensional embeddings to produce lower-dimensional, interpretable representations. SAEs are trained to accurately reconstruct the original embeddings from a learned low-dimensional latent representation that is constrained to be \emph{sparse}, encouraging the model to capture the information in the original embeddings using a small number of active components; empirically, this has been found to yield a low-dimensional representation whose components align with distinct, human-interpretable concepts present in the data~\cite{cunningham2023sparse,bricken2023towards,templeton2024scaling,hypothesaes, o2024disentangling}. Our SAE uses a latent representation with dimension $M=32$; we enforce sparsity using a \texttt{Top-K} operator that sets all activations besides the $K=4$ largest activations to zero~\cite{hypothesaes}. These hyperparameter values are chosen based on advice from past work for datasets of our size \cite{hypothesaes}, and empirically yield interpretable and non-redundant dimensions on our dataset. Models are trained using the HypotheSAEs library.\footnote{\href{https://github.com/rmovva/HypotheSAEs}{https://github.com/rmovva/HypotheSAEs}}

\subsubsection*{LLM Interpretation}
\label{sec:methods-framework-interp}

The goal of the LLM interpretation step is to produce a natural language description of the theme each SAE dimension encodes. We use the \texttt{GPT-4o} model to generate candidate descriptions for each SAE dimension by asking it to compare (1) ten free-text responses with the most positive scores along that dimension to (2) ten randomly-sampled free-text responses with zero activations along that dimension. We prompt the model to identify a feature that is present in the positive samples but absent in the negative samples. We repeat this process three times (temperature set to 0.7), yielding three candidate natural language interpretations for each dimension. Each candidate description takes the form of a short, human-interpretable theme (e.g., ``mentions cultural food as important to identity"). The exact prompt is shown in Figure \ref{fig:neuron-interpretation-prompt}. 

Following \citeauthor{hypothesaes} (\citeyear{hypothesaes}), we select the final interpretation for each SAE dimension as the candidate with the highest \emph{fidelity}. Fidelity is defined as follows. For a given candidate interpretation, we sample 100 responses with high activation along the SAE dimension and 100 responses with zero activation, treating these as true positives and true negatives, respectively. We then prompt an LLM (\texttt{GPT-4o-mini}) to annotate whether each response expresses the candidate interpretation, using the prompt shown in Figure \ref{fig:annotate-prompt}. We define a candidate interpretation's fidelity as the F1 score of this classifier. Intuitively, fidelity measures how well the candidate interpretation discriminates between the responses that highly activate an SAE dimension and those that do not.  

We perform several validations to verify that the LLM interpretations accurately reflect the underlying SAE dimensions. In particular, we manually assess whether each selected interpretation provides a reasonable summary of the responses that most highly activate the dimension. For each dimension, we manually review the ten responses with the highest positive activations and ten randomly sampled responses with zero activations---similar to the information provided to the LLM during the interpretation step---and assign a binary label denoting whether the proposed interpretation accurately describes the positive activation responses. For example, the SAE dimension interpreted as ``mentions cultural food as important to identity” is most highly activated by responses containing text such as ``traditional home-cooked meals passed down through the family” and ``we regularly consume Chinese cuisine, including rice and dumplings"; in contrast, responses that do not activate this dimension do not reference food or cultural practices. Based on this comparison, we label the interpretation as accurate. We conduct this assessment for a random subset of eight dimensions for each identity axis (race, gender, and sexual orientation); in 92\% (22/24) of cases, the interpretations provide accurate summaries of their highest-activating responses. 
Furthermore, the median fidelity (F1 score) across the selected interpretations is 0.78, indicating good agreement \cite{hypothesaes}. As a final protection against inaccurate interpretations, we remove a small number of themes with F1 scores below 0.50 (since these interpretations may not accurately reflect the original data). Collectively, these validations and filtering steps ensure that our interpretations accurately match the underlying SAE dimensions. We also remove themes that capture superficial writing details (e.g., ``uses single-word or very short descriptions of race/ethnicity”) rather than meaningful information about identity. We retain 26 themes for race, 27 for gender, and 28 for sexual orientation (Tables \ref{tab:race_interpretation_fidelity}-\ref{tab:sexual_orientation_interpretation_fidelity}). 

To assess whether the SAE dimensions adequately capture the range of themes expressed in the free-text responses, we conduct a qualitative coverage validation. We randomly sample 100 free-text responses drawn across race, gender, and sexual orientation questions. For each response, a human annotator assesses whether it contains any salient identity-related themes not represented among the 32 LLM-generated themes corresponding to the SAE dimensions. Responses expressing such themes are recorded as potential gaps. This validation is designed to assess whether our computational framework systematically overlooks important identity-related content in the data. We find that the themes our framework misses occur rarely (as expected, because with $M=32$ our framework prioritizes commonly-occurring themes) and that many of these themes can be recovered simply by increasing the value of $M$ (at the cost of some additional redundancy in the learned themes). Appendix $\S$\ref{supp:manual-coverage-val} fully describes these results. 

\subsubsection*{LLM Annotation}
\label{sec:methods-framework-annotate}

The goal of this step is to annotate each free-text response for whether it contains each theme.
We use \texttt{GPT-4.1-mini} for this task, following prior work demonstrating that LLMs can serve as reliable annotators in qualitative coding and related computational social science settings \cite{than2025updating, asirvatham2026gpt}. 
The LLM is prompted separately for each theme-response pair and outputs a binary label indicating whether the theme is present (1) or absent (0) in the response. The annotation prompt is identical to the one used to measure fidelity in the LLM interpretation step (Figure \ref{fig:annotate-prompt}).
This procedure creates a binary matrix whose rows are free-text responses, whose columns are themes, and whose entries indicate whether a given response displays a given theme. This matrix is the final structured output of the computational framework and is used in the downstream quantitative analyses.

We select the LLM and annotation prompt by comparing several candidate models' outputs to human annotations on a randomly sampled subset of responses; we randomly select eight themes for each identity axis (race, gender, and sexual orientation; 24 themes total) and, for each theme, randomly sample 100 free-text responses from the full dataset. Human annotations are produced independently by a trained researcher using the same annotation prompt, with the annotator blinded to the LLM's annotation.  
We evaluate annotation agreement using the following two metrics. First, we measure interrater agreement between each model and a human annotator using Cohen’s Kappa ($\kappa$), summarized as the median $\kappa$ across themes and identity axes. Second, we assess whether models systematically over- or under-annotate theme presence by comparing the proportion of responses labeled as positive by each model to the corresponding proportion labeled by the human annotator. We find that a \texttt{GPT-4.1-mini} and identity-specific annotation prompt configuration performs best overall; we therefore use it for all annotation tasks (see Table \ref{tab:annotation-model-comp} for full comparison results).

\subsection{Regression analysis of self-reported health and social outcomes}
\label{sec:nestedf}

We conduct linear regressions to assess whether free-text identity themes provide additional explanatory power for self-reported health and social outcomes beyond standardized identity categories. We consider six outcomes (described in \S\ref{sec:methods-outcomes}): mental health, physical health, life satisfaction, everyday discrimination, income, and identity importance. Because these outcomes are measured on different scales (e.g., life satisfaction on a 0-10 scale and mental health on a 5-point ordinal scale), all outcome variables ($Y$) are z-scored prior to analysis---that is, each outcome is transformed to have mean zero and standard deviation one within the analytic sample. This normalization is a standard transformation \cite{angrist2009mostly} that allows regression coefficients to be interpreted in standard deviation units, making them comparable across outcomes. Categorical outcomes are first mapped to ordered numeric scales prior to standardization. All independent variables are binary indicators and are not rescaled.

For each identity axis (race, gender, and sexual orientation) and each outcome, we perform a nested F-test comparing (1) a base model predicting the outcome using only standardized identity categories as covariates with (2) a full model that also includes free-text themes. 

The base model is described by the following equation: 

\begin{equation}
Y_i = \beta_0 + \sum_j \beta_j D_{ij} + \varepsilon_i,
\end{equation}

\noindent where $Y_i$ denotes the normalized outcome for respondent $i$, $\beta_0$ is an intercept term, $D_{ij}$ is a binary indicator equal to 1 if respondent $i$ selected standardized category $j$ for the relevant identity axis, $\beta_j$ is the regression coefficient for that category $j$, and $\epsilon_i$ is a noise term. For example, in the race specification, $D$ includes indicators for White, Black or African American, American Indian or Alaska Native, Asian, Middle Eastern or North African, Native Hawaiian or Other Pacific Islander, Hispanic or Latino, and Some Other Race. For race and sexual orientation, respondents could select multiple categories; for gender, categories are mutually exclusive (and thus we omit the reference group ``Cisgender Man" to avoid perfect collinearity, as is standard \cite{greene2018econometric}).

The full model augments the base specification with indicators for free-text identity themes:

\begin{equation}
Y_i = \beta_0 + \sum_j \beta_j D_{ij} + \sum_k \gamma_k Z_{ik} + \varepsilon_i,
\end{equation}

\noindent where $Z_{ik}$ is a binary indicator equal to 1 if respondent $i$’s free-text response activates theme $k$, and $\gamma_k$ is the regression coefficient on theme $k$. 

Our nested F-test evaluates whether the inclusion of free-text themes in the full model significantly improves model fit relative to the base model. All F-test $p$-values are corrected for multiple hypothesis testing using the Benjamini-Hochberg procedure with false discovery rate set to $FDR=0.05$ \cite{benjamini1995controlling}. Table \ref{tab:nested-f-results} reports adjusted $R^2$ values for the full and base models. 

To better understand which \emph{specific} free-text themes are associated with life outcomes, for each identity-outcome pair where adding free-text themes significantly improved model fit in the nested F-test, we estimate follow-up regression models that include each theme individually, alongside controls for the standardized identity categories for the corresponding identity axis. That is, for each free-text theme $k$, we estimate the following model: 

\begin{equation}
Y_i = \beta_0 + \sum_j \beta_j D_{ij} + \gamma_k Z_{ik} + \varepsilon_i,
\end{equation}

This approach helps clarify which individual themes drive the improvement in model fit observed in the nested F-tests. For example, we estimate a model where the outcome is self-reported mental health and the covariates include a binary indicator for whether the free-text response mentions ``feelings of not belonging”, together with indicators for the respondent’s standardized race categories. We estimate models for each theme individually because themes may be collinear, complicating interpretation of multivariate models. We report point estimates and 95\% confidence intervals (uncorrected for multiple hypotheses) for these models in Figures \ref{fig:race-theme-predictiveness}-\ref{fig:sexual-orientation-theme-predictiveness}. Due to the large number of themes, these analyses should be viewed as exploratory, as discussed in the main text.
\backmatter
\appendix
\makeatletter
\renewcommand{\@seccntformat}[1]{Appendix~\csname the#1\endcsname\quad}
\makeatother
\newpage

\renewcommand{\thefigure}{S\arabic{figure}}
\setcounter{figure}{0}
\makeatletter
\renewcommand{\fnum@figure}{\textbf{Fig.~\thefigure}}
\makeatother

\renewcommand{\thetable}{S\arabic{table}}
\setcounter{table}{0}
\makeatletter
\renewcommand{\fnum@table}{\textbf{Table~\thetable}}
\makeatother

\renewcommand{\thesubsection}{\thesection.\arabic{subsection}}
\renewcommand{\thesubsubsection}{\thesection.\arabic{subsection}.\arabic{subsubsection}} 

\section{Distribution of responses to the survey}
\label{supp:word_distribution}

\begin{figure}[bh!]
    \centering
    \includegraphics[width=\linewidth]{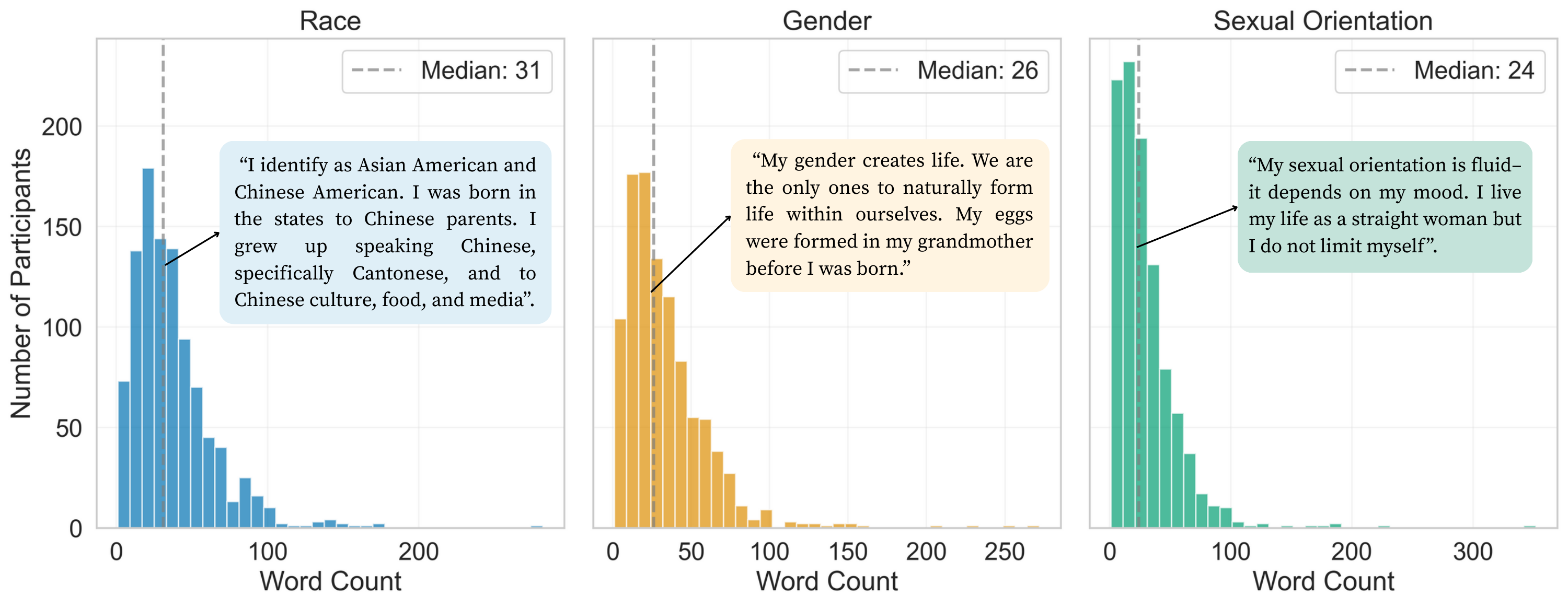}
    \caption{\textbf{Self-described identity: Distribution of word counts in participants’ free-text responses about their race, gender, and sexual orientation.} Dashed lines indicate median word count for each identity axis. Annotated quotes highlight the richness in information the median respondent is able to provide.}
    \label{fig:word_count_distribution}
\end{figure}

\begin{figure}[bh!]
    \centering
    \includegraphics[width=\linewidth]{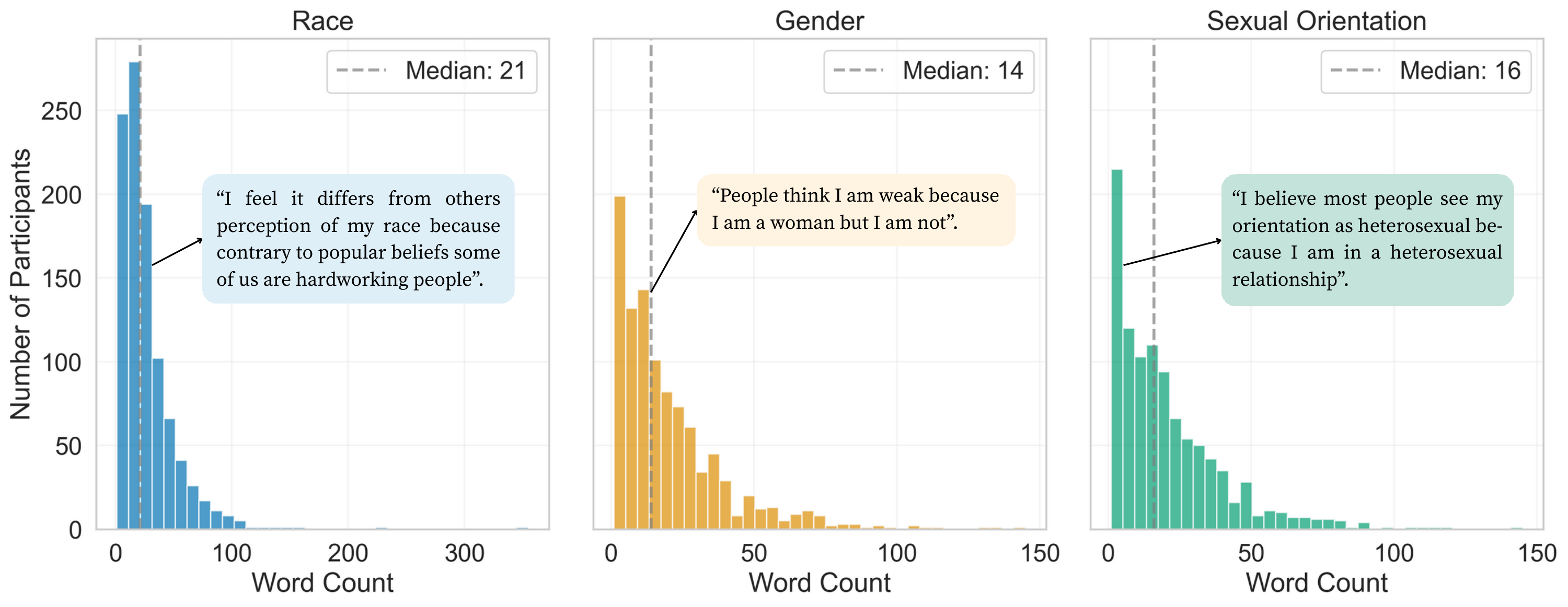}
    \caption{\textbf{Perceived identity: Distribution of word counts in participants’ free-text responses about their perceived race, gender, and sexual orientation.} Dashed lines indicate median word count for each identity axis. Annotated quotes highlight the richness in information the median respondent is able to provide.}
    \label{fig:perception_word_count_distribution}
\end{figure}

\noindent \textbf{Word count distribution:} Most participants wrote relatively concise responses to the race, gender, and sexual orientation free-text responses, as shown in Figure \ref{fig:word_count_distribution}. The median word count was 31 for race, 26 for gender, and 24 for sexual orientation. While a few individuals wrote longer responses (maximum: 282 for race, 272 for gender, and 352 for sexual orientation), the vast majority of responses were under 50 words, corresponding to the 75-85th percentile across all three axes. These distributions are similarly shaped---right-skewed with a long tail---and suggest that, despite the 2-3 sentence prompt, most participants did not write extensive narratives. 
These patterns, alongside our themes, underscore that our approach is operationalizable on responses of varying lengths. Figure \ref{fig:perception_word_count_distribution} presents word counts for the perceived identity free-text responses, which tended to be shorter than the self-described identity responses. The median word count was 21 for race, 14 for gender, and 16 for sexual orientation, again suggesting that participants in our sample tended to have more to say about their race than their gender and sexual orientation.

\noindent \textbf{Demographic distribution:} We used quota sampling on Prolific to ensure broad demographic coverage across race, gender, and sexual orientation, as described in $\S\ref{sec:methods-sample}$. While not nationally representative, the sample was constructed to oversample minority identities and capture diversity in identity expression. Table \ref{tab:demographic_breakdown} describes the marginal categories represented across the full sample.

\begin{table}[htbp]
\centering
\caption{\textbf{Demographic breakdown of the full sample (N = 1,004).} Percentages indicate the share of participants who selected each identity category. Because race and sexual orientation were multiselect items, totals for these categories exceed 100\%.}
\label{tab:demographic_breakdown}
\begin{tabular}{lc}
\toprule
Category & Multiselect (\%) \\
\midrule
\textbf{Race and/or Ethnicity} & \\
American Indian or Alaska Native & 6.8\% \\
Asian & 15.4\% \\
Black or African American & 22.2\% \\
Hispanic or Latino & 16.8\% \\
Middle Eastern or North African & 4.4\% \\
Native Hawaiian or Pacific Islander & 1.0\% \\
Some Other Race & 1.4\% \\
White & 53.8\% \\
\addlinespace
\textbf{Gender} & \\
Cisgender, Man & 28.6\% \\
Cisgender, Woman & 49.5\% \\
Cisgender, Another gender & 5.4\% \\
Transgender, Man & 2.1\% \\
Transgender, Woman & 1.8\% \\
Transgender, Another gender & 9.8\% \\
Prefer not to answer & 2.9\% \\
\addlinespace
\textbf{Sexual Orientation} & \\
Asexual or aromantic & 11.8\% \\
Bisexual & 19.2\% \\
Demisexual & 5.3\% \\
Gay & 5.8\% \\
Lesbian & 6.7\% \\
Other sexual identity or orientation & 2.2\% \\
Pansexual & 9.9\% \\
Prefer not to answer & 0.6\% \\
Queer & 12.7\% \\
Questioning & 3.2\% \\
Sexually fluid & 2.6\% \\
Straight or heterosexual & 52.3\% \\
\bottomrule
\end{tabular}
\end{table}

\newpage
\section{Additional details on `Most respondents, especially minorities, believe that their free-text responses
add important context'}
\label{supp:minority_share_more}

\begin{figure}[ht!]
    \centering
    \includegraphics[width=\linewidth]{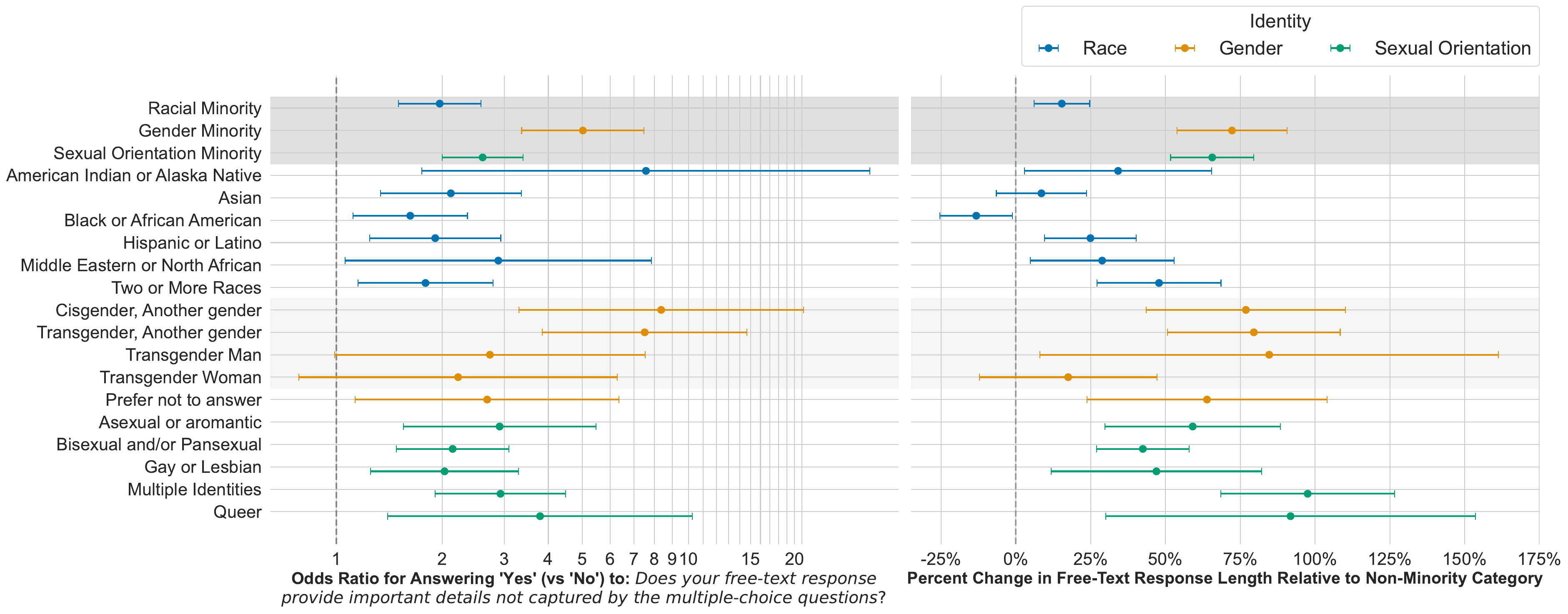}
    \caption{\textbf{Respondents with minority identities report greater added value from free text and write longer responses than individuals with non-minority identities.} Across both panels, reference (non-minority) categories are \emph{White} for race, both \emph{Cisgender Man} and \emph{Cisgender Woman} for gender, and \emph{Straight or heterosexual} for sexual orientation; all comparisons are computed within identity axis.
\textbf{Left:} Odds ratios from identity-specific logistic regressions predicting answering ``Yes'' to: \emph{Do you feel that your free-text response provides important details about your \texttt{[identity]} that were not captured by the multiple-choice questions?}
\textbf{Right:} Percent change in free-text word count relative to the corresponding reference category (within each identity axis).
Categories with $\leq$15 responses are omitted from category-specific regressions due to unstable estimates but are included in the ``Minority'' vs.\ ``Not Minority'' regression.}
    \label{fig:minority_say_more}
\end{figure}

We find that respondents with minority identities had significantly higher odds of reporting that these responses conveyed something important about who they are and consistently wrote longer responses to free-text identity questions. This pattern holds across race, gender, and sexual orientation, though the magnitude of the effect varies both across and within identity axes. To distinguish between minority and non-minority categories, we defined the minority category as \emph{not White} for race, \emph{not Cisgender Man or Cisgender Woman} for gender, and \emph{not Straight or heterosexual} for sexual orientation. Appendix $\S$\ref{sec:mutually-exclusive-cat} provides details on how we grouped mutually exclusive categories for race, gender, and sexual orientation.

As mentioned in $\S$\ref{sec:min-write-more}, race, gender, and sexual orientation minorities all had significantly higher odds of reporting that their free-text response contained important details, relative to the non-minority group. Specifically, race minorities had $2.0\times$ higher odds, gender minorities $5.0\times$ higher odds, and sexual orientation minorities $2.6\times$ higher odds of saying their free-text response conveyed something meaningful. We also observe meaningful variation within identity axes. Figure \ref{fig:minority_say_more} (left) shows that participants of all minority categories, including Asian, Black or African American, transgender, and gay or lesbian, had higher odds of saying their free-text responses conveyed important details about who they are. Estimated odds ratios ranged from $1.6\times$ to over $8\times$.

In Figure \ref{fig:minority_say_more} (right), we compare the relative length of free-text identity responses across categories. Again, we observe meaningful heterogeneity within each identity axis. Among race categories, American Indian or Alaska Native, Hispanic or Latino, Middle Eastern or North African, and multiracial respondents wrote significantly more than White respondents. Asian and Native Hawaiian or Pacific Islander individuals did not differ significantly, and Black or African American respondents wrote slightly fewer words on average (a small but statistically significant difference). Within the gender responses, minorities besides transgender women wrote significantly more than cisgender men and women. For sexual orientation, minority categories---including asexual, bisexual, gay or lesbian, and queer respondents---wrote more than heterosexual respondents. 
These findings suggest that respondents with minority identities were often among those who wrote the most and valued the opportunity to describe themselves in their own words.

\clearpage
\section{Additional details on the computational framework}
\label{supp:computational_framework}

We include below the full prompts used in our computational framework for interpreting SAE-derived dimensions as natural language themes and annotating whether a given free-text response expresses a given theme. Figure \ref{fig:neuron-interpretation-prompt} shows the prompt used for interpretation, and Figure \ref{fig:annotate-prompt} shows the prompt used for annotation.

\begin{figure*}[ht!]
    \centering
    \begin{tcolorbox}[enhanced, width=\linewidth, boxrule=0.8mm, colback=gray!5, colframe=gray!60, 
    title=Interpretation Prompt, fonttitle=\bfseries\ttfamily\large, coltitle=black]

    \scriptsize \tt
    You are a machine learning researcher who has trained a neural network on a text dataset. You are trying to understand what text features cause a specific neuron in the neural network to fire.\newline
    
    You are given two sets of SAMPLES: POSITIVE SAMPLES that strongly activate the neuron, and NEGATIVE SAMPLES from the same distribution that do not activate the neuron.
    Your goal is to identify a feature that is present in the positive samples but absent in the negative samples.\newline
    
    All of the texts are responses to the question:\newline
    In at least 2-3 sentences, how would you describe your \{identity\}?\newline
    Features should describe a specific aspect of the response. For example:\newline
    - "mentions relationship to ..."\newline
    - "self-describes as ...."\newline
    - "discusses ... they ..."\newline
    
    POSITIVE SAMPLES:\newline
    ----------------\newline
    \{positive\_texts\}\newline
    ----------------\newline
    
    NEGATIVE SAMPLES:\newline
    ----------------\newline
    \{negative\_texts\}\newline
    ----------------\newline
    
    Rules about the feature you identify:\newline
    - The feature should be objective, focusing on concrete attributes rather than abstract concepts.\newline
    - The feature should be present in the positive samples and absent in the negative samples. Do not output a generic feature which also appears in negative samples.\newline
    - The feature should be as specific as possible, while still applying to all of the positive samples. For example, if all of the positive samples mention Golden or Labrador retrievers, then the feature should be "mentions retriever dogs", not "mentions dogs" or "mentions Golden retrievers".\newline
    
    Do not output anything else. Your response should be formatted exactly as shown in the examples above.\newline
    Please suggest exactly one description, starting with "-" and surrounded by quotes "". Your response is:
    - "
    \end{tcolorbox}
    \caption{\textbf{Prompt used to interpret high-activating responses within an SAE dimension, adapted from \citeauthor{hypothesaes} (\citeyear{hypothesaes}).} Unlike the original general-purpose prompt, this version is tailored for identity-specific responses and provides specific formatting suggestions to standardize the output features.}
    \label{fig:neuron-interpretation-prompt}
\end{figure*}

\begin{figure*}[ht]
    \centering
    \begin{tcolorbox}[enhanced, width=\linewidth, boxrule=0.8mm, colback=gray!5, colframe=gray!60, 
    title=Annotation Prompt, fonttitle=\bfseries\ttfamily\large, coltitle=black]

    \tiny \tt
    Check whether the TEXT satisfies a PROPERTY. Respond with Yes or No with an explanation that discusses the evidence from the TEXT (at most a sentence). When uncertain, output No.\newline
    
    Example 1:\newline
    PROPERTY: "mentions specific regions or countries of ancestral origin"\newline
    TEXT: "I describe myself as half white/caucasion and half Mexican American."\newline
    Output: Yes. The TEXT mentions the respondent is Mexican American, which specifies the country of Mexico.\newline
    
    Example 2:\newline
    PROPERTY: "mentions feelings of not fully belonging or being out of place related to race/ethnicity"\newline
    TEXT: "Racially I am white. I am a white passing person. I am ethnically Jewish and that ethnicity has shaped my perception of the world"\newline
    Output: No. The TEXT mentions that race/ethnicity affects their perception, but it does NOT IMPLY that they feel out of place or like they do not belong.\newline
    
    Example 3:\newline
    PROPERTY: "mentions feelings of not fully belonging or being out of place related to race/ethnicity"\newline
    TEXT: "Middle Eastern, I felt very separated from my culture and ethnicity in rural southern Pennsylvania"\newline
    Output: Yes. Feeling "separated” from one’s culture and ethnicity is an example of not fully belonging or being out of place.\newline
    
    Example 4:\newline
    PROPERTY: "mentions childhood experiences related to gender roles or expectations”\newline
    TEXT: "I have not had anything significant (good or bad) happen to me because I am a woman"\newline
    Output: No, the TEXT explicitly states the absence of significant gender-related experiences and does not mention childhood or gender role expectations.\newline
    
    Example 5:\newline
    PROPERTY: "mentions experiences or milestones related to navigating societal expectations or breaking barriers tied to gender identity"\newline
    TEXT: "I am a male who who was assigned male at birth. I questioned my gender identity during my time figuring out my sexual orientation"\newline
    Output: No. While the TEXT implies nonbinary identities, the TEXT does not explicitly mention any experiences, challenges, or milestones related to navigating societal expectations.\newline
    
    Example 6:\newline
    PROPERTY: "mentions uncertainty or questioning about their sexual orientation"\newline
    TEXT: "I consider myself pansexual: attracted to people, not gender; though the majority of my sexual partners have been female."\newline
    Output: No. The TEXT expresses a clear and specific identification with pansexuality and does not mention uncertainty or questioning about sexual orientation.\newline
    
    Example 7:\newline
    PROPERTY: "explicitly self-describes as straight using the exact word ‘straight’”\newline
    TEXT: "I am straight. I have only dated and been attracted to men"\newline
    Output: Yes. the TEXT satisfies the PROPERTY if it explicitly includes the word "straight”. The word "straight” indeed appears in the TEXT.\newline
    
    Example 8:\newline
    PROPERTY: "mentions difficulty or hesitation in forming romantic or sexual connections"\newline
    TEXT: "I am asexual, in that sex and romance are of absolutely no importance to me. I have been married for decades and not once have I felt any sexual desire or romantic notions or attraction to anyone."\newline
    Output: No. The TEXT expresses confidence and clarity in their asexual identity, without indicating difficulty or hesitation in forming connections.\newline
    
    Now complete the following example - Respond with Yes or No with an explanation that discusses the evidence from the TEXT. When uncertain, output No.\newline
    PROPERTY: "\{hypothesis\}"\newline
    TEXT: "\{text\}"\newline
    Output:
    
    \end{tcolorbox}
    \caption{\textbf{Prompt used to annotate whether a response satisfies a theme.} The structure is similar to \citeauthor{hypothesaes} (\citeyear{hypothesaes}), but we replace their general-purpose examples with identity-specific few-shot cases to better capture the nuances of race, gender, and sexual orientation.}
    \label{fig:annotate-prompt}
\end{figure*}

The prompt shown in Figure \ref{fig:annotate-prompt} was adapted to support annotation of identity-related free-text responses. To select an annotation model configuration, we evaluated several OpenAI model variants using both this identity-specific prompt and a baseline prompt drawn from prior work on SAE interpretation \cite{hypothesaes}. Specifically, we compared \texttt{GPT-4.1} and \texttt{GPT-4.1-mini} across model-prompt combinations and assessed performance relative to human annotations using three criteria: overall annotation agreement, consistency of agreement across race, gender, and sexual orientation themes, and how closely the model matches humans in how often it assigns a theme (positive annotation rate). 
Based on these criteria, we selected \texttt{GPT-4.1-mini} with the identity-specific prompt for use in our final analyses.

\begin{figure}[h]
    \centering
    \includegraphics[width=\linewidth]{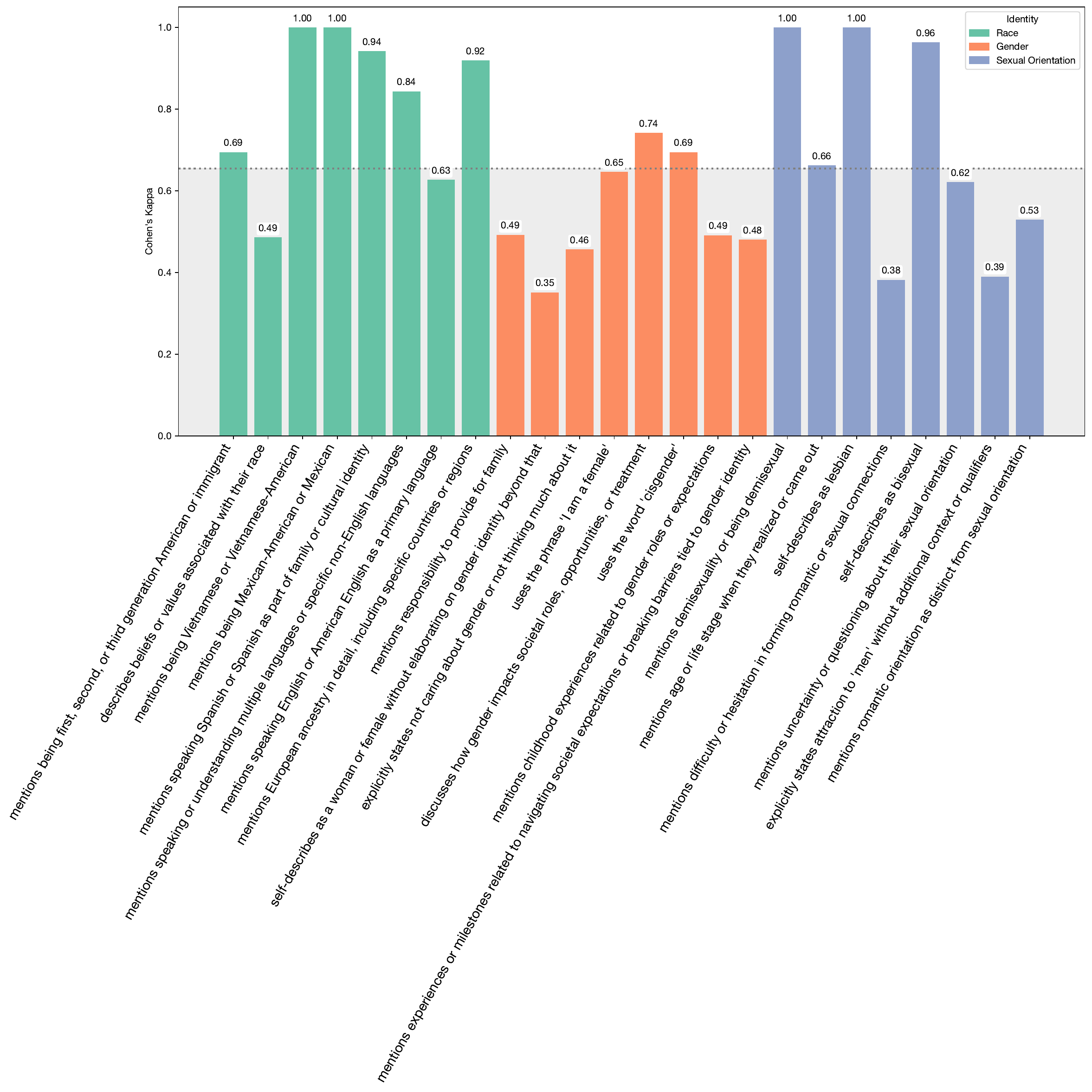}
    \caption{\textbf{Reliability of \texttt{GPT-4.1-mini} across identity axes.} Distribution of Cohen's Kappa ($\kappa$) scores for the \texttt{GPT-4.1-mini} and identity-specific prompt configuration used in our final analyses. Each bar represents a unique theme within the race, gender, and sexual orientation categories. The median score of 0.65 (horizontal line) indicates substantial agreement with human labels.}
    \label{fig:kappa-distribution}
\end{figure}

Table \ref{tab:annotation-model-comp} reports Cohen’s Kappa ($\kappa$) agreement between model and human annotations across the evaluated configurations, along with each model’s overall positive annotation rate. Human annotations were produced by a trained researcher who manually labeled theme presence for a random sample of responses across identity axes, with the researcher blinded to the LLM's annotation.
While differences in average $\kappa$ were modest across configurations, \texttt{GPT-4.1-mini} with the identity-specific prompt exhibited consistently strong agreement across all three identity axes. To assess whether models systematically over- or under-annotate theme presence, we compared each model’s proportion of positive annotations to that of a human annotator on the same set of responses. For the selected configuration, the average difference in positive annotation rate across race, gender, and sexual orientation themes was 0.1 percentage points, with the model’s overall positive rate (13.1\%) closely matching the human annotator’s rate (13\%), indicating no systematic bias toward over- or under-annotation. Figure \ref{fig:kappa-distribution} further illustrates the distribution of $\kappa$ scores for the selected configuration across themes, with a median agreement of $\kappa=0.65$.

\subsection{Manual coverage validation of free-text themes}
\label{supp:manual-coverage-val}

To assess \emph{recall}---i.e. whether our computationally identified free-text themes miss salient patterns in respondents’ free-text self-descriptions---we manually review 100 responses per identity axis and annotate them for recurring motifs not captured by our free-text themes. We use ``motif” to refer to a potentially generalizable concept that appears across multiple responses in the sample (as opposed to idiosyncratic personal details). Overall, most motifs missing from our free-text themes occur rarely (typically in 1--10\% of responses), which is consistent with their not emerging as distinct dimensions when the SAE is constrained to learn only $M=32$ dimensions. In other words, our computational framework prioritizes representing frequent, semantically coherent patterns; less common motifs are often absorbed into broader themes or not surfaced at this dimensionality. We identify two main types of missing motifs. 

First, many seemingly ``missing” motifs reflect differences in representation: respondents use labels or descriptions that map onto themes we already capture, but at different levels of abstraction. For race, respondents sometimes name identity categories or framings that are adjacent to existing themes: for example, Hispanic/Latino identification (10\%) is a broader motif that partially overlaps with the more specific theme our framework identifies, ``mentions speaking Spanish or Spanish as part of family or cultural identity"; conversely, mentions of German heritage (8\%), are a more specific motif captured by the broader theme our framework identifies, ``mentions European ancestry in detail, including specific countries or regions". 
We also observe broader expressions of racial or ethnic pride across categories (3\%), which relate to our more specific theme, ``mentions pride in Black or African American identity". 
For gender, mentions of gender expression (15\%) function as a broader umbrella encompassing our themes which reference stereotypically feminine activities or items (e.g., makeup, fashion) and stereotypically masculine activities or traits (e.g., sports, fixing things, provider/protector roles).
For sexual orientation, references to attraction based on specific characteristics or traits (2\%) refine our theme ``mentions attraction not based on gender or sex". In these cases, manual validation indicates that the underlying content is already represented in our theme set, but a human coder might label it at a different granularity---e.g., collapsing several of our specific themes into a single broader motif, or naming an adjacent identity category directly.

Second, the remaining missing motifs reflect content that our computational framework does not surface at $M=32$, but that can often be recovered simply by increasing SAE dimensionality. Some missing motifs relate to respondents discussing their identity alongside other identity axes or contextual background characteristics---e.g., gender responses that reference race (4\%) or sexual orientation (9\%), and race responses that mention urban or rural upbringing (1\%) or current location (5\%). Others reflect meaningful identity-relevant themes that occur too rarely to be captured at $M=32$. For race, this includes mentions of discrimination (5\%), privilege (5\%), skin color or physical features (6\%), and religion---particularly as it relates to cultural values (8\%).  In gender, we see explicit statements that there are only two genders (3\%), as well as references to feminism (1\%) and fear or discomfort discussing gender (1\%). In sexual orientation, we observe mentions of dating---often framed in terms of the gender(s) that respondents date (9\%). As shown in Tables~\ref{tab:m_64_race_table}-\ref{tab:m_64_sexual_orientation_table}, increasing to $M=64$ recovers many of these motifs---including those related to experiences of discrimination or racism, white privilege, being raised in a religious environment, belief in a binary gender system, and a process of self-discovery in sexual orientation. The trade-off is that higher-dimensional representations can produce more redundant themes and finer splits of existing concepts---for example, the $M=32$-level theme ``mentions specific regions or countries of ancestral origin" subdivides into many ancestry-specific themes (e.g., German, Italian, Chinese) at $M=64$. This highlights a balance between capturing fine-grained content and avoiding redundancy. We emphasize that different analytic goals may motivate different choices of $M$, reinforcing the necessity of qualitative analysis to complement the quantitative methods proposed here.

\begin{table}
    \centering
    \resizebox{\linewidth}{!}{%
    \begin{tabular}{llccccccc}
    \toprule
    Model & Prompt & Race $\kappa$ & Gender $\kappa$ & Sexual Orient. $\kappa$ & Human Pos. (\%) & LLM Pos. (\%) & $\Delta$ Pos. (pp) \\
    \midrule
    GPT-4.1-mini & Baseline & 0.754 & 0.552 & 0.609 & 13.0 & 16.3 & +3.3 \\
    GPT-4.1 & Baseline & 0.685 & 0.559 & 0.741 & 13.0 & 10.2 & -2.8 \\
    GPT-4.1-mini & Identity-Specific & 0.771 & 0.544 & 0.666 & 13.0 & 13.1 & +0.1 \\
    GPT-4.1 & Identity-Specific & 0.683 & 0.547 & 0.778 & 13.0 & 9.0 & -4.0 \\
    \bottomrule
    \end{tabular}}
    \caption{\textbf{Annotation agreement and positive annotation rates across annotation model configurations.} Cohen’s Kappa ($\kappa$) compares each model’s annotations to a human annotator. \emph{Human Pos.} and \emph{LLM Pos.} report average positive annotation rates expressed as percentages, where the positive rate denotes the fraction of responses labeled as activating a given theme in response to the annotation prompt (Figure \ref{fig:annotate-prompt}). $\Delta$ Pos. reports the difference between model and human positive rates in percentage points. We select the (GPT-4.1-mini, Identity-Specific) configuration for subsequent analyses, as it achieves reasonably high agreement across identity axes while most closely matching the human positive annotation rate.}
    \label{tab:annotation-model-comp}
\end{table}

\clearpage

\section{Computationally identified free-text themes for self-described identity}\label{supp:self-described-themes}

\subsection{Detailed presentation of free-text themes}

Table \ref{tab:race_interpretation_fidelity} shows the 32 themes initially generated from the race free-text responses. We excluded themes with fidelity (F1 scores) below 0.50 or those describing writing style, resulting in 26 interpretable themes. The excluded themes are marked with a strikethrough in the table. Table \ref{tab:gender_interpretation_fidelity} presents the same information for the gender free-text responses. The excluded themes are marked with a strikethrough, and ultimately we keep 27 interpretable themes. Table \ref{tab:sexual_orientation_interpretation_fidelity} shows the 28 themes extracted from sexual orientation free-text responses. 

\begin{table}[ht]
\centering
\renewcommand{\arraystretch}{1}
\scriptsize
\begin{tabular}{r|p{12cm}|c}
\toprule
 & \textbf{Race Theme} & \textbf{Fidelity} \\
\midrule
1 & mentions celebrating Christian holidays & 0.78 \\
2 & \st{uses the phrase `I would describe my race' or a close variation of it} & 0.91 \\
3 & mentions speaking English or American English as a primary language & 0.90 \\
4 & mentions being first, second, or third generation American or immigrant & 0.75 \\
5 & mentions pride in Black or African American identity & 0.88 \\
6 & \st{uses the phrase `I am' followed by a race or ethnicity descriptor without elaboration} & 0.75 \\
7 & self-describes as mixed or multiracial & 0.79 \\
8 & explicitly self-describes as white without mentioning other ethnicities or cultural heritages & 0.86 \\
9 & mentions specific regions or countries of ancestral origin & 0.69 \\
10 & mentions European ancestry or descent explicitly & 0.64 \\
11 & mentions being Black or African American & 0.98 \\
12 & mentions speaking or understanding multiple languages or specific non-English languages & 0.71 \\
13 & mentions being Native American or belonging to a Native American tribe & 0.97 \\
14 & self-describes as South Asian or mentions a South Asian country or ethnicity & 0.75 \\
15 & \st{uses single-word or very short descriptions of race/ethnicity (1-3 words)} & 0.69 \\
16 & mentions feelings of not fully belonging or being out of place related to race/ethnicity & 0.67 \\
17 & mentions being Mexican-American or Mexican & 0.99 \\
18 & explicitly states a lack of cultural or ethnic identity or traditions & 0.97 \\
19 & mentions European ancestry in detail, including specific countries or regions & 0.68 \\
20 & mentions parentage explicitly (e.g., `my mom is X and my dad is Y') & 0.86 \\
21 & mentions the languages spoken by themselves or their family & 0.78 \\
22 & \st{explicitly uses the phrase `identify as' or `identify with' to describe their race or ethnicity} & 0.74 \\
23 & \st{mentions speaking Arabic language or dialects} & 0.46 \\
24 & describes beliefs or values associated with their race & 0.70 \\
25 & mentions whiteness or being white & 0.61 \\
26 & \st{mentions celebrating Lunar New Year or similar cultural holidays} & 0.39 \\
27 & mentions being Vietnamese or Vietnamese-American & 0.51 \\
28 & mentions being born in the United States or America & 0.82 \\
29 & mentions speaking Spanish or Spanish as part of family or cultural identity & 0.80 \\
30 & mentions traditional or cultural food as an important aspect of their identity & 0.62 \\
31 & mentions African heritage or African ancestry & 0.72 \\
32 & self-describes as white American & 0.76 \\
\bottomrule
\end{tabular}
\caption{\textbf{Computationally identified free-text themes for SAE dimensions trained on race-related free-text response embeddings and their associated interpretation fidelity (F1 score).} Low-fidelity themes (F1 $< 0.50$) and themes reflecting writing style rather than semantic content are excluded from analysis; excluded themes are indicated with strikethroughs.}
\label{tab:race_interpretation_fidelity}
\end{table}

\clearpage

\begin{table}
\centering
\renewcommand{\arraystretch}{1}
\scriptsize
\begin{tabular}{r|p{12cm}|c}
\toprule
& \textbf{Gender Theme} & \textbf{Fidelity} \\
\midrule
1 & uses the exact word `male' & 0.91 \\
2 & mentions childhood experiences related to gender roles or expectations & 0.78 \\
3 & mentions responsibility to provide for family & 0.55 \\
4 & \st{mentions use of multiple pronouns or experimenting with pronouns} & 0.46 \\
5 & \st{mentions physical appearance or specific physical traits} & 0.38 \\
6 & mentions giving birth or having children as a significant aspect of their identity & 0.86 \\
7 & \st{single-word self-description of gender} & 0.56 \\
8 & explicitly states never questioning their gender identity & 0.82 \\
9 & explicitly states being born a specific gender and continuing to identify with that same gender & 0.70 \\
10 & mentions how gender identity influences decisions, safety, or interactions in life & 0.72 \\
11 & self-describes as male or man and mentions alignment with gender assigned at birth & 0.90 \\
12 & \st{uses the phrase `I would describe my gender identity as ...'} & 0.77 \\
13 & mentions enjoyment or participation in activities or items stereotypically associated with femininity (e.g., makeup, fashion, dressing up, heels, etc.) & 0.57 \\
14 & mentions traditionally masculine activities or traits such as sports, fixing things, outdoor activities, or being a provider/protector & 0.50 \\
15 & uses the word `cisgender' & 0.98 \\
16 & explicitly states being heterosexual & 0.90 \\
17 & mentions being non-binary or nonbinary & 0.81 \\
18 & mentions fluidity or fluctuation in gender identity & 0.78 \\
19 & uses the phrase `I am a female' & 0.81 \\
20 & \st{mentions being assigned a gender at birth} & 0.38 \\
21 & mentions experiences of questioning or struggling with their gender identity & 0.75 \\
22 & mentions experiences or milestones related to navigating societal expectations or breaking barriers tied to gender identity & 0.65 \\
23 & mentions roles or responsibilities within a family or household & 0.56 \\
24 & self-describes as a woman or female without elaborating on gender identity beyond that & 0.76 \\
25 & mentions being male or a man in a direct and explicit manner & 0.86 \\
26 & explicitly rejects or feels disconnected from the concept of gender as a meaningful personal identity & 0.71 \\
27 & explicitly states not caring about gender or not thinking much about it & 0.91 \\
28 & explicitly identifies as male or a variant of male & 0.77 \\
29 & discusses how gender impacts societal roles, opportunities, or treatment & 0.85 \\
30 & explicitly states `I am a woman' & 0.93 \\
31 & mentions being born female and identifying as female & 0.80 \\
32 & explicitly self-describes as non-binary or uses the term `non-binary' to describe their gender identity & 0.98 \\
\bottomrule
\end{tabular}
\caption{\textbf{Computationally identified free-text themes for SAE dimensions trained on gender-related free-text response embeddings and their associated interpretation fidelity (F1 score).} Low-fidelity themes (F1 $< 0.50$) and themes reflecting writing style rather than semantic content are excluded from analysis; excluded themes are indicated with strikethroughs.}
\label{tab:gender_interpretation_fidelity}
\end{table}

\clearpage
\begin{table}
\centering
\renewcommand{\arraystretch}{1}
\scriptsize
\begin{tabular}{r|p{12cm}|c}
\toprule
 & \textbf{Sexual Orientation Theme} & \textbf{Fidelity} \\
\midrule
1 & refuses to provide a clear or specific answer about sexual orientation & 0.64 \\
2 & explicitly states attraction to `men' without additional context or qualifiers & 0.91 \\
3 & mentions demisexuality or being demisexual & 0.89 \\
4 & mentions being married & 0.91 \\
5 & explicitly self-describes as asexual or aromantic & 0.99 \\
6 & explicitly self-describes as straight or heterosexual without additional context or elaboration & 0.80 \\
7 & explicitly self-describes as straight using the exact word `straight' & 0.99 \\
8 & discusses how sexual orientation influences personal growth, relationships, or key life milestones & 0.68 \\
9 & uses the term `queer' to describe their sexual orientation & 0.71 \\
10 & explicitly states never questioning their sexual orientation & 0.85 \\
11 & mentions uncertainty or questioning about their sexual orientation & 0.84 \\
12 & mentions that their sexual orientation does not significantly impact their life & 0.91 \\
13 & mentions `opposite sex' or `opposite gender' & 0.78 \\
14 & mentions attraction to more than two genders or a spectrum of genders & 0.82 \\
15 & mentions attraction specifically to the opposite sex & 0.55 \\
16 & mentions difficulty or hesitation in forming romantic or sexual connections & 0.65 \\
17 & self-describes as gay & 0.84 \\
18 & mentions romantic orientation as distinct from sexual orientation & 0.72 \\
19 & \st{uses the phrase `I would describe'} & 0.81 \\
20 & \st{mentions how sexual orientation influences their interactions or relationships with others} & 0.45 \\
21 & mentions the word `straight' & 0.50 \\
22 & mentions age or life stage when they realized or came out & 0.80 \\
23 & discusses how their sexual orientation influences their personal relationships or life milestones & 0.51 \\
24 & mentions attraction to transgender or non-binary individuals & 0.87 \\
25 & \st{mentions religion or God in relation to their orientation} & 0.21 \\
26 & self-describes as lesbian & 0.90 \\
27 & mentions pansexuality or pansexual as a descriptor & 1.00 \\
28 & self-describes as bisexual & 0.89 \\
29 & mentions attraction not based on gender or sex & 0.81 \\
30 & mentions discomfort or rejection of labels for their sexual orientation & 0.73 \\
31 & explicitly uses the term `heterosexual' & 0.96 \\
32 & \st{uses a single word or very brief phrase to describe sexual orientation without elaboration} & 0.73 \\
\bottomrule
\end{tabular}
\caption{\textbf{Computationally identified free-text themes for SAE dimensions trained on sexual orientation-related free-text response embeddings and their associated interpretation fidelity (F1 score).} Low-fidelity themes (F1 $< 0.50$) and themes reflecting writing style rather than semantic content are excluded from analysis; excluded themes are indicated with strikethroughs.}
\label{tab:sexual_orientation_interpretation_fidelity}
\end{table}

\clearpage
\subsection{Comparison of computationally identified free-text themes to standardized categories}
\label{supp:category-details}

We provide additional results for race, gender, and sexual orientation that examine how free-text themes align with or span across standardized categories, building on $\S$\ref{sec:present-themes}.
We also show that our findings remain robust upon computing alternative $R^2$ metrics.

\subsubsection*{Grouping of standardized categories}
\label{sec:mutually-exclusive-cat}

\begin{figure}[b]
    \centering
    \includegraphics[width=\linewidth]{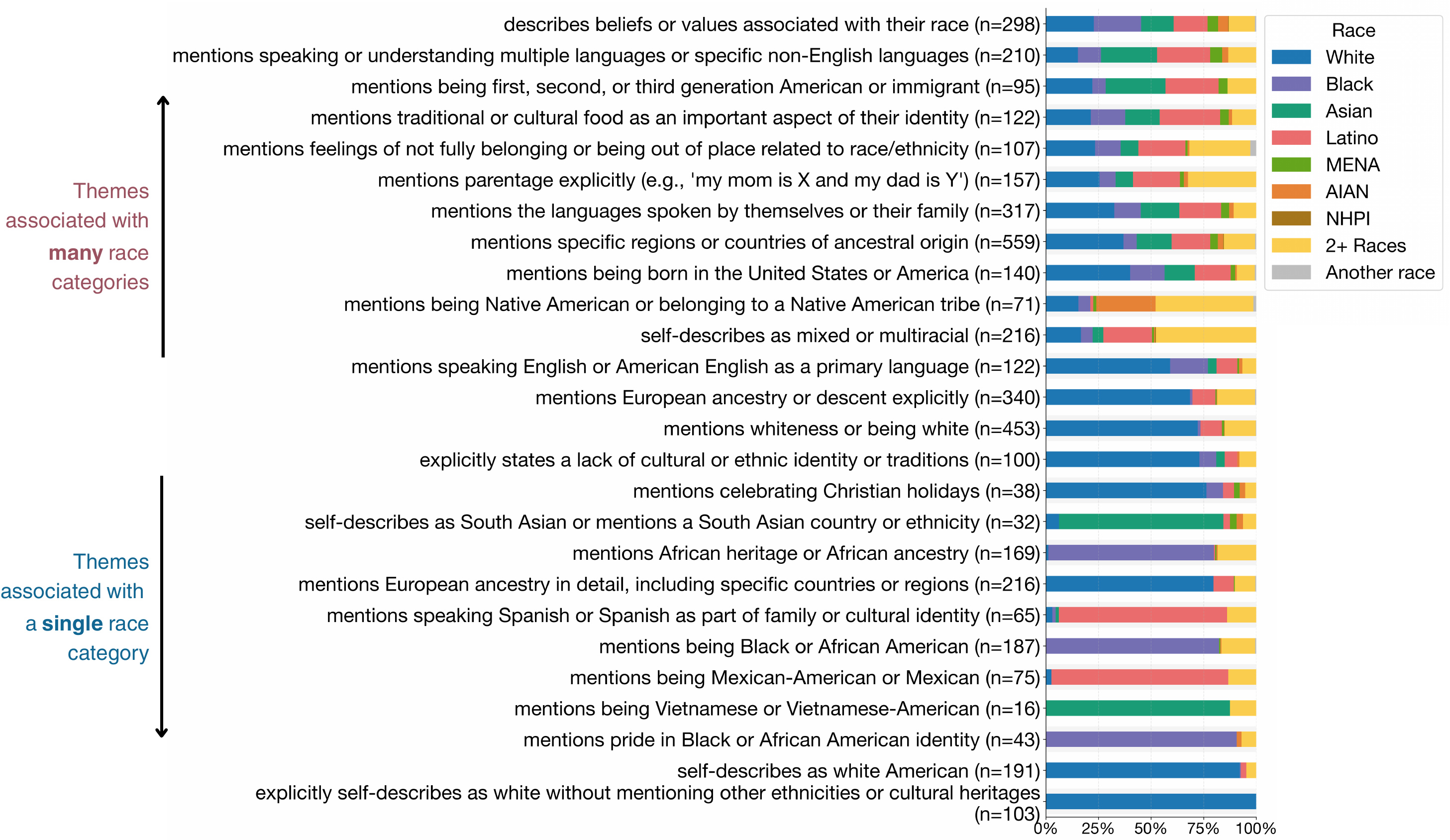}
    \caption{\textbf{Some identity themes align with standardized racial categories, but many cut across categories}. 
   Each row represents a theme, with colored bars representing the standardized race categories of respondents whose free-text responses contain the theme. Themes are ordered by the maximum proportion of responses (among those mentioning the theme) that come from any single race category---in other words, by the highest value among the race-specific proportions for each theme. Thus, themes at the top cut across categories and are not strongly associated with any particular category, while themes at the bottom are predominantly expressed by respondents from specific race categories.}
    \label{fig:race-themes-barchart}
\end{figure}

To compare theme prevalence across standardized categories, we summarize 
themes using mutually exclusive groupings derived from the original survey 
responses. These groupings are introduced \emph{only} to support the visualizations and aggregate comparisons in Figures \ref{fig:all-themes}, \ref{fig:minority_say_more}, and \ref{fig:race-themes-barchart}-\ref{fig:sexual-orientation-themes}; all other analyses in this paper use the original survey responses directly. We follow prior conventions in demographic and public health research in constructing these categories \cite{mays2003classification, ct2022classification, revesz2024revisions, TrevorProject2021Gay, TrevorProjectBisexuality, HRC2023Queer}, and acknowledge that any categorical scheme involves simplifications that will 
not fully capture every individual's identity.

For race, we construct mutually exclusive race categories from the original multi-select options (which were derived from U.S. Office of Management and Budget’s standards \cite{revesz2024revisions}). Respondents who select a single category are assigned to that category. For those who select multiple categories, we follow established federal 
reporting conventions \cite{revesz2024revisions}, while acknowledging that 
no single standard exists for classifying multiracial respondents 
\cite{mays2003classification, ct2022classification}. Participants who select both ``Hispanic and/or Latino” and ``White” are categorized as \emph{Hispanic and/or Latino}, reflecting recent shifts toward treating Hispanic or Latino as a standalone response category \cite{revesz2024revisions}. Participants who select “White” and any other minority label, or select two or more minority categories, are categorized as \emph{Two or More Races}. The final nine options include American Indian or Alaska Native, Asian, Black or African American, Hispanic or Latino, Middle Eastern or North African, Native Hawaiian or Pacific Islander, Some Other Race, Two or More Races, and White.  

For gender, we construct categories based on the cross-tabulation of current gender identity (e.g., ``Man", ``Woman", ``Some other way") \cite{amaya2020adapting} and transgender status (``Yes", ``No", ``Prefer not to answer"). Respondents who select ``Prefer not to answer” for transgender status and any gender identity are grouped into a single \emph{Prefer not to answer} category. This yields seven mutually exclusive gender categories: Cisgender, Man; Cisgender, Woman; Cisgender, Some other way; Transgender, Man; Transgender, Woman; Transgender, Some other way; and Prefer not to answer.

For sexual orientation, we construct groupings by 
drawing on terminology and definitions from The Trevor Project and the Human Rights Campaign  \cite{TrevorProject2021Gay, TrevorProjectBisexuality, HRC2023Queer}, while 
recognizing that collapsing distinct identities into shared categories 
(e.g., Bisexual and Pansexual) may not reflect how all individuals 
understand their own identities. Responses of “Gay” and “Lesbian” are grouped into the analytical category \emph{Gay or Lesbian} \cite{TrevorProject2021Gay}, and “Bisexual” and “Pansexual” are grouped as \emph{Bisexual and/or Pansexual} \cite{TrevorProjectBisexuality}. Respondents who selected “Queer” alongside another minority identity are categorized under the more specific label, recognizing that “Queer” often serves as an umbrella term \cite{HRC2023Queer}. Respondents who select two or more options after these categorizations are assigned to the \emph{Multiple Identities} category. The final set of 11 categories includes Asexual or aromantic, Bisexual and/or Pansexual, Demisexual, Gay or Lesbian, Multiple Identities, Other sexual identity or orientation, Prefer not to answer, Queer, Questioning, Sexually fluid, and Straight or heterosexual.

\subsubsection*{Analysis of free-text themes by categories }

As noted in $\S$\ref{sec:present-themes}, while some themes align closely with a single category, many do not. For race, themes like ``mentions being born in the United States or America" is expressed by respondents across six race categories and has an $R^2$ of only 0.01. We also note that $R^2$ can be low even if a theme is predominantly expressed by members of one race group (Table \ref{tab:race_r_squared}). For example, while the theme ``mentions being Vietnamese or Vietnamese-American" is expressed primarily by Asian respondents, its $R^2$ remains low (0.09) because it is not expressed by \emph{all} Asian respondents; rather, it captures a subgroup concealed by the coarser standardized race category.

\begin{figure}[]
    \centering
    \includegraphics[width=\linewidth]{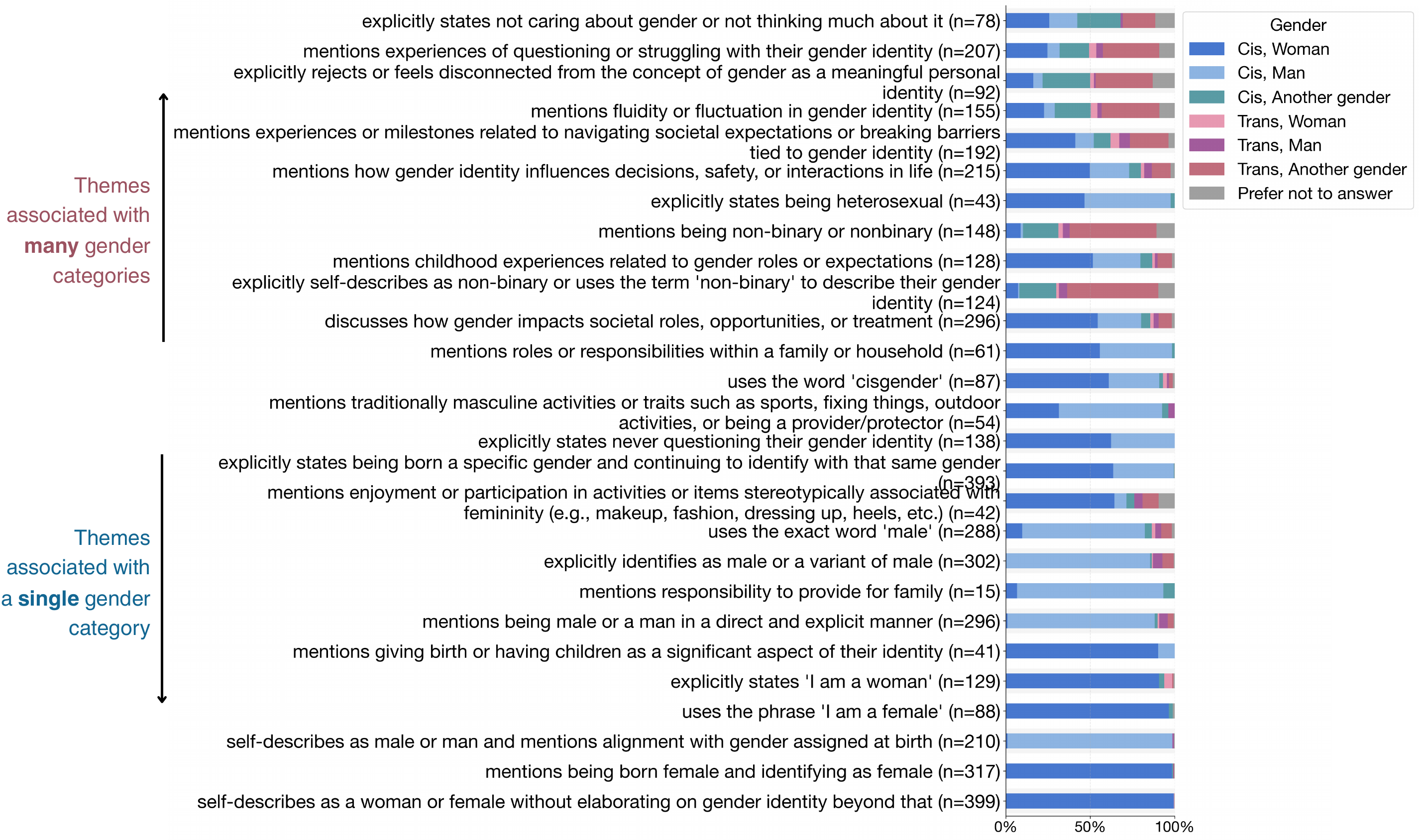}
    \caption{\textbf{Some identity themes align with standardized gender categories, but many cut across categories}. Each row represents a theme, with colored bars representing the standardized gender categories of respondents whose free-text responses contain the theme. Themes are ordered by the maximum proportion of responses (among those mentioning the theme) that come from any single gender category---in other words, by the highest value among the gender-specific proportions for each theme. Thus, themes at the top cut across categories and are not strongly associated with any particular category, while themes at the bottom are predominantly expressed by respondents from specific gender categories.}
    \label{fig:gender-themes}
\end{figure}

For gender, the themes learned from free text also reveal both category-specific patterns and narratives that cut across gender categories (Figure \ref{fig:gender-themes}). At one end of the spectrum, themes align closely with standardized gender categories and yield high explained variance (e.g., ``self-describes as male or man and mentions alignment with gender assigned at birth'' ($R^2 = 0.62$) and ``mentions being born female and identifying as female'' ($R^2=0.44$). We also observe themes that capture gender-specific social roles, particularly among cisgender individuals. For example, “mentions giving birth or having children...'' is almost exclusively expressed by cisgender women ($R^2 = 0.03$), while “mentions traditionally masculine activities or traits...'' ($R^2=0.03$) aligns closely with cisgender men. Themes reflecting conventional roles, such as “mentions responsibility to provide for family” ($R^2 = 0.03$), appear across both cisgender men and women. The low $R^2$ values (Table \ref{tab:gender_r_squared}) for these themes indicate that while they are associated with cisgender identities, they are not uniformly expressed by everyone in those groups. Finally, many themes cut across gender categories, particularly those expressed by minority groups. Respondents from multiple minority categories describe their identity using themes like ``mentions childhood experiences related to gender roles,'' ``mentions experiences of questioning or struggling with their gender identity,'' and ``mentions fluidity or fluctuation in gender identity". 

\begin{figure}[t]
    \centering
    \includegraphics[width=\linewidth]{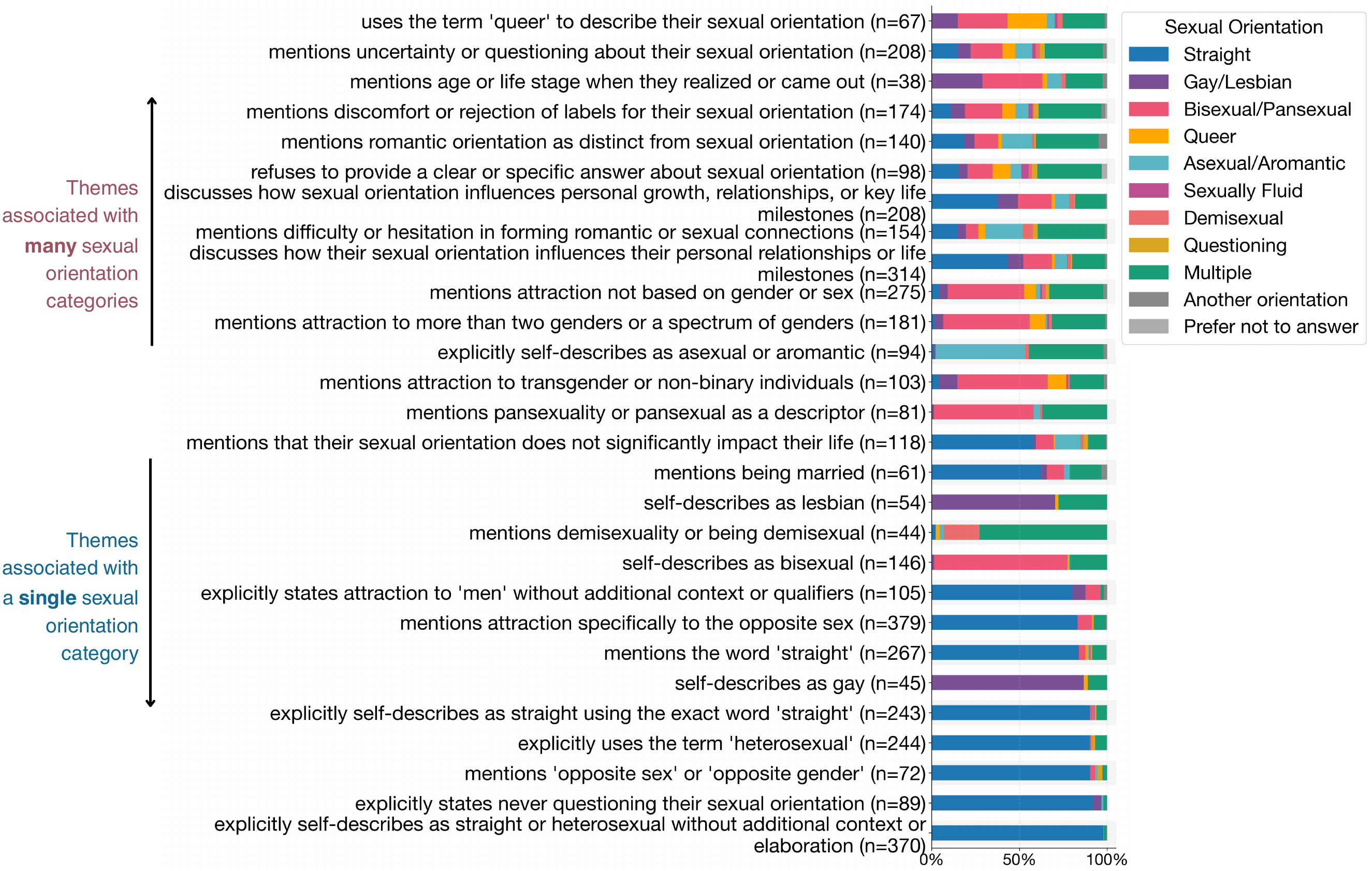}
    \caption{\textbf{Some identity themes align with standardized sexual orientation categories, but many cut across categories}. Each row represents a theme, with colored bars representing the standardized sexual orientation categories of respondents whose free-text responses contain the theme. Themes are ordered by the maximum proportion of responses (among those mentioning the theme) that come from any single sexual orientation category---in other words, by the highest value among the sexual orientation-specific proportions for each theme. Thus, themes at the top cut across categories and are not strongly associated with any particular category, while themes at the bottom are predominantly expressed by respondents from specific sexual orientation categories.}
    \label{fig:sexual-orientation-themes}
\end{figure}

For sexual orientation, many minority-aligned themes span multiple categories and have low $R^2$ values (Figure \ref{fig:sexual-orientation-themes}, Table \ref{tab:sexual_orientation_r_squared}). These include themes such as ``mentions discomfort or rejection of labels for their sexual orientation” ($R^2=0.17$), ``mentions uncertainty or questioning” ($R^2=0.17$), and ``mentions romantic orientation as distinct from sexual orientation” ($R^2=0.17$). Additionally, themes like ``mentions age or life stage when they realized or came out” ($R^2=0.06$) or ``refuses to provide a clear or specific answer” ($R^2=0.11$) reflect lived experiences that are rarely captured in standard labels. Compared to the gender themes, Figure \ref{fig:sexual-orientation-themes} shows that fewer sexual orientation themes are dominated by the non-minority group, `Straight or heterosexual'.

\subsubsection*{Fraction of variance in themes explained by standardized category ($R^2$)}
\label{supp:themes_r2_robust}

To confirm the robustness of this analysis, Table \ref{tab:median_r_squared} compares our primary $R^2$ metric with alternative pseudo-$R^2$ formulations (e.g., McFadden's and logit-based pseudo-$R^2$), demonstrating that the findings are not dependent on the specific metric chosen. In each alternative formulation, free-text themes are not well-predicted by standardized categories. 

\begin{table}[h]
\caption{Fraction of variance in each race theme ($R^2$) explained by the standardized categories, sorted by ascending $R^2$.}
\label{tab:race_r_squared}
\renewcommand{\arraystretch}{1}
\scriptsize
\begin{tabular}{l|r}
\toprule
 \textbf{Race Theme} & $R^2$ \\
\midrule
mentions being born in the United States or America & 0.01 \\
mentions celebrating Christian holidays & 0.03 \\
mentions speaking English or American English as a primary language & 0.03 \\
mentions traditional or cultural food as an important aspect of their identity & 0.04 \\
explicitly states a lack of cultural or ethnic identity or traditions & 0.05 \\
mentions being first, second, or third generation American or immigrant & 0.05 \\
mentions feelings of not fully belonging or being out of place related to race/ethnicity & 0.05 \\
mentions the languages spoken by themselves or their family & 0.05 \\
describes beliefs or values associated with their race & 0.06 \\
mentions parentage explicitly (e.g., `my mom is X and my dad is Y') & 0.08 \\
mentions being Vietnamese or Vietnamese-American & 0.09 \\
self-describes as South Asian or mentions a South Asian country or ethnicity & 0.13 \\
mentions pride in Black or African American identity & 0.15 \\
explicitly self-describes as white without mentioning other ethnicities or cultural heritages & 0.15 \\
mentions speaking or understanding multiple languages or specific non-English languages & 0.15 \\
mentions specific regions or countries of ancestral origin & 0.18 \\
mentions European ancestry in detail, including specific countries or regions & 0.24 \\
self-describes as white American & 0.26 \\
mentions speaking Spanish or Spanish as part of family or cultural identity & 0.32 \\
mentions European ancestry or descent explicitly & 0.34 \\
self-describes as mixed or multiracial & 0.36 \\
mentions being Mexican-American or Mexican & 0.39 \\
mentions being Native American or belonging to a Native American tribe & 0.57 \\
mentions whiteness or being white & 0.59 \\
mentions African heritage or African ancestry & 0.65 \\
mentions being Black or African American & 0.77 \\
\bottomrule
\end{tabular}
\end{table}

\clearpage

\begin{table}
\caption{Fraction of variance in each gender theme ($R^2$) explained by the standardized categories, sorted by ascending $R^2$.}
\label{tab:gender_r_squared}
\renewcommand{\arraystretch}{1}
\scriptsize
\begin{tabular}{p{13cm}|c}
\toprule
 \textbf{Gender Theme} & $R^2$ \\
\midrule
mentions childhood experiences related to gender roles or expectations & 0.00 \\
uses the word `cisgender' & 0.01 \\
discusses how gender impacts societal roles, opportunities, or treatment & 0.01 \\
explicitly states being heterosexual & 0.02 \\
mentions roles or responsibilities within a family or household & 0.02 \\
mentions how gender identity influences decisions, safety, or interactions in life & 0.02 \\
mentions responsibility to provide for family & 0.03 \\
mentions giving birth or having children as a significant aspect of their identity & 0.03 \\
mentions traditionally masculine activities or traits such as sports, fixing things, outdoor activities, or being a provider/protector & 0.03 \\
mentions enjoyment or participation in activities or items stereotypically associated with femininity (e.g., makeup, fashion, dressing up, heels, etc.) & 0.04 \\
explicitly states never questioning their gender identity & 0.04 \\
uses the phrase `I am a female' & 0.09 \\
explicitly states not caring about gender or not thinking much about it & 0.12 \\
explicitly states `I am a woman' & 0.12 \\
mentions experiences or milestones related to navigating societal expectations or breaking barriers tied to gender identity & 0.12 \\
explicitly states being born a specific gender and continuing to identify with that same gender & 0.17 \\
explicitly rejects or feels disconnected from the concept of gender as a meaningful personal identity & 0.25 \\
mentions fluidity or fluctuation in gender identity & 0.29 \\
mentions experiences of questioning or struggling with their gender identity & 0.35 \\
uses the exact word `male' & 0.41 \\
mentions being born female and identifying as female & 0.44 \\
explicitly self-describes as non-binary or uses the term `non-binary' to describe their gender identity & 0.48 \\
mentions being non-binary or nonbinary & 0.55 \\
self-describes as male or man and mentions alignment with gender assigned at birth & 0.62 \\
self-describes as a woman or female without elaborating on gender identity beyond that & 0.66 \\
explicitly identifies as male or a variant of male & 0.75 \\
mentions being male or a man in a direct and explicit manner & 0.76 \\
\bottomrule
\end{tabular}
\end{table}

\begin{table}
\caption{Fraction of variance in each sexual orientation theme ($R^2$) explained by the standardized categories, sorted by ascending $R^2$.}
\label{tab:sexual_orientation_r_squared}
\renewcommand{\arraystretch}{1}
\scriptsize
\begin{tabular}{lr}
\toprule
 \textbf{Sexual Orientation Theme} & $R^2$ \\
\midrule
mentions being married & 0.02 \\
discusses how sexual orientation influences personal growth, relationships, or key life milestones & 0.03 \\
mentions that their sexual orientation does not significantly impact their life & 0.03 \\
discusses how their sexual orientation influences their personal relationships or life milestones & 0.04 \\
mentions `opposite sex' or `opposite gender' & 0.05 \\
explicitly states attraction to `men' without additional context or qualifiers & 0.05 \\
mentions age or life stage when they realized or came out & 0.06 \\
explicitly states never questioning their sexual orientation & 0.07 \\
refuses to provide a clear or specific answer about sexual orientation & 0.11 \\
mentions romantic orientation as distinct from sexual orientation & 0.17 \\
mentions discomfort or rejection of labels for their sexual orientation & 0.17 \\
mentions uncertainty or questioning about their sexual orientation & 0.17 \\
mentions attraction to transgender or non-binary individuals & 0.18 \\
mentions the word `straight' & 0.21 \\
explicitly self-describes as straight using the exact word `straight' & 0.24 \\
explicitly uses the term `heterosexual' & 0.25 \\
mentions difficulty or hesitation in forming romantic or sexual connections & 0.28 \\
mentions attraction specifically to the opposite sex & 0.32 \\
uses the term `queer' to describe their sexual orientation & 0.46 \\
mentions attraction to more than two genders or a spectrum of genders & 0.48 \\
mentions attraction not based on gender or sex & 0.51 \\
explicitly self-describes as straight or heterosexual without additional context or elaboration & 0.54 \\
mentions demisexuality or being demisexual & 0.58 \\
self-describes as gay & 0.61 \\
self-describes as bisexual & 0.63 \\
mentions pansexuality or pansexual as a descriptor & 0.63 \\
explicitly self-describes as asexual or aromantic & 0.67 \\
self-describes as lesbian & 0.78 \\
\bottomrule
\end{tabular}
\end{table}

\begin{table}[h]
\caption{\textbf{Median $R^2$ values between themes and standardized categories under alternative metrics.} We assess how much variation in theme activation is explained by standardized categories. Standard and adjusted $R^2$ are computed from linear probability models with binary theme activation as the outcome, while McFadden and Cox-Snell $R^2$ are computed from logistic regressions with the same outcome. All values are reported as medians across themes within each identity axis.}
\label{tab:median_r_squared}
\begin{tabular*}{\textwidth}{@{\extracolsep{\fill}}lrrr}
\toprule
 & \textbf{Race} & \textbf{Gender} & \textbf{Sexual Orientation} \\
\midrule
$R^2$ & 0.15 & 0.12 & 0.22 \\
Adjusted $R^2$ & 0.14 & 0.11 & 0.22 \\
$R^2$ (McFadden) & 0.31 & 0.18 & 0.26 \\
$R^2$ (Cox-Snell) & 0.13 & 0.11 & 0.20 \\
\bottomrule
\end{tabular*}
\end{table}

\clearpage
\subsection{Additional details on `Free-text themes reveal within-group heterogeneity'}

\begin{figure}[b]
    \centering
    \includegraphics[width=\linewidth]{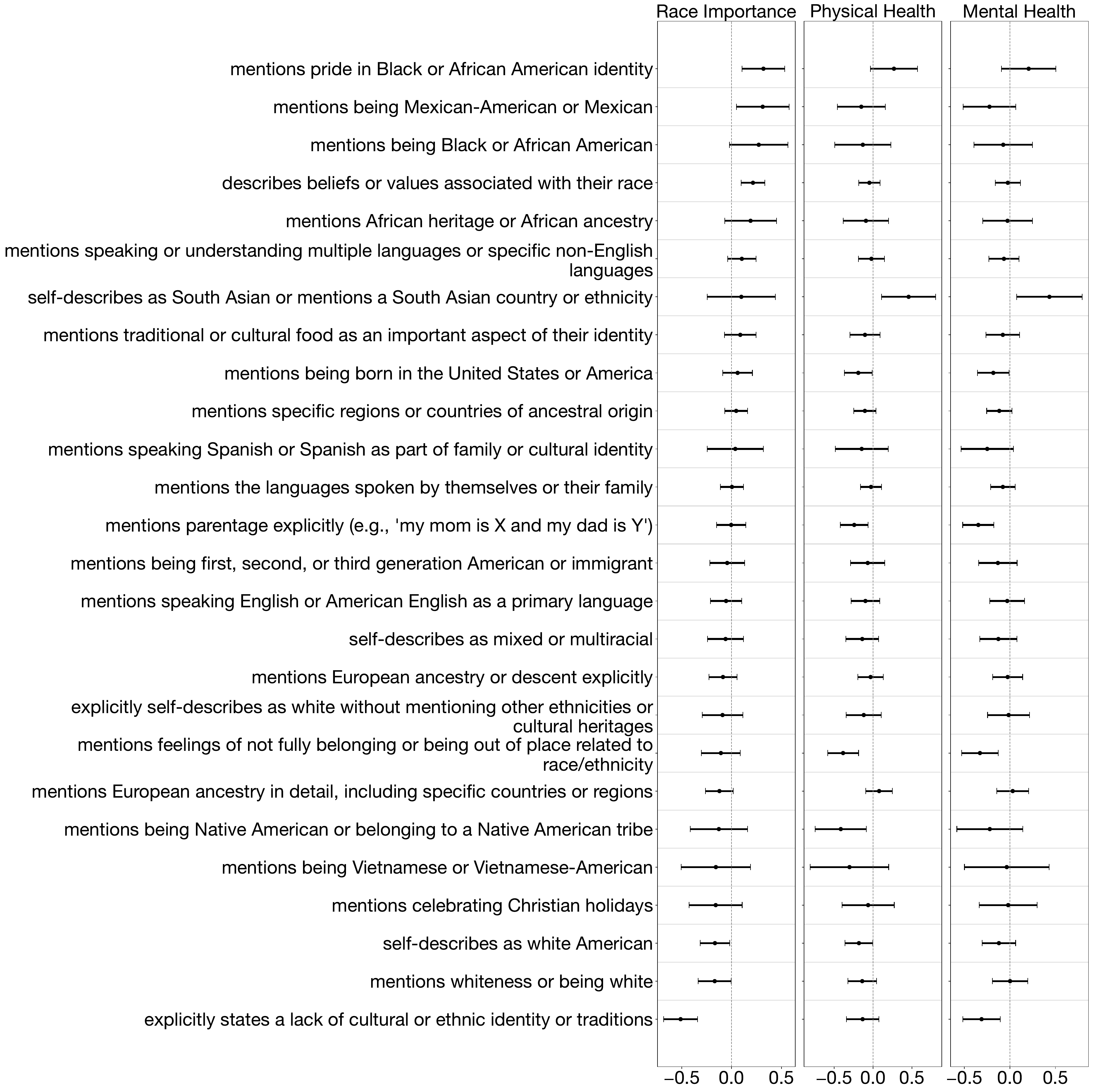}
    \caption{Regression coefficients for free-text race themes, when controlling for standardized race categories, in regressions to predict the three outcomes---race importance, physical health, and mental health---for which free-text themes significantly improved model fit after multiple hypothesis correction. We estimate models for each theme individually because themes may be collinear, complicating interpretation of multivariate models. Both point estimates and 95\% confidence intervals are plotted (uncorrected for multiple hypotheses). We report HC3 heteroskedasticity-consistent standard errors. Themes are ordered by the estimated coefficient for Race Importance (largest to smallest).}
    \label{fig:race-theme-predictiveness}
\end{figure}

\begin{figure}
    \centering
    \includegraphics[width=\linewidth]{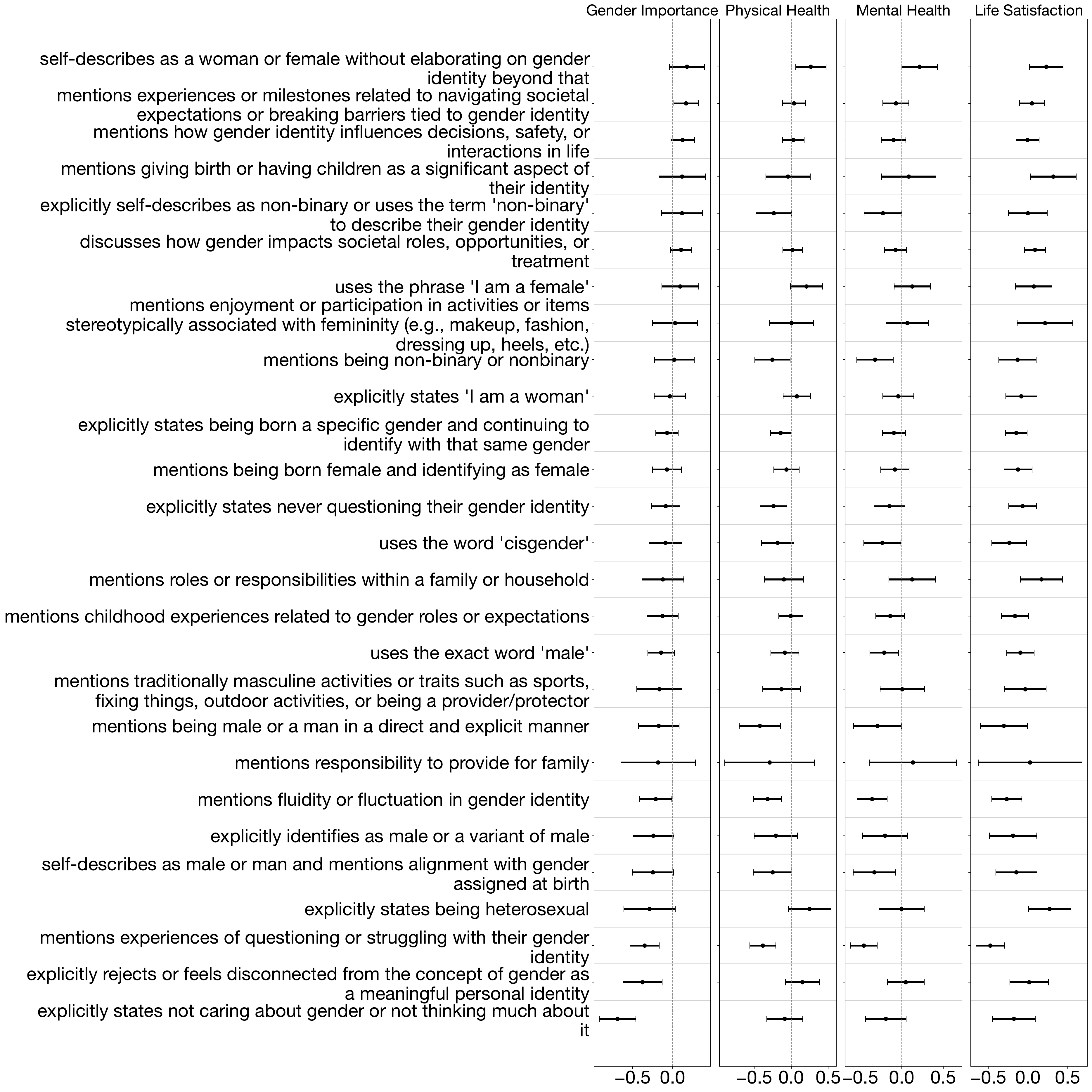}
    \caption{Regression coefficients for free-text gender themes, when controlling for standardized gender categories, in regressions to predict the four outcomes---gender importance, physical health, mental health, and life satisfaction---for which free-text themes significantly improved model fit after multiple hypothesis correction. We estimate models for each theme individually because themes may be collinear, complicating interpretation of multivariate models. Both point estimates and 95\% confidence intervals are plotted (uncorrected for multiple hypotheses). We report HC3 heteroskedasticity-consistent standard errors. Themes are ordered by the estimated coefficient for Gender Importance (largest to smallest).}
    \label{fig:gender-theme-predictiveness}
\end{figure}

\begin{figure}
    \centering
    \includegraphics[width=\linewidth]{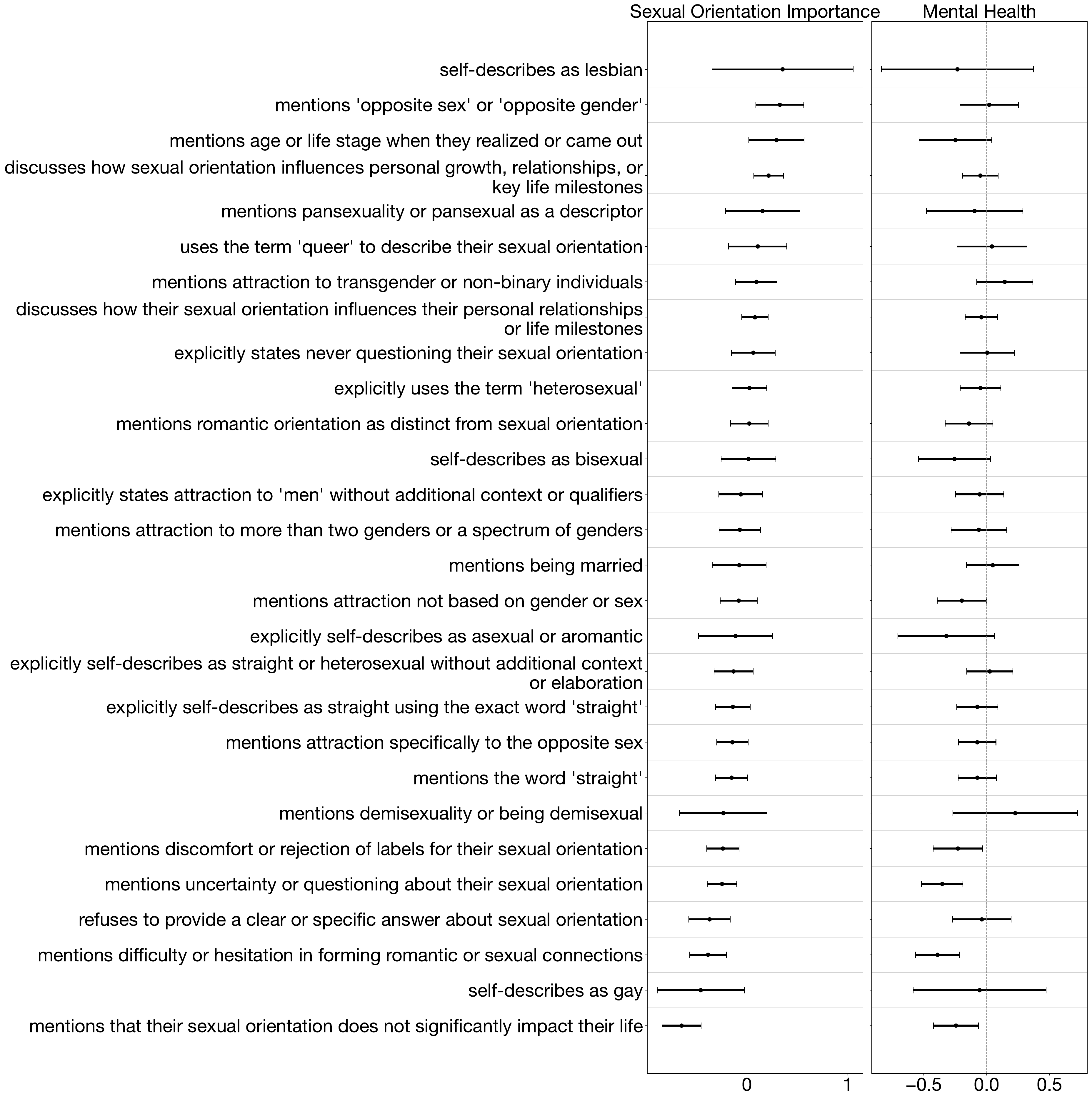}
    \caption{Regression coefficients for free-text sexual orientation themes, when controlling for standardized sexual orientation categories, in regressions to predict the two outcomes---sexual orientation importance and mental health---for which free-text themes significantly improved model fit after multiple hypothesis correction. We estimate models for each theme individually because themes may be collinear, complicating interpretation of multivariate models. Both point estimates and 95\% confidence intervals are plotted (uncorrected for multiple hypotheses). We report HC3 heteroskedasticity-consistent standard errors. Themes are ordered by the estimated coefficient for Sexual Orientation Importance (largest to smallest).}
    \label{fig:sexual-orientation-theme-predictiveness}
\end{figure}

Here we provide additional details supporting the 
analyses which examine whether free-text themes can help explain additional variation in key health and social measures. We report full regression results for all outcome-identity pairs and visualize the associations between free-text themes and life outcomes after controlling for standardized identity categories. Table \ref{tab:nested-f-results} presents the complete nested model comparison results, while Figures \ref{fig:race-theme-predictiveness}-\ref{fig:sexual-orientation-theme-predictiveness} show theme-level associations for race, gender, and sexual orientation, respectively.

\begin{table}[htbp]
\centering
\resizebox{0.8\textwidth}{!}{
\begin{tabular}{lll}
\toprule
\textbf{Identity} & \textbf{Outcome} & \textbf{Relative $R^2$ Increase ($\times$)} \\
\midrule
\multirow{6}{*}{Race}
& Mental Health & 1.5$\times$ (0.04 $\rightarrow$ 0.06)$^{*}$ \\
& Physical Health & 1.9$\times$ (0.02 $\rightarrow$ 0.05)$^{*}$ \\
& Life Satisfaction & 1.9$\times$ (0.01 $\rightarrow$ 0.03) \\
& Everyday Discrimination & 0.8$\times$ (0.02 $\rightarrow$ 0.02) \\
& Income & 1.5$\times$ (0.02 $\rightarrow$ 0.04) \\
& Identity Importance & 1.1$\times$ (0.30 $\rightarrow$ 0.33)$^{***}$ \\
\midrule
\multirow{6}{*}{Gender}
& Mental Health & 1.4$\times$ (0.07 $\rightarrow$ 0.10)$^{***}$ \\
& Physical Health & 1.7$\times$ (0.05 $\rightarrow$ 0.08)$^{***}$ \\
& Life Satisfaction & 1.5$\times$ (0.06 $\rightarrow$ 0.10)$^{***}$ \\
& Everyday Discrimination & 1.1$\times$ (0.03 $\rightarrow$ 0.03) \\
& Income & 1.3$\times$ (0.04 $\rightarrow$ 0.05) \\
& Identity Importance & 1.8$\times$ (0.05 $\rightarrow$ 0.09)$^{***}$ \\
\midrule
\multirow{6}{*}{Sexual Orientation}
& Mental Health & 1.3$\times$ (0.09 $\rightarrow$ 0.11)$^{*}$ \\
& Physical Health & 1.1$\times$ (0.08 $\rightarrow$ 0.09) \\
& Life Satisfaction & 1.2$\times$ (0.08 $\rightarrow$ 0.09) \\
& Everyday Discrimination & 1.2$\times$ (0.04 $\rightarrow$ 0.05) \\
& Income & 1.1$\times$ (0.05 $\rightarrow$ 0.06) \\
& Identity Importance & 3.1$\times$ (0.04 $\rightarrow$ 0.11)$^{***}$ \\
\bottomrule
\end{tabular}}
\caption{\textbf{Free-text themes reveal within-group heterogeneity.} For each identity-outcome pair (rows) we compute the adjusted $R^2$ for two regression models: one that predicts the outcome from the standardized categories alone; and one that additionally includes free-text themes. Each row reports the ratio of adjusted $R^2$ from the two models, with numbers greater than one indicating that the free-text themes increase adjusted $R^2$; the adjusted $R^2$ for both models are reported in parentheses. Stars denote whether there is a statistically significant increase in predictive power; * indicates \( p < 0.05 \); ** \( p < 0.01 \); *** \( p < 0.001 \); all p-values are computed using F-tests for improvement in model fit with Benjamini-Hochberg multiple hypothesis correction ($FDR=0.05$) \cite{benjamini1995controlling}.}
\label{tab:nested-f-results}
\end{table}

We find that race, gender, and sexual orientation themes consistently predict identity importance. Participants expressing themes like ``mentions being Mexican-American or Mexican'', ``mentions pride in Black or African American identity'', and ``describes beliefs or values associated with their race'' were strongly associated with race importance (Figure \ref{fig:race-theme-predictiveness}). In contrast, the strongest negatively associated theme was ``explicitly states a lack of cultural or ethnic identity and traditions''. For gender, a cluster of related themes were negatively associated with identity importance: themes related to gender fluidity, questioning gender identity, and rejecting the concept of gender (Figure \ref{fig:gender-theme-predictiveness}). Similarly, negatively associated sexual orientation themes mentioned that sexual orientation did not significantly impact their life, rejected labels for sexual orientation, or refused to provide a clear answer (Figure \ref{fig:sexual-orientation-theme-predictiveness}). Meanwhile, themes mentioning gender and sexual orientation's influence on life milestones were positively associated with each identity axis' importance. For all three identities---race, gender, and sexual orientation---themes involving explicit mentions of being in the non-minority category were negatively associated with identity importance. 

\clearpage
\section{Computationally identified free-text themes for perceived identity}
\label{supp:perceived_identity}

Here we provide additional details supporting the analyses which examine how individuals describe discrepancies between their self-identified identity and how they believe others perceive them. As mentioned in $\S$\ref{sec:methods-perceive}, we analyze a subset of responses to the question \emph{``How does your self-identified [identity] compare to how you believe others perceive your [identity]?''} from respondents who reported a difference between their self-described identity and how they believe their identity is perceived using our computational framework. We apply the framework separately to perceived identity responses for race, gender, and sexual orientation, using SAEs with parameters $(M=32, K=4)$ for each identity axis. Tables \ref{tab:perceived_race_table}-\ref{tab:perceived_sexual_orientation_table} report the resulting interpretable perceived identity themes for race, gender, and sexual orientation, respectively. These tables are included for reference only and not used in downstream analysis; strikethroughs for excluded themes are therefore not shown.

\subsection{Detailed presentation of perceived identity free-text themes}

\begin{table}[h]
\caption{Computationally identified free-text themes for SAE dimensions trained on perceived race-related free text response embeddings and their associated fidelity (F1 score).}
\label{tab:perceived_race_table}
\begin{tabular}{r|p{12cm}|c}
\toprule
 &  \textbf{Perceived Race Theme} &  \textbf{Fidelity} \\
\midrule
1 & mentions connection (or lack thereof) to European heritage or traditions & 0.548 \\
2 & mentions being black or African & 0.916 \\
3 & mentions feeling disconnected from or not fully immersed in their cultural heritage & 0.804 \\
4 & mentions being perceived as white & 0.800 \\
5 & mentions equality or equal treatment & 0.436 \\
6 & mentions uncertainty or lack of knowledge about how others perceive their race/ethnicity & 0.630 \\
7 & mentions alignment or discrepancy between self-identified and perceived race/ethnicity & 0.569 \\
8 & expresses indifference or uncertainty about race or how it is perceived by others & 0.576 \\
9 & mentions Native American tribes or heritage & 0.148 \\
10 & mentions physical appearance or features, such as skin color, body shape, or facial traits & 0.887 \\
11 & mentions the assumption or stereotype that all Asians are Chinese & 0.734 \\
12 & mentions being Latino, Hispanic, or Mexican and how others perceive that identity & 0.940 \\
13 & expresses uncertainty or similarity without elaboration & 0.652 \\
14 & self-identifies with more detailed or nuanced terms for their race or ethnicity than how others perceive them & 0.669 \\
15 & expresses pride in their race, ethnicity, or culture & 0.744 \\
16 & describes differences in perception without specifying detailed personal experiences or examples & 0.479 \\
17 & describes their appearance as racially or ethnically ambiguous & 0.864 \\
18 & mentions being perceived as generically white or European without further specificity & 0.844 \\
19 & mentions not thinking about their race or ethnicity much & 0.658 \\
20 & mentions experiences of discrimination or cultural misunderstanding & 0.336 \\
21 & expresses uncertainty or lack of clarity about their race or ethnicity & 0.555 \\
22 & mentions stereotypes about their race or ethnicity & 0.707 \\
23 & discusses Black identity or perception of Black people & 0.920 \\
24 & mentions the complexity or nuances of their identity being overlooked or misinterpreted by others & 0.487 \\
25 & discusses being perceived or identified as American or not American & 0.710 \\
26 & mentions being perceived as Jewish or references Jewish identity & 0.462 \\
27 & uses brief and vague responses without detailed explanations or context & 0.748 \\
28 & discusses how others stereotype their group in a negative way & 0.540 \\
29 & mentions perceptions of racism or privilege associated with their race or ethnicity & 0.566 \\
30 & mentions assumptions others make about their ethnicity based on appearance or name & 0.632 \\
31 & mentions lack of concern or misunderstanding by others about their race or ethnicity & 0.422 \\
32 & expresses uncertainty or lack of concern about race/ethnicity perception & 0.387 \\
\bottomrule
\end{tabular}
\end{table}

\begin{table}[h]
\caption{Computationally identified free-text themes for SAE dimensions trained on perceived gender-related free text response embeddings and their associated fidelity (F1 score).}
\label{tab:perceived_gender_table}
\begin{tabular}{r|p{12cm}|c}
\toprule
 &  \textbf{Perceived Gender Theme} &  \textbf{Fidelity} \\
\midrule
1 & mentions being perceived with mixed or multiple gender identities & 0.810 \\
2 & mentions societal or cultural expectations related to gender & 0.679 \\
3 & discusses societal perceptions or stereotypes specifically about women & 0.561 \\
4 & mentions societal perceptions or treatment of women & 0.749 \\
5 & mentions discomfort or negative emotions regarding the topic & 0.316 \\
6 & expresses uncertainty or avoidance in response & 0.480 \\
7 & mentions challenges or discomfort related to being perceived as a woman & 0.561 \\
8 & mentions being perceived as cisgender or misgendered & 0.631 \\
9 & mentions being perceived as a specific binary gender (male or female) & 0.726 \\
10 & mentions societal expectations or stereotypes about masculinity & 0.699 \\
11 & expresses uncertainty or lack of knowledge about the topic & 0.395 \\
12 & mentions alignment or similarity between self-perception and others' perception & 0.525 \\
13 & mentions a strong mismatch between self-perception of gender and how others perceive their gender & 0.648 \\
14 & discusses rejection or critique of labels or categorizations related to gender identity & 0.663 \\
15 & mentions hiding or concealing their gender identity & 0.816 \\
16 & discusses the perception of women as weak and contrasts it with their own or women's actual strength or capabilities & 0.588 \\
17 & mentions how others perceive their gender using terms like `man', `woman', `male', `female', or similar descriptors & 0.534 \\
18 & mentions similarity or difference between self-identified gender identity and others' perception & 0.364 \\
19 & discusses being perceived as nonbinary or struggles with nonbinary identity being understood & 0.693 \\
20 & mentions alignment between self-identified gender and others' perception of gender & 0.645 \\
21 & mentions being a tomboy or non-typical gender presentation & 0.330 \\
22 & expresses disregard or indifference towards others' perceptions of their gender identity & 0.443 \\
23 & uses concise or brief responses without elaboration & 0.466 \\
24 & mentions cultural or racial identity & 0.113 \\
25 & challenges or critiques traditional gender roles or stereotypes & 0.594 \\
26 & mentions how others perceive them as conforming to or diverging from societal or traditional gender expectations & 0.427 \\
27 & mentions a mismatch between how they feel internally about their gender and how others perceive their gender & 0.611 \\
28 & mentions being perceived as female or a girl & 0.650 \\
29 & expresses uncertainty or lack of knowledge about the topic & 0.531 \\
30 & mentions assumptions or perceptions about sexual orientation or behavior & 0.525 \\
31 & mentions physical attributes like body shape, voice, facial hair, or specific features when describing how others perceive their gender identity & 0.753 \\
32 & expresses indifference or lack of strong emotional engagement regarding the topic & 0.375 \\
\bottomrule
\end{tabular}
\end{table}

\begin{table}[h]
\caption{Computationally identified free-text themes for SAE dimensions trained on perceived sexual orientation-related free text response embeddings and their associated fidelity (F1 score).}
\label{tab:perceived_sexual_orientation_table}
\begin{tabular}{r|p{12cm}|c}
\toprule
 &  \textbf{Perceived Sexual Orientation Theme} &  \textbf{Fidelity} \\
\midrule
1 & avoids explicitly identifying or discussing specific sexual orientations & 0.486 \\
2 & expresses indifference or lack of concern about others' perceptions of their sexual orientation & 0.748 \\
3 & mentions connection or comparison between sexual orientation and gender identity & 0.699 \\
4 & mentions how others assume they are heterosexual due to appearances or societal norms & 0.652 \\
5 & mentions being perceived as straight due to being married to a man & 0.828 \\
6 & mentions being perceived as lesbian or bisexual & 0.732 \\
7 & expresses uncertainty with the phrase `not sure' or a synonym & 0.723 \\
8 & expresses uncertainty or lack of knowledge explicitly & 0.723 \\
9 & mentions others' views about heterosexuality or straightness & 0.517 \\
10 & mentions telling or not telling others about their sexual orientation & 0.820 \\
11 & mentions how others misperceive their sexual orientation & 0.487 \\
12 & mentions alignment between self-identified sexual orientation and others' perception & 0.550 \\
13 & mentions assumptions made by others about their sexual orientation & 0.571 \\
14 & expresses uncertainty or lack of knowledge about how others perceive their sexual orientation & 0.490 \\
15 & expresses uncertainty or lack of knowledge about their sexual orientation or how others perceive it & 0.450 \\
16 & mentions being in or perceived as being in a relationship that contrasts with their self-identified sexual orientation & 0.649 \\
17 & mentions bisexuality or being attracted to multiple genders & 0.747 \\
18 & mentions being perceived as gay or a different sexual orientation than self-identified & 0.566 \\
19 & expresses uncertainty or lack of knowledge & 0.590 \\
20 & mentions asexuality or related terms (e.g., asexual, demisexual, ace) & 0.823 \\
21 & mentions perceptions or assumptions based on stereotypes & 0.420 \\
22 & mentions being transgender or t4t & 0.113 \\
23 & expresses uncertainty or lack of knowledge about others' perceptions & 0.514 \\
24 & expresses uncertainty or lack of knowledge about how others perceive their sexual orientation & 0.307 \\
25 & mentions uncertainty or doubt about how others perceive their sexual orientation & 0.360 \\
26 & mentions how others perceive them as different or unusual & 0.550 \\
27 & mentions the concept of `difference' or `different' & 0.753 \\
28 & discusses misconceptions or stereotypes about bisexuality & 0.666 \\
29 & explicitly states that their self-identified sexual orientation is the same as how others perceive it & 0.077 \\
30 & expresses indifference or positive emotion about their sexual orientation & 0.355 \\
31 & mentions similarity or sameness in perception & 0.534 \\
32 & uses vague or ambiguous language to describe orientation or perception & 0.246 \\
\bottomrule
\end{tabular}
\end{table}
\clearpage

\section{Comparison to baseline methods}\label{supp:baseline-comp}

We compare our framework to three widely used baselines: LDA \cite{blei2003latent}; BERTopic \cite{grootendorst2022bertopic}; and TopicGPT \cite{pham-etal-2024-topicgpt}. In \S\ref{supp:baseline_hyperparams}, we describe the hyperparameters that we use to run each baseline; for all baselines, we use standard hyperparameters. In \S\ref{supp:baseline_results}, we compare the themes each baseline recovers to the free-text themes our framework recovers.\footnote{A related approach is \citeauthor{carlson2025making} (\citeyear{carlson2025making}), which uses pretrained SAEs \cite{lieberum2024gemma} to extract themes from text data. We do not compare to this method because it relies on pretrained SAEs, and thus generates an extremely large number of themes (more than 10,000) that reflect concepts that commonly occur in general text corpora---not necessarily those that matter for a specific survey context. The large number of non-specific themes would be challenging to analyze in our context. The focus of \citeauthor{carlson2025making} (\citeyear{carlson2025making}) is also somewhat different from our own---in particular, it focuses on rigorous multiple-hypothesis testing with very large numbers of themes, which is a less relevant consideration in our setting \cite{carlson2025making}.}

\subsection{Hyperparameters}
\label{supp:baseline_hyperparams}

\noindent \textbf{LDA (\citeauthor{blei2003latent}, \citeyear{blei2003latent})}. 
LDA produces a fixed number of themes---each represented as a ranked list of words---and assigns each document a mixture over these themes.
We use a Gibbs-sampling implementation of LDA via the \texttt{tomotopy} library, setting the number of themes to match the 32 SAE dimensions used in our computational framework. Following recommendations by \citeauthor{antoniak2022topic} (\citeyear{antoniak2022topic}) and \citeauthor{schofield2017pulling} (\citeyear{schofield2017pulling}), we train on text with stopwords and lemmatization intact, applying only lowercasing and the removal of punctuation and numerical characters. We use symmetric Dirichlet priors ($\alpha=1.0, \beta=0.1$) and run the sampler for 2,000 iterations, as used by \citeauthor{pham-etal-2024-topicgpt} (\citeyear{pham-etal-2024-topicgpt}). Consistent with \citeauthor{schofield2017pulling} (\citeyear{schofield2017pulling}), who find that removing stopwords is largely useful for interpreting the results, we apply a post-training filtering step to remove the top 50 high-frequency stopwords (e.g., ``i", ``and", ``a", ...) from the final theme word lists.

\medskip

\noindent \textbf{BERTopic (\citeauthor{grootendorst2022bertopic}, \citeyear{grootendorst2022bertopic})}.
BERTopic produces a variable number of themes---each represented by a set of terms weighted by how unique they are to the cluster---and assigns each document to a single theme. We run BERTopic on the same pretrained OpenAI text embeddings used in the \textsc{In Your Own Words} framework ($\S$\ref{sec:methods-framework-sae}), with UMAP for dimensionality reduction, HDBSCAN for clustering, and c-TF-IDF for theme representation. This aligns with prior work (e.g., \cite{hypothesaes, pham-etal-2024-topicgpt}).
All components are left at their default settings except for the HDBSCAN minimum cluster size hyperparameter, which controls theme granularity.\footnote{\href{https://github.com/MaartenGr/BERTopic}{https://github.com/MaartenGr/BERTopic}} 
We set the minimum cluster size to 7, which generates a comparable number of themes (23-29 per identity axis) to the number of free-text themes retained by our computational framework after filtering (26-28 per axis). After fitting BERTopic, we fine-tune the themes with an LLM label, using BERTopic's built-in LLM representation function.
\medskip

\noindent \textbf{TopicGPT (\citeauthor{pham-etal-2024-topicgpt}, \citeyear{pham-etal-2024-topicgpt})}.
TopicGPT produces a variable number of themes---each represented by a natural language label and a descriptive summary generated by an LLM---and assigns each document a set of relevant themes based on prompt-based reasoning.
We apply the official TopicGPT implementation to the raw free-text responses using its default settings, as defined in the provided prompt templates.\footnote{\href{https://github.com/chtmp223/topicGPT}{https://github.com/chtmp223/topicGPT}}. By default, TopicGPT employs an early stopping criterion during theme generation: it incrementally proposes themes from a subset of the corpus and halts when few new themes are discovered, rather than processing the entire dataset. To assess whether this early stopping behavior affects the number or range of themes produced in our setting, we additionally run TopicGPT without early stopping, allowing the model to process all 1,004 responses when generating themes. We also apply TopicGPT’s ``subtopic generation" functionality (which we refer to as generating subthemes), producing a set of nested subthemes beneath each high-level theme. 

\subsection{Baseline Results}
\label{supp:baseline_results}

\noindent \textbf{LDA}. We report LDA themes in Tables \ref{tab:lda_race_topics}-\ref{tab:lda_sexual_orientation_topics}. The LDA themes exhibit several limitations relative to our framework.
First, a large fraction of themes occur extremely rarely. As shown by the \# Responses column in Table \ref{tab:lda_race_topics}, while a small number of dominant themes capture hundreds of responses (e.g., themes 10, 16, and 25), many themes occur very rarely. In particular, 13 out of 32 race themes are not the primary theme for \emph{any} respondent. In contrast, our framework produces a more continuous distribution of responses across free-text themes, avoiding the extreme concentration and sparsity observed in LDA.
Second, multiple rare themes capture random word associations rather than coherent identity concepts. For example, some themes mix unrelated top terms (e.g., ``bengali" and ``iceland" both appear among top words in race theme 3), while others are dominated by semantically disconnected words (e.g., ``guess", ``instead", ``days", ``adopted", ``racist" in race theme 23) that do not cohere into an interpretable concept.
Third, many themes are represented by uncommon words. For example, in race theme 1, five of the ten highest-probability words occur only once in the race free-text responses; a similar pattern appears in themes 3, 9, 12, 18, 21, 22, 26, and 31.
Finally, because LDA represents themes as unordered word lists inferred from word co-occurrence, the resulting word lists are often difficult to interpret for both humans and LLMs.
In contrast, our framework produces explicit annotations that indicate whether a given response expresses a particular free-text theme, making it straightforward to assess whether a theme meaningfully represents the responses it captures. 

\medskip

\noindent \textbf{BERTopic.} We report BERTopic themes in Tables \ref{tab:bertopic_race_topics}-\ref{tab:bertopic_sexual_orientation_topics}. While BERTopic produces coherent clusters of semantically similar responses---often recovering major racial and ethnicity categories (e.g., ``Asian American Identity and Cultural Practices'', ``Black American Identity and Heritage'')---we observe several limitations when applying it to our data. 
First, because BERTopic relies on single-theme assignment (hard clustering), it struggles to capture nuanced identity concepts that span categories and frequently co-occur within the same response. In the race axis, for example, BERTopic does not surface concepts identified by our framework, such as ``mentions feelings of not fully belonging'' or ``mentions being first-, second-, or third-generation American or immigrant'', because these themes are typically discussed by responses that also mention racial or ethnicity categories. 
Single-theme assignment also obscures responses with multiple overlapping identities; for example, our sexual orientation data shows that 20 out of 74 responses mentioning ``queer" also specifically mentioned ``bisexual", a relationship that a single topic model cannot represent.
Second, BERTopic classifies 289 (29\%) of the race responses, 175 (17\%) of the gender responses, and 299 (30\%) of the sexual orientation responses as noise, and does not assign them any themes: this means that these responses are effectively ignored and cannot be used in downstream analysis.
In contrast, our framework can assign multiple free-text themes per response and explicitly represent theme co-occurrence, allowing a richer set of recurring, experiential dimensions---such as belonging, language use, and migration---to be captured even when respondents also mention standardized identity categories.
Rather than discarding responses as ``noise", we include all responses in the structured output, explicitly marking those that do not strongly express any theme instead of excluding them.
\medskip

\noindent \textbf{TopicGPT.} We report TopicGPT themes and subthemes \emph{without} early stopping (i.e., the model processes the full dataset) in Tables \ref{tab:topicgpt_race}-\ref{tab:topicgpt_sexual_orientation}. 
TopicGPT produces themes and subthemes that are either broad and generic or primarily recover explicit identity categories.
Under the default configuration \emph{with} early stopping (i.e., theme generation stops once 100 consecutive iterations yield no new themes), TopicGPT produces only two themes for race: ``Cultural Identity" and ``Indigenous Peoples"; six themes for gender: ``Gender Identity", ``Gender Roles", ``Asexuality", ``Race", ``Health", and ``Outdoor Activities"; and a single theme for sexual orientation: ``Sexuality". While the theme ``Sexuality" contains six subthemes which closely mirror standardized identity labels (e.g., ``Heterosexuality", ``Homosexuality", ``Bisexuality"), both race and gender subthemes are more difficult to interpret and do not consistently correspond to meaningful identity-related dimensions. Under early stopping, TopicGPT is also sensitive to minor setup choices: shuffling the order of responses changes the specific themes produced, though they remain broad and weakly relevant; and testing several identity-specific prompt variants does not achieve more consistent themes and subthemes.

\emph{Without} early stopping, the resulting theme structure is highly uneven across identity axes. For race, TopicGPT produces only two themes (``Cultural Identity" and ``Agriculture"), with ``Cultural Identity" assigned to 99\% of responses, indicating that nearly the entire dataset is collapsed into a single, overly broad theme. For gender, TopicGPT generates eleven themes, but the dominant theme (``Gender Identity") is assigned to 98\% of responses, again resulting in minimal differentiation across responses. In contrast, for sexual orientation, TopicGPT produces over 50 themes, many of which occur infrequently, overlap in content, and redundantly encode similar identity labels (e.g., ``Sexuality", ``Sexual Orientation").
Although TopicGPT’s subtheme generation can recover specific identity categories within the highest-frequency themes, this added hierarchical structure does not fix its core issue in our setting: the theme-subtheme hierarchy is not consistent across the race, gender, and sexual orientation responses. Instead, it introduces additional redundancy, with some concepts appearing as both themes and subthemes (e.g., ``Pansexuality" appears as a subtheme under ``Sexuality", but also appears as a redundant standalone theme with 6.8\% of responses). 
TopicGPT also produces themes that are weakly aligned with the free-text responses (e.g.,``Agriculture" for race; ``Social Activities" and ``Outdoor Activities" for gender). Together, these issues make TopicGPT difficult to use for systematic analysis or to adapt to specific analytic goals.
In contrast, our computational framework produces a stable set of interpretable free-text themes, with explicit control over their granularity. This supports downstream analysis without the uneven coverage and sensitivity to response ordering we observe in TopicGPT.

\begin{table}
\caption{\textbf{Top Words for LDA Model Learned from Race Free-Text Responses.} Themes were learned using Gibbs sampling on raw text with stopwords and lemmatization intact. The ``\# Responses" column indicates the number of responses where that theme was the primary (highest probability) assignment. ``Top 10 Race Words" displays the most probable terms for each theme after a post-training filter removed the top 50 high-frequency stopwords for interpretability.}
\label{tab:lda_race_topics}
\begin{tabular}{rrp{0.75\textwidth}}
\toprule
 & \textbf{\# Responses} & \textbf{Top 10 Race Words} \\
\midrule
1 & 0 & appearance, someone, colonizers, system, language, identify, advantage, colonized, history--, unequitable \\
2 & 0 & disorder, narratives, lighthearted, american, white, family, race, english, ethnicity, black \\
3 & 0 & bengali, iceland, bangladeshi, speaking, southern, geneology, icelandic, icelantic, varieties, organizing \\
4 & 31 & side, mother, father, dad, mom, hispanic, half, italian, puerto, family \\
5 & 5 & christmas, holidays, celebrate, thanksgiving, year, new, lunar, chinese, vietnamese, easter \\
6 & 28 & black, african, american, africa, race, experiences, community, skin, americans, caribbean \\
7 & 0 & narratives, american, white, family, race, english, ethnicity, black, speak, culture \\
8 & 0 & holiday, events, custom, farsi, grew, east, fluently, gender, siblings, iran \\
9 & 0 & spirituality, lao, belgian, saxon, financially, expanding, style, fufu, eats, appointments \\
10 & 229 & family, our, speak, english, food, language, traditions, all, people, like \\
11 & 0 & filipinos, ashanti, love, church, occasions, ghanian, while, jesus, specifically, every \\
12 & 1 & when, ’re, freedom, obviously, throughout, factor, which, person, persevere, scraps \\
13 & 0 & jewish, ashkenazi, arab, complex, argentinean, yiddish, since, judaism, jews, religiously \\
14 & 0 & slavic, ukrainian, russian, west, thing, negro, stay, specifically, experience, gather \\
15 & 2 & middle, eastern, arabic, lebanese, arab, east, greek, egyptian, albanian, polish \\
16 & 209 & white, culture, race, feel, american, any, people, really, they, traditions \\
17 & 0 & ones, easier, university, everyday, property, fully, taught, your, day, know \\
18 & 1 & tongan, respect, choctaw, than, around, marriage, till, females, subjects, ebb \\
19 & 1 & another, each, light, major, humans, vehicle, bills, together, backgrounds, fair \\
20 & 110 & describe, race, white, american, ethnicity, hispanic, mexican, mixed, identify, asian \\
21 & 1 & music, temple, event, scottish, playing, there, burning, incense, smell, praying \\
22 & 1 & loves, our, father, drawl, fields, dogs, hot, blue, rhythm, teaching \\
23 & 0 & overly, guess, instead, days, adopted, racist, mine, however, although, automatically \\
24 & 44 & cultural, heritage, traditions, identity, history, black, ethnicity, experiences, shaped, community \\
25 & 195 & white, american, european, caucasian, english, irish, german, heritage, family, ancestry \\
26 & 0 & japanese, osage, world, without, de, history, no, rituals, power, multiethnic \\
27 & 1 & asian, korean, chinese, mandarin, japanese, culturally, china, though, taiwanese, thai \\
28 & 0 & then, vietnamese, guest, heard, eyes, fish, sauce, maintaining, mother, ozarks \\
29 & 2 & native, tribe, indigenous, tribal, land, ancestors, away, tribes, cherokee, our \\
30 & 1 & children, woman, gaelic, now, her, having, stories, carry, bilingual, conservative \\
31 & 1 & dutch, hard, matter, christmas, describe, language, yourself, rely, trusted, geographic \\
32 & 141 & born, american, raised, grew, states, speak, parents, mexican, were, united \\
\bottomrule
\end{tabular}
\end{table}

\begin{table}
\caption{\textbf{Top Words for LDA Model Learned from Gender Free-Text Responses.} Themes were learned using Gibbs sampling on raw text with stopwords and lemmatization intact. The ``\# Responses" column indicates the number of responses where that theme was the primary (highest probability) assignment. ``Top 10 Gender Words" displays the most probable terms for each theme after a post-training filter removed the top 50 high-frequency stopwords for interpretability.}
\label{tab:lda_gender_topics}
\begin{tabular}{rrp{0.75\textwidth}}
\toprule
 & \textbf{\# Responses} & \textbf{Top 10 Gender Words} \\
\midrule
1 & 13 & can, family, women, men, our, hard, they, when, believe, you \\
2 & 1 & black, higher, when, based, women, femme, blackness, internal, color, respected \\
3 & 1 & androgyny, work, nothing, physical, connection, desire, stronger, gain, muscle, exude \\
4 & 1 & side, mujer, child, american, towards, fashion, sort, lean, clothes, an \\
5 & 1 & lesbian, butch, role, stemme, influences, prefer, presentation, word, fem, black \\
6 & 1 & answer, spirit, dual, aspects, sports, future, choice, options, below, odd \\
7 & 3 & pronouns, they, use, she, her, them, he, binary, him, people \\
8 & 386 & female, male, woman, born, identify, man, never, cis, felt, cisgender \\
9 & 31 & life, gender, identity, any, had, influences, due, their, work, milestones \\
10 & 0 & strict, ways, men, some, defined, meant, supposed, story, designed, ground \\
11 & 0 & gay, color, guy, early, towards, internal, stuck, alter, esteem, factors \\
12 & 61 & gender, identity, expectations, life, roles, experiences, societal, world, shaped, masculinity \\
13 & 0 & feminine, wear, makeup, dress, typically, wearing, clothing, present, ovaries, use \\
14 & 0 & you, your, should, yourself, nt, god, crap, 're, new, here \\
15 & 5 & did, felt, grew, growing, age, boy, life, never, started, young \\
16 & 0 & contexts, made, childhood, describe, n’t, frilly, confident, detach, girlhood, engrained \\
17 & 0 & sexual, enough, eyes, get, long, hair, old, choices, stay, desire \\
18 & 0 & going, sexless, answer, alone, form, short, case, falls, course, acknowledgment \\
19 & 0 & among, gender, female, male, woman, identity, identify, feel, born, man \\
20 & 1 & things, relationships, doing, subculture, induce, girl, compensate, displays, overt, flaunt \\
21 & 0 & prefer, work, birthday, girls, hunting, enjoy, colored, types, requires, party \\
22 & 0 & mix, ethnicity, guys, groups, wish, better, inside, lead, labor, surgeries \\
23 & 0 & among, gender, female, male, woman, identity, identify, feel, born, man \\
24 & 19 & gender, birth, assigned, identity, sex, female, male, cisgender, there, may \\
25 & 0 & masculinity, lot, instinct, yang, yin, depictions, dichotomous, intent, exists, deconstruction \\
26 & 0 & context, age, young, inconsistent, usa, laziness, pure, division, ingrained, disturbed \\
27 & 479 & gender, feel, like, identity, man, think, binary, non, masculine, identify \\
28 & 0 & job, accepted, knowing, running, far, actual, fall, surveys, expectations, their \\
29 & 0 & surgery, years, socially, transitioned, hrt, null, friends, traditionally, testosterone, forced \\
30 & 1 & menopause, wanted, gone, ’ve, menstrual, truly, intelligence, inclined, interest, cycle \\
31 & 0 & people, latinx, independent, any, bunch, employer, recent, watched, stood, customers \\
32 & 0 & among, gender, female, male, woman, identity, identify, feel, born, man \\
\bottomrule
\end{tabular}
\end{table}

\begin{table}
\caption{\textbf{Top Words for LDA Model Learned from Sexual Orientation Free-Text Responses.} Themes were learned using Gibbs sampling on raw text with stopwords and lemmatization intact. The ``\# Responses" column indicates the number of responses where that theme was the primary (highest probability) assignment. ``Top 10 Sexual Orientation Words" displays the most probable terms for each theme after a post-training filter removed the top 50 high-frequency stopwords for interpretability.}
\label{tab:lda_sexual_orientation_topics}
\begin{tabular}{rrp{0.75\textwidth}}
\toprule
 & \textbf{\# Responses} & \textbf{Top 10 Sexual Orientation Words} \\
\midrule
1 & 339 & asexual, feel, really, like, sexual, attraction, think, if, 've, can \\
2 & 1 & rainbow, christian, no, he, focusing, sunglasses, pins, covered, telling, human \\
3 & 0 & base, sexual, attracted, straight, heterosexual, orientation, men, women, people, like \\
4 & 1 & no, back, problem, effeminate, 're, hope, nowadays, fits, gay, identity \\
5 & 0 & intimacy, getting, trauma, sexuality, labels, beyond, must, last, kept, earlier \\
6 & 0 & base, sexual, attracted, straight, heterosexual, orientation, men, women, people, like \\
7 & 2 & should, getting, bionery, go, biologically, milestone, feel, seem, still, control \\
8 & 0 & many, loved, penis, child, bring, him, job, races, their, attempts \\
9 & 0 & nuanced, could, conform, leads, had, never, exf3pt, serial, disgust, vaginas \\
10 & 0 & here, want, great, worry, panromatic, hypersexual, ol', thnere, rampant, provided \\
11 & 0 & these, may, coming, does, extremely, platform, digital, irl, attachment, ban \\
12 & 54 & sex, opposite, any, sexual, same, relationships, heterosexual, romantic, no, had \\
13 & 0 & base, sexual, attracted, straight, heterosexual, orientation, men, women, people, like \\
14 & 0 & 'll, what, logical, proactively, everybody, erotica, het, surgery, idk, convenience \\
15 & 1 & several, stop, simply, medication, an, mild, male, ocd, bout, seduce \\
16 & 0 & base, sexual, attracted, straight, heterosexual, orientation, men, women, people, like \\
17 & 0 & identified, future, decided, late, fantasizing, what, impacts, excites, stake, crosses \\
18 & 0 & base, sexual, attracted, straight, heterosexual, orientation, men, women, people, like \\
19 & 1 & aroace, slightly, e.g., repulsed, biologically, clear, neutral, gray, identity, became \\
20 & 115 & attracted, people, bisexual, gender, queer, pansexual, identify, men, genders, gay \\
21 & 334 & straight, heterosexual, attracted, men, women, woman, man, like, male, female \\
22 & 0 & causes, soon, aegosexual, change, marry, its, mainly, structure, pansexual, dating \\
23 & 0 & instead, died, physically, edging, casual, phases, perfectly, bones, arts, literature \\
24 & 1 & culture, touch, media, dating, social, teen, brings, can, some, admire \\
25 & 58 & orientation, sexual, life, straight, describe, influence, influences, due, any, many \\
26 & 1 & our, however, room, aka, expression, conservative, red, now, unlikely, research \\
27 & 0 & every, choose, general, seen, though, while, opened, prerogative, connecting, sexy \\
28 & 63 & sexual, relationships, orientation, experiences, others, identify, romantic, identity, life, gender \\
29 & 2 & ace, n’t, spectrum, world, off, guys, turned, focused, when, relating \\
30 & 0 & toward, honestly, hot, practice, closest, since, meaning, intercourse, recognized, referred \\
31 & 0 & tendencies, roles, implications, equal, criticism, consist, traditionally, living, curiosity, follow \\
32 & 31 & had, did, gay, out, years, thought, never, bisexual, realized, felt \\
\bottomrule
\end{tabular}
\end{table}


\begin{table}[h]
\caption{\textbf{BERTopic Representations Learned from Race Free-Text Responses.} The ``\# Responses" column counts responses assigned to each unique cluster. Because the underlying clustering is noise-aware, 289 responses failing to meet the density threshold for any theme (representing roughly 29\% of the corpus) are classified as noise and excluded from this table for clarity.}
\label{tab:bertopic_race_topics}
\begin{tabular}{rrp{0.6\textwidth}}
\toprule
 & \textbf{\# Responses} & \textbf{Theme Representation} \\
\midrule
1 & 178 & Black American Identity and Heritage \\
2 & 64 & Asian American Identity and Cultural Practices \\
3 & 44 & European Ancestry of White Americans \\
4 & 39 & Cultural Heritage and Ethnic Identity in American Families \\
5 & 39 & White American English Speakers \\
6 & 32 & racial and ethnic self-identification \\
7 & 28 & Native American Identity and Cultural Heritage \\
8 & 27 & Middle Eastern Arab Identity and Heritage \\
9 & 27 & Mexican-American Identity and Language Experience \\
10 & 25 & South Asian Cultural Identities in America \\
11 & 25 & American Family Holiday Celebrations \\
12 & 23 & Cultural Identity and Ethnic Traditions \\
13 & 21 & Experiences of White Privilege and Identity \\
14 & 20 & Filipino American Identity and Language Preservation \\
15 & 20 & Identity and Ethnic Heritage in Family Background \\
16 & 18 & Complex Identities of Ashkenazi and Mixed Ethnicity \\
17 & 14 & White Racial Classification and Its Contexts \\
18 & 12 & Ancestral Heritage and Identity \\
19 & 11 & White American Identity and Cultural Detachment \\
20 & 11 & Cultural Traditions and Identity of White Americans \\
21 & 10 & Mixed Ethnic Background and Cultural Identity \\
22 & 10 & Reflections on Race and Racism in Society \\
23 & 9 & Mixed Race Cultural Identity \\
24 & 8 & Racial and Ethnic Identity and Its Impact on Personal Experiences \\
\bottomrule
\end{tabular}
\end{table}

\begin{table}
\caption{\textbf{BERTopic Representations Learned from Gender Free-Text Responses.} The ``\# Responses" column counts responses assigned to each unique cluster. Because the underlying clustering is noise-aware, 175 responses failing to meet the density threshold for any theme (representing roughly 17\% of the corpus) are classified as noise and excluded from this table for clarity.}
\label{tab:bertopic_gender_topics}
\begin{tabular}{rrp{0.6\textwidth}}
\toprule
 & \textbf{\# Responses} & \textbf{Theme Representation} \\
\midrule
1 & 237 & Gender Identity and Expression \\
2 & 67 & Impact of Gender Identity on Women's Life Experiences \\
3 & 63 & Cisgender Female Identity \\
4 & 59 & Gender Identity and Biological Sex \\
5 & 54 & Male Identity and Biological Sex \\
6 & 48 & Female Gender Identity \\
7 & 40 & Heterosexual Identity and Gender \\
8 & 35 & Male Gender Identity and Masculinity \\
9 & 26 & Male Identity and Outdoor Lifestyle \\
10 & 25 & Gender Identity and Self-Perception \\
11 & 23 & Female Identity and Motherhood \\
12 & 22 & cisgender male identity \\
13 & 21 & Female Empowerment and Sensitivity \\
14 & 17 & Female Gender Identity Consistency \\
15 & 17 & Feminine Identity and Preferences \\
16 & 16 & Male Identity and Family Responsibility \\
17 & 16 & Male Dominance and Advantage in Social Contexts \\
18 & 10 & Male Gender Identity in Native Traditions \\
19 & 10 & Challenges and Experiences of Identifying as a Woman \\
20 & 9 & Female Identity and Motherhood \\
21 & 7 & Gender Identity and Stability Among Males \\
22 & 7 & Gender Identity and Its Impact on Life \\
\bottomrule
\end{tabular}
\end{table}

\begin{table}
\caption{\textbf{BERTopic Representations Learned from Sexual Orientation Free-Text Responses.} The ``\# Responses" column counts responses assigned to each unique cluster. Because the underlying clustering is noise-aware, 299 responses failing to meet the density threshold for any theme (representing roughly 30\% of the corpus) are classified as noise and excluded from this table for clarity.}
\label{tab:bertopic_sexual_orientation_topics}
\begin{tabular}{rrp{0.6\textwidth}}
\toprule
 & \textbf{\# Responses} & \textbf{Theme Representation} \\
\midrule
1 & 96 & Heterosexual Attraction to Men \\
2 & 87 & Asexual and Aromantic Experiences \\
3 & 48 & Bisexual Attraction to Men and Women \\
4 & 37 & Heterosexual Identity \\
5 & 33 & Demisexual and Emotional Attraction \\
6 & 33 & Influence of Heterosexuality on Personal Life \\
7 & 29 & Queer Sexuality and Identity \\
8 & 24 & Lesbian Identity and Attraction \\
9 & 24 & Lesbian Identity and Coming Out Experiences \\
10 & 24 & Identification as Gay Man \\
11 & 23 & heterosexual marriage and relationships \\
12 & 20 & Exploration of Pansexuality and Attraction Across Genders \\
13 & 20 & Sexual Orientation and Identity \\
14 & 20 & Experiences of Bisexuality and Family Acceptance \\
15 & 18 & Exploration of Sexual Orientation Beyond Heterosexuality \\
16 & 17 & Heterosexual Identity and Orientation Certainty \\
17 & 17 & Heterosexual Orientation of Women \\
18 & 16 & Understanding Straight Identity and Its Influence on Relationships \\
19 & 16 & The Role of Sexual Orientation in Shaping Identity and Relationships \\
20 & 16 & Navigating Sexuality and Labels with Fluidity \\
21 & 14 & Pansexual Attraction and Identity \\
22 & 12 & Heterosexual Relationships and Experiences with Men \\
23 & 11 & Understanding Bisexuality and Attraction \\
24 & 11 & Experiences of Straight Male Sexual Orientation and Relationships \\
25 & 10 & Heterosexuality and Cultural Perspectives on Sexual Issues \\
26 & 10 & Sexual Orientation - Heterosexual Descriptions \\
27 & 10 & Attraction Spectrum and Gender Identity Exploration \\
28 & 9 & Straight Active Black \\
\bottomrule
\end{tabular}
\end{table}


\begin{table}[htbp]
\centering
\caption{\textbf{TopicGPT Hierarchical Themes Learned from Race Free-Text Responses.} [1] denotes top-level theme; [2] indicates subthemes grouped under the preceding themes. Counts indicate the number of responses in which each theme was identified; responses may contribute to multiple themes.}
\label{tab:topicgpt_race}
\begin{tabularx}{\textwidth}{l r >{\RaggedRight\arraybackslash}X}
\toprule
\textbf{Level} & \textbf{Count} & \textbf{Theme / Subtheme Name \& Description} \\
\midrule
\textbf{[1]} & \textbf{989} & \textbf{Cultural Identity:} Mentions the significance of cultural heritage and personal identity in shaping experiences and perspectives. \\
\addlinespace[0.4em]
\hspace{1em} [2] & 1 & $\llcorner$ \textit{Race and Ethnicity:} How individuals identify with race/ethnicity, including cultural heritage and experiences. \\
\hspace{1em} [2] & 1 & $\llcorner$ \textit{Language and Communication:} Role of language spoken at home and its impact on cultural connection. \\
\hspace{1em} [2] & 1 & $\llcorner$ \textit{Cultural Practices and Traditions:} Cultural practices (food, celebrations, customs) shaping identity. \\
\midrule
\textbf{[1]} & \textbf{2} & \textbf{Agriculture:} Mentions the context of living in a rural farming community. \\
\bottomrule
\end{tabularx}
\end{table}

\begin{table}[htbp]
\centering
\caption{\textbf{TopicGPT Hierarchical Themes Learned from Gender Free-Text Responses.} [1] denotes top-level themes; [2] indicates subthemes grouped under the preceding theme. To improve readability, only themes appearing in ten or more responses are shown individually; all themes with fewer than ten occurrences are grouped under ``Other Themes.” Counts indicate the number of responses in which each theme was identified; responses may contribute to multiple theme.}
\label{tab:topicgpt_gender}
\begin{tabularx}{\textwidth}{l r >{\RaggedRight\arraybackslash}X}
\toprule
\textbf{Level} & \textbf{Count} & \textbf{Theme / Subtheme Name \& Description} \\
\midrule
\textbf{[1]} & \textbf{981} & \textbf{Gender Identity:} Discusses the experiences and perspectives shaped by one's gender identity, including the diversity of expressions and identities beyond traditional norms. \\
\addlinespace[0.4em]
\hspace{1em} [2] & 1 & $\llcorner$ \textit{Cisgender:} Refers to individuals whose gender identity matches the sex they were assigned at birth. \\
\hspace{1em} [2] & 1 & $\llcorner$ \textit{Transgender:} Refers to individuals whose gender identity differs from the sex they were assigned at birth. \\
\hspace{1em} [2] & 1 & $\llcorner$ \textit{Non-binary:} Refers to gender identities that do not fit within the traditional binary of male and female. \\
\hspace{1em} [2] & 1 & $\llcorner$ \textit{Gender Fluid:} Refers to a gender identity that is not fixed and can change over time or depending on the situation. \\
\hspace{1em} [2] & 1 & $\llcorner$ \textit{Agender:} Refers to individuals who do not identify with any gender. \\
\hspace{1em} [2] & 1 & $\llcorner$ \textit{Genderqueer:} Refers to a gender identity that is not exclusively male or female and may encompass a combination of genders. \\
\hspace{1em} [2] & 1 & $\llcorner$ \textit{Gender Expression:} Refers to the external manifestation of gender, expressed through behavior, clothing, haircut, voice, and other forms of presentation. \\
\midrule
\textbf{[1]} & 15 & \textbf{Social Activities:} Mentions going out and having fun with friends. \\
\addlinespace[0.4em]
\hspace{1em} [2] & 1 & $\llcorner$ \textit{Gender Identity:} Discusses how gender identity influences social interactions, personal experiences, and societal roles. \\
\hspace{1em} [2] & 1 & $\llcorner$ \textit{Socializing and Leisure:} Involves activities related to spending time with friends and engaging in leisure activities. \\
\hspace{1em} [2] & 1 & $\llcorner$ \textit{Gender Roles and Expectations:} Explores societal expectations and roles based on gender, including challenges and stereotypes. \\
\midrule
\textbf{[1]} & 11 & \textbf{Outdoor Activities:} Mentions experiences in the wilderness and exposure to nature. \\
\addlinespace[0.4em]
\hspace{1em} [2] & 1 & $\llcorner$ \textit{Wilderness Activities:} Involves spending time in natural environments, such as camping and fishing. \\
\hspace{1em} [2] & 1 & $\llcorner$ \textit{Sports:} Involves physical activities and games, often competitive, such as playing or watching sports. \\
\hspace{1em} [2] & 1 & $\llcorner$ \textit{Hunting and Fishing:} Involves activities related to pursuing and capturing or killing animals for food or sport. \\
\midrule
\textbf{[1]} & 19 & \textbf{Work:} Mentions working in a male-dominated industry, highlighting gender dynamics in the workplace. \\
\addlinespace[0.4em]
\hspace{1em} [2] & 1 & $\llcorner$ \textit{Gender Identity in the Workplace:} Discusses how gender identity influences experiences, challenges, and discrimination in professional settings. \\
\hspace{1em} [2] & 1 & $\llcorner$ \textit{Gender Roles and Expectations:} Explores societal expectations and traditional roles associated with gender, particularly in relation to work and family responsibilities. \\
\midrule
\textbf{[1]} & 30 & \textbf{Family:} Mentions motherhood and the joy of giving birth to children. \\
\addlinespace[0.4em]
\hspace{1em} [2] & 1 & $\llcorner$ \textit{Gender Roles:} Discusses expectations and societal norms associated with being male or female, including roles such as provider, caregiver, and societal perceptions. \\
\hspace{1em} [2] & 1 & $\llcorner$ \textit{Parenthood:} Focuses on the experience and significance of being a parent, including giving birth, raising children, and the impact of parenthood on identity. \\
\hspace{1em} [2] & 1 & $\llcorner$ \textit{Gender Identity:} Explores personal identification with gender, including experiences of being female or male, and how these identities influence personal and social experiences. \\
\midrule
\textbf{[1]} & 16 & \textbf{Cultural Identity:} Mentions the experiences and values related to cultural heritage and identity within the African diaspora. \\
\addlinespace[0.4em]
\hspace{1em} [2] & 1 & $\llcorner$ \textit{Racial Identity:} Discusses the influence of race and ethnicity on personal and cultural experiences. \\
\hspace{1em} [2] & 1 & $\llcorner$ \textit{Gender Identity:} Explores how gender identity shapes personal experiences and societal interactions. \\
\midrule
\textbf{[1]} & 19 & \textbf{Sexual Orientation:} Discusses aspects of sexual attraction and asexuality, as well as companionship beyond sexual relationships. \\
\addlinespace[0.4em]
\hspace{1em} [2] & 1 & $\llcorner$ \textit{Gender Identity:} Discusses how individuals identify their gender, including cisgender, transgender, non-binary, and other identities. \\
\hspace{1em} [2] & 1 & $\llcorner$ \textit{Sexual Orientation:} Covers the spectrum of sexual attraction and orientation, including heterosexual, homosexual, bisexual, pansexual, and asexual orientations. \\
\midrule
\textbf{[1]} & --- & \textbf{Other Themes:} Health (Count: 9), Fashion (Count: 8), Education (Count: 4), and Political Climate (Count: 1). \\

\bottomrule
\end{tabularx}
\end{table}


\FloatBarrier        
{\small
\begin{longtable}{l r >{\RaggedRight\arraybackslash}p{0.72\textwidth}}
\caption{\textbf{TopicGPT Hierarchical Themes Learned from Sexual Orientation Free-Text Responses.} [1] denotes top-level themes; [2] indicates subthemes grouped under the preceding theme. To improve readability, only themes appearing in 100 or more responses are shown individually; all themes with fewer than 100 occurrences are grouped under ``Other Themes.”  Counts indicate the number of responses in which each theme was identified; responses may contribute to multiple themes.}
\label{tab:topicgpt_sexual_orientation} \\
\toprule
\textbf{Level} & \textbf{Count} & \textbf{Theme / Subtheme Name \& Description} \\
\midrule
\endfirsthead

\caption[]{(continued)} \\
\toprule
\textbf{Level} & \textbf{Count} & \textbf{Theme / Subtheme Name \& Description} \\
\midrule
\endhead

\midrule
\multicolumn{3}{r}{\textit{continued on next page}} \\
\endfoot

\bottomrule
\endlastfoot

\textbf{[1]} & \textbf{240} & \textbf{Sexuality:} Discusses the complexities of sexual orientation and the nuances of attraction and relationships. \\
\addlinespace[0.25em]
\hspace{1em} [2] & 1 & $\llcorner$ \textit{Heterosexuality:} Refers to attraction to the opposite gender. \\
\hspace{1em} [2] & 1 & $\llcorner$ \textit{Homosexuality:} Refers to attraction to the same gender. \\
\hspace{1em} [2] & 1 & $\llcorner$ \textit{Bisexuality:} Refers to attraction to both men and women. \\
\hspace{1em} [2] & 1 & $\llcorner$ \textit{Pansexuality:} Refers to attraction to people regardless of gender. \\
\hspace{1em} [2] & 1 & $\llcorner$ \textit{Asexuality:} Refers to little or no sexual attraction to others. \\
\hspace{1em} [2] & 1 & $\llcorner$ \textit{Questioning:} Refers to individuals exploring or uncertain about their sexual orientation. \\
\midrule

\textbf{[1]} & \textbf{126} & \textbf{Asexuality:} Refers to a sexual orientation characterized by a lack of sexual attraction to others. \\
\addlinespace[0.25em]
\hspace{1em} [2] & 1 & $\llcorner$ \textit{Aromanticism:} Mentions lack of romantic attraction, often associated with asexuality. \\
\hspace{1em} [2] & 1 & $\llcorner$ \textit{Demisexuality:} Mentions sexual attraction only after forming strong emotional bonds. \\
\hspace{1em} [2] & 1 & $\llcorner$ \textit{Grey-Asexuality:} Mentions experiencing sexual attraction infrequently or under specific circumstances. \\
\hspace{1em} [2] & 1 & $\llcorner$ \textit{Biromanticism:} Mentions romantic attraction to multiple genders without sexual attraction. \\
\hspace{1em} [2] & 1 & $\llcorner$ \textit{Sex-Positive Asexuality:} Mentions engaging in sexual activities without experiencing sexual attraction. \\
\midrule

\textbf{[1]} & \textbf{160} & \textbf{Bisexuality:} Mentions attraction to multiple genders, indicating a sexual orientation. \\
\addlinespace[0.25em]
\hspace{1em} [2] & 1 & $\llcorner$ \textit{Sexual Fluidity:} Discusses the changing or fluid nature of sexual orientation over time. \\
\hspace{1em} [2] & 1 & $\llcorner$ \textit{Romantic Preferences:} Mentions specific romantic preferences or inclinations within bisexuality. \\
\hspace{1em} [2] & 1 & $\llcorner$ \textit{Gender Spectrum:} Refers to attraction to multiple genders beyond the binary male and female. \\
\hspace{1em} [2] & 1 & $\llcorner$ \textit{Asexuality and Bisexuality:} Discusses the intersection of asexuality with bisexuality. \\
\hspace{1em} [2] & 1 & $\llcorner$ \textit{Queer Identity:} Refers to identifying as queer in addition to or instead of bisexual. \\
\hspace{1em} [2] & 1 & $\llcorner$ \textit{Coming Out and Identity:} Discusses experiences related to coming out and embracing bisexual identity. \\
\midrule

\textbf{[1]} & \textbf{330} & \textbf{Heterosexuality:} Refers specifically to the sexual orientation of being attracted to the opposite sex. \\
\addlinespace[0.25em]
\hspace{1em} [2] & 1 & $\llcorner$ \textit{Romantic Relationships:} Explores how societal norms impact perceptions of heterosexuality. \\
\hspace{1em} [2] & 1 & $\llcorner$ \textit{Sexual Orientation Identity:} Discusses personal identification and understanding of one's sexual orientation. \\
\hspace{1em} [2] & 1 & $\llcorner$ \textit{Societal and Cultural Influences:} Explores how societal norms impact perceptions of heterosexuality. \\
\midrule

\textbf{[1]} & \textbf{205} & \textbf{Relationships:} Mentions the connections and interactions between individuals. \\
\addlinespace[0.25em]
\hspace{1em} [2] & 1 & $\llcorner$ \textit{Sexual Orientation:} Discusses how orientation influences relationships and personal experiences. \\
\hspace{1em} [2] & 1 & $\llcorner$ \textit{Romantic Relationships:} Explores the nature of romantic relationships, including preferences, experiences, and societal influences on romantic connections. \\
\midrule

\textbf{[1]} & --- & \textbf{Other Themes:} Queer (Count: 72), Sexual Orientation (Count: 69), Pansexuality (Count: 68), Female (Count: 56), Lesbian (Count: 56), Identity (Count: 50), Love (Count: 48), Gay (Count: 43), Demisexuality (Count: 43), Fluidity (Count: 42), Romanticism (Count: 40), Male (Count: 40), Gender (Count: 36), Family (Count: 32), Aromanticism (Count: 26), Panromanticism (Count: 22), Homosexuality (Count: 17), Community (Count: 16), Transgender (Count: 12), Emotional Connection (Count: 12), Discrimination (Count: 9), Polyamory (Count: 7), Attraction (Count: 7), Masculinity (Count: 6), Nonbinary (Count: 6), Monogamy (Count: 5), Compulsory Heterosexuality (Count: 4), Bullying (Count: 4), LGBTQ Community (Count: 4), Heteroromanticism (Count: 4), Intelligence (Count: 3), Awareness (Count: 3), Greysexuality (Count: 3), Dating (Count: 3), Trauma (Count: 3), Greyromanticism (Count: 2), Biromanticism (Count: 2), Androsexuality (Count: 2), Trade (Count: 2), Sapiosexuality (Count: 2), Mental Health (Count: 2), Gynesexuality (Count: 1), Aesthetic Attraction (Count: 1), Isolation (Count: 1), Hypersexuality (Count: 1), Fraysexual (Count: 1), Aegosexual (Count: 1), Homoromanticism (Count: 1), Autism (Count: 1), Diversity (Count: 1), Sapphic (Count: 1), Heteroflexibility (Count: 1), Polysexuality (Count: 1), Demiromanticism (Count: 1), Pride (Count: 1).

\end{longtable}
}
\FloatBarrier

\clearpage
\section{Varying the number of computationally identified free-text themes}
\label{supp:compare-m-k}

We systematically compare results across different values of the SAE dimensionality $M$, while holding the sparsity parameter $K$ fixed. As a reminder, these hyperparameters jointly control the granularity of the concepts learned by the SAE: $M$ specifies the number of latent dimensions (or components) available to represent distinct patterns across the dataset, while $K$ limits how many of those dimensions may be active for any given response. Following prior work, we fix $K=4$, which has been shown to provide a reasonable balance between interpretability and expressiveness for similar tasks \cite{hypothesaes}.

Varying $M$ allows us to assess how the granularity of free-text themes changes as the SAE is permitted to allocate more dimensions to latent patterns, allowing for more fine-grained themes. Consistent with the intuition provided by \citeauthor{hypothesaes} (\citeyear{hypothesaes}), smaller values of $M$ encourage the SAE to represent broader, higher-level patterns, while larger values of $M$ allow the model to allocate separate dimensions to more specific or niche patterns. Empirically, we observe that with fewer dimensions the SAE primarily recovers high-level themes, whereas increasing $M$ yields a mixture of high-level and more fine-grained themes. Tables \ref{tab:m_16_race_table}-\ref{tab:m_16_sexual_orientation_table} show themes learned at $M=16$, $K=4$, while Tables \ref{tab:m_64_race_table}-\ref{tab:m_64_sexual_orientation_table} show themes learned at $M=64$, $K=4$. These tables are included for reference only and not used in downstream analysis; strikethroughs for excluded themes are therefore not shown. 

As discussed in Appendix \S\ref{supp:manual-coverage-val}, several themes that are not isolated at $M=32$ emerge more explicitly when the SAE dimensionality is increased (e.g., $M=64$), illustrating the same trade-off between granularity and redundancy described in prior work.

\begin{table}[h]
\caption{Computationally identified free-text themes for SAE dimensions ($M=16$, $K=4$) trained on race-related free text response embeddings and their associated fidelity (F1 score).}
\label{tab:m_16_race_table}
\begin{tabular}{r|p{12cm}|c}
\toprule
 & \textbf{Race Theme} & \textbf{Fidelity} \\
\midrule
1 & mentions whiteness or being white & 0.359 \\
2 & self-describes as white & 0.990 \\
3 & mentions cultural traditions or practices & 0.940 \\
4 & explicitly self-describes as Black or African American & 0.910 \\
5 & mentions collective or community-oriented values & 0.699 \\
6 & mentions speaking an Asian language & 0.723 \\
7 & mentions Spanish language or bilingualism & 0.795 \\
8 & uses single-word or extremely brief responses to describe race/ethnicity & 0.744 \\
9 & mentions not thinking about their race or ethnicity much & 0.853 \\
10 & mentions skin color or complexion explicitly & 0.462 \\
11 & mentions being mixed race or having mixed ancestry & 0.818 \\
12 & mentions family or cultural traditions/practices & 0.528 \\
13 & mentions family or ancestors in a personal or cultural context & 0.610 \\
14 & mentions being black or African American & 0.990 \\
15 & mentions personal or familial connection to multiple ethnicities or cultures & 0.260 \\
16 & explicitly self-identifies race and ethnicity in distinct terms (e.g., `My race is... and my ethnicity is...') & 0.561 \\
\bottomrule
\end{tabular}
\end{table}

\begin{table}[h]
\caption{Computationally identified free-text themes for SAE dimensions ($M=16$, $K=4$) trained on gender-related free text response embeddings and their associated fidelity (F1 score).}
\label{tab:m_16_gender_table}
\begin{tabular}{r|p{12cm}|c}
\toprule
 & \textbf{Gender Theme} & \textbf{Fidelity} \\
\midrule
1 & uses the term `female' or `females' explicitly to describe gender & 0.397 \\
2 & mentions physical appearance or traits about self & 0.407 \\
3 & is a single word response & 0.438 \\
4 & mentions being nonbinary or genderfluid & 0.910 \\
5 & includes a definitive statement about gender identity without discussing ambiguity or questioning & 0.561 \\
6 & self-describes as cisgender male or man & 0.950 \\
7 & self-describes as feminine or mentions femininity & 0.843 \\
8 & uses single words or very short phrases to describe gender identity & 0.497 \\
9 & expresses belief in a strict binary concept of gender & 0.619 \\
10 & expresses uncertainty or questioning about their gender identity & 0.597 \\
11 & explicitly states alignment between birth-assigned sex and current gender identity & 0.764 \\
12 & self-describes as a man or male & 0.918 \\
13 & discusses experiences of navigating societal expectations or norms related to gender & 0.938 \\
14 & mentions lack of significant impact or influence of gender on their life & 0.621 \\
15 & explicitly states alignment between gender identity and sex assigned at birth & 0.490 \\
16 & describes gender identity as aligned with biological sex or birth gender without significant questioning or exploration & 0.577 \\
\bottomrule
\end{tabular}
\end{table}

\begin{table}[h]
\caption{Computationally identified free-text themes for SAE dimensions ($M=16$, $K=4$) trained on sexual orientation-related free text response embeddings and their associated fidelity (F1 score).}
\label{tab:m_16_sexual_orientation_table}
\begin{tabular}{r|p{12cm}|c}
\toprule
 & \textbf{Sexual Orientation Theme} & \textbf{Fidelity} \\
\midrule
1 & expresses uncertainty or refusal to describe their sexual orientation & 0.615 \\
2 & mentions experimentation or exploration of sexuality & 0.486 \\
3 & mentions fluidity or changes in sexual orientation over time & 0.370 \\
4 & discusses how sexual orientation influences relationships or life experiences & 0.815 \\
5 & mentions bisexual or queer as part of their sexual orientation & 0.890 \\
6 & mentions comfort or natural feeling about sexual orientation & 0.525 \\
7 & mentions discomfort or reluctance in discussing sexual orientation & 0.367 \\
8 & mentions personal milestones or transformative moments related to sexual orientation & 0.417 \\
9 & responses are extremely brief, often one or two words, and lack elaboration & 0.843 \\
10 & explicitly self-describes as straight or heterosexual & 0.958 \\
11 & expresses uncertainty or confusion about their sexual orientation or labels & 0.367 \\
12 & uses the word `straight' without additional elaboration & 0.577 \\
13 & self-describes as a woman attracted to women & 0.723 \\
14 & responses are single words or very short phrases & 0.497 \\
15 & mentions lack of sexual attraction to real people & 1.000 \\
16 & explicitly self-describes as a specific sexual orientation without elaborating on fluidity or complexity & 0.884 \\
\bottomrule
\end{tabular}
\end{table}

\begin{table}[h]
{\renewcommand{\arraystretch}{0.95}
\caption{Computationally identified free-text themes for SAE dimensions ($M$=64, $K$=4) trained on race-related free text response embeddings and their associated fidelity (F1 score).}
\label{tab:m_64_race_table}
\begin{tabular}{r|p{12cm}|c}
\toprule
 & \textbf{Race Theme} & \textbf{Fidelity} \\
\midrule
1 & mentions African American identity or culture & 0.769 \\
2 & self-describes using a single concise phrase or label without additional context or elaboration & 0.679 \\
3 & mentions specific ancestry percentages or fractions & 0.584 \\
4 & mentions lack of specific traditions or customs related to their race or ethnicity & 0.767 \\
5 & mentions Arabic language or dialect & 0.507 \\
6 & mentions indigenous roots or ancestry & 0.686 \\
7 & uses abstract or vague language to describe race/ethnicity or identity & 0.466 \\
8 & mentions Italian heritage or cultural elements & 0.759 \\
9 & self-describes as ordinary, plain, or generic with regards to race or ethnicity & 0.950 \\
10 & mentions African or African American identity, heritage, or culture explicitly & 0.799 \\
11 & mentions Southern or rural cultural attributes (e.g., Southern food, accents, traditions, or rural life) & 0.723 \\
12 & mentions gender or gender identity explicitly & 0.936 \\
13 & mentions being Mexican American or having Mexican heritage & 0.970 \\
14 & self-describes as white or includes `white' in their racial/ethnic identity & 0.642 \\
15 & mentions being Vietnamese or Vietnamese-American & 0.507 \\
16 & mentions being mixed race or biracial & 0.843 \\
17 & explicitly describes the cultural or ethnic traditions or practices associated with their race or ethnicity & 0.590 \\
18 & self-describes as Black or African American & 0.920 \\
19 & describes race using collective traits, values, or actions of the group & 0.679 \\
20 & mentions being South Asian or a specific South Asian ethnicity & 0.734 \\
21 & mentions German ancestry & 0.884 \\
22 & mentions Puerto Rican heritage & 0.438 \\
23 & mentions Asian or Pacific Islander identity & 0.769 \\
24 & mentions ancestors or ancestry in detail, including migration or origins & 0.734 \\
25 & mentions celebration of American holidays such as Christmas, Thanksgiving, or Independence Day & 0.830 \\
26 & mentions speaking only English at home & 0.667 \\
27 & mentions privilege associated with being white & 0.830 \\
28 & mentions Scottish or Irish heritage & 0.960 \\
29 & mentions European ancestry in detail & 0.898 \\
30 & mentions speaking or learning multiple languages & 0.484 \\
31 & mentions Filipino ethnicity or culture & 0.750 \\
32 & mentions being American or having American identity explicitly & 0.630 \\
33 & mentions traditions or practices related to holidays and food & 0.609 \\
34 & describes themselves explicitly as `white' & 0.789 \\
35 & mentions being Korean or engaging in Korean culture & 0.551 \\
36 & mentions being raised in a religious household or community & 0.581 \\
37 & mentions Native American tribal affiliation & 0.938 \\
38 & mentions taking a DNA test or ancestry analysis & 0.507 \\
39 & mentions lack of strong connection or significance of race/ethnicity to identity & 0.721 \\
40 & explicitly self-describes as `white' & 0.589 \\
41 & self-describes as black or African American & 0.808 \\
42 & mentions lack of connection to cultural heritage or traditions & 0.818 \\
43 & mentions activities involving food, such as BBQ, soul food, or cooking & 0.369 \\
44 & mentions bilingualism or the use of multiple languages & 0.663 \\
45 & mentions speaking English fluently or as a primary language & 0.940 \\
46 & self-describes as mixed race or biracial & 0.824 \\
47 & emphasizes family orientation and traditional values & 0.769 \\
48 & mentions Africa or being African & 0.900 \\
49 & mentions speaking a non-English language at home or with family & 0.519 \\
50 & mentions food as a significant aspect of cultural identity & 0.840 \\
51 & mentions rich cultural heritage or traditions & 0.622 \\
52 & mentions languages spoken in their family or heritage & 0.808 \\
53 & mentions Native American culture, traditions, or identity & 0.626 \\
54 & mentions Jewish holidays or customs & 0.246 \\
55 & mentions mixed European heritage or ancestry & 0.604 \\
56 & mentions being born in the United States or America & 0.864 \\
57 & mentions celebrating Lunar New Year or cultural new year celebrations & 0.361 \\
58 & mentions experiences of discrimination or racism & 0.774 \\
59 & mentions being Chinese or of Chinese descent & 0.734 \\
60 & mentions mixed racial or ethnic identity & 0.582 \\
61 & self-describes as Black American or related terms (e.g., Black, Black American, ADOS) & 0.929 \\
62 & mentions experiences of being of mixed race or ethnicity and feeling disconnected from one or more of their cultural backgrounds & 0.775 \\
63 & mentions Spanish as a spoken language or cultural element & 0.843 \\
64 & mentions specific ancestral or geographic origins in addition to race/ethnicity & 0.471 \\
\bottomrule
\end{tabular}}
\end{table}

\begin{table}[h]
{\renewcommand{\arraystretch}{0.95}
\caption{Computationally identified free-text themes for SAE dimensions ($M$=64, $K$=4) trained on gender-related free text response embeddings and their associated fidelity (F1 score).}
\label{tab:m_64_gender_table}
\begin{tabular}{r|p{12cm}|c}
\toprule
& \textbf{Gender Theme} & \textbf{Fidelity} \\
\midrule
1 & mentions using multiple pronouns or being comfortable with multiple pronouns & 0.893 \\
2 & mentions racial or ethnic identity explicitly & 0.630 \\
3 & explicitly self-describes as cisgender & 0.626 \\
4 & mentions being born male and still identifying as male & 0.867 \\
5 & mentions how gender impacts their life & 0.452 \\
6 & discusses challenges or discrimination faced in professional or societal contexts due to gender & 0.636 \\
7 & mentions enjoyment of traditionally feminine activities or traits & 0.565 \\
8 & uses the term nonbinary or non-binary to describe their gender identity & 0.950 \\
9 & self-describes as genderqueer & 0.686 \\
10 & expresses indifference or flexibility about how others perceive their gender or which pronouns are used & 0.782 \\
11 & discusses challenges or societal expectations specifically related to being a woman & 0.849 \\
12 & explicitly mentions never questioning their gender identity & 0.754 \\
13 & mentions lifelong consistency in gender identity & 0.759 \\
14 & uses concise statements to describe their gender without elaboration or context & 0.809 \\
15 & self-describes as cisgender or cis woman & 0.828 \\
16 & mentions discomfort or questioning of gender identity from a young age & 0.734 \\
17 & mentions responsibility to provide or take care of family & 0.524 \\
18 & discusses societal expectations or roles related to masculinity & 0.795 \\
19 & explicitly states sexual orientation (e.g., heterosexual, bisexual, gay) & 0.990 \\
20 & mentions alignment with masculinity or masculine traits & 0.762 \\
21 & mentions advantages or benefits related to their gender identity & 0.489 \\
22 & mentions milestones related to womanhood, such as motherhood, giving birth, or navigating societal expectations specific to women & 0.546 \\
23 & mentions being born a specific gender and still identifying as that gender & 0.565 \\
24 & mentions medical transition or plans for medical transition & 0.571 \\
25 & single-word responses & 0.359 \\
26 & mentions experiences or challenges related to societal or workplace discrimination based on gender & 0.884 \\
27 & expresses confusion or uncertainty about their gender identity & 0.715 \\
28 & mentions giving birth or ability to bear children & 0.330 \\
29 & mentions biological or physical characteristics such as chromosomes, reproductive organs, or secondary sex characteristics to describe gender identity & 0.869 \\
30 & describes gender identity as straightforward or simple & 0.638 \\
31 & mentions giving birth or having children & 0.925 \\
32 & mentions gender presentation as distinct or differing from gender identity & 0.799 \\
33 & mentions gender identity as fluid or changing over time & 0.864 \\
34 & mentions being assigned female at birth & 0.713 \\
35 & discusses how gender identity aligns or interacts with societal norms, expectations, or roles & 0.627 \\
36 & explicitly self-describes as cisgender man or cis male & 0.790 \\
37 & mentions emotional characteristics or behaviors & 0.407 \\
38 & uses the phrase `I am a female' or `I'm a female' & 0.820 \\
39 & uses the term `nonbinary' or `agender' to describe their gender identity & 0.920 \\
40 & mentions nonbinary or non-binary as their gender identity & 0.889 \\
41 & single-word responses referring only to gender & 0.658 \\
42 & uses the term `cisgender' or `cis' to describe their gender identity & 0.970 \\
43 & mentions a sense of detachment or complexity regarding gender identity & 0.548 \\
44 & mentions being born a woman and explicitly repeats or emphasizes being born a woman & 0.770 \\
45 & mentions lack of significant influence of gender on their life & 0.796 \\
46 & mentions struggles or discomfort with femininity or societal expectations of femininity & 0.650 \\
47 & expresses apathy, confusion, or rejection toward the concept of gender & 0.860 \\
48 & explicitly states `born female' or `born a female' & 0.864 \\
49 & mentions hobbies or activities associated with traditional masculinity & 0.303 \\
50 & mentions expectations or roles tied to gender within family or societal context & 0.789 \\
51 & explicitly states `I identify as [a gender]' & 0.817 \\
52 & mentions belief in a binary gender system based on being born as either male or female & 0.675 \\
53 & describes personal qualities or physical appearance in detail & 0.273 \\
54 & explicitly self-describes as male or man & 0.759 \\
55 & mentions how gender identity influences personal decisions or life milestones & 0.538 \\
56 & expresses belief in the existence of only two genders & 0.594 \\
57 & explicitly states being heterosexual or straight & 0.635 \\
58 & explicitly self-describes as `woman' or `female' without additional qualifiers or complexities & 0.759 \\
59 & mentions being heterosexual & 0.635 \\
60 & mentions sports or activities traditionally associated with men & 0.275 \\
61 & mentions ability to create or bring life (e.g., bearing children, conceiving, procreating, etc.) & 0.436 \\
62 & mentions providing or taking responsibility for others & 0.568 \\
63 & mentions transitioning or undergoing medical/physical changes related to gender & 0.816 \\
64 & mentions masculinity or masculine traits & 0.629 \\
\bottomrule
\end{tabular}}
\end{table}

\begin{table}[h]
{\renewcommand{\arraystretch}{0.95}
\caption{Computationally identified free-text themes for SAE dimensions ($M$=64, $K$=4) trained on sexual orientation-related free text response embeddings and their associated fidelity (F1 score).}
\label{tab:m_64_sexual_orientation_table}
\begin{tabular}{r|p{12cm}|c}
\toprule
 & \textbf{Sexual Orientation Theme} & \textbf{Fidelity} \\
\midrule
1 & mentions pansexual or pan-romantic & 0.958 \\
2 & discusses how sexual orientation shapes identity, relationships, or personal growth & 0.357 \\
3 & mentions attraction to or relationships with nonbinary or transgender individuals & 0.838 \\
4 & uses the term `heterosexual' & 0.889 \\
5 & mentions that sexual orientation does not significantly influence their life & 0.844 \\
6 & mentions questioning or uncertainty about their sexual or romantic orientation & 0.508 \\
7 & mentions lack of sexual attraction or interest in sex & 0.916 \\
8 & explicitly states `I like men' & 0.744 \\
9 & mentions being a virgin or not having sexual experiences & 0.212 \\
10 & mentions how sexual orientation influences their life & 0.148 \\
11 & explicitly mentions having always felt straight or being straight since birth & 0.748 \\
12 & describes a process of self-discovery or change in understanding of their sexual orientation & 0.644 \\
13 & mentions discomfort, judgment, or privacy concerns related to discussing sexual orientation & 0.615 \\
14 & expresses discomfort or dissatisfaction with labeling their sexual orientation & 0.878 \\
15 & mentions asexuality or being on the asexual spectrum & 0.969 \\
16 & mentions attraction to men & 0.723 \\
17 & mentions attraction to femininity or masculinity as a spectrum or fluid concept & 0.750 \\
18 & mentions curiosity or experimentation with same-sex relationships & 0.679 \\
19 & discusses separation between romantic and sexual orientation & 0.666 \\
20 & uses the phrase `I would describe my sexual orientation as ...' & 0.893 \\
21 & mentions asexuality or being asexual & 0.635 \\
22 & uses the word `straight' without additional qualifiers or elaboration & 0.439 \\
23 & explicitly uses the phrase `My sexual orientation is ...' & 0.893 \\
24 & mentions pansexual identity specifically & 0.896 \\
25 & mentions gender identity or transition explicitly & 0.758 \\
26 & uses the term `straight' without elaboration or additional context & 0.864 \\
27 & mentions God or creation in the context of sexual orientation & 0.214 \\
28 & expresses confusion or uncertainty about their sexual orientation & 0.706 \\
29 & uses the phrase `I am straight' & 0.840 \\
30 & mentions significant milestones or life events related to their sexual orientation & 0.695 \\
31 & explicitly mentions a preference for masculine or feminine traits in potential partners & 0.747 \\
32 & mentions emotional connection as a key aspect of attraction or relationships & 0.422 \\
33 & explicitly states being exclusively attracted to one gender & 0.702 \\
34 & mentions ease or lack of hardship due to sexual orientation & 0.829 \\
35 & self-describes as gay or a gay man & 0.867 \\
36 & mentions being straight or heterosexual & 0.659 \\
37 & contains single-word responses or extremely brief descriptions & 0.803 \\
38 & self-describes as lesbian & 0.980 \\
39 & mentions being married and the gender of their spouse & 0.938 \\
40 & explicitly states interest in the opposite sex using the phrase `only interested' & 0.817 \\
41 & describes the process or experience of coming out & 0.790 \\
42 & expresses discomfort, uncertainty, or refusal to disclose their sexual orientation & 0.584 \\
43 & mentions never questioning or changing their sexual orientation & 0.699 \\
44 & discusses personal experiences or practices related to sexual intimacy or relationships in detail & 0.624 \\
45 & explicitly mentions `opposite sex' or `opposite gender' & 0.969 \\
46 & explicitly states attraction to females & 0.573 \\
47 & explicitly self-describes as bisexual & 0.958 \\
48 & discusses societal or external impacts of their sexual orientation & 0.795 \\
49 & explicitly denies any past or present attraction or experimentation with the same sex & 0.741 \\
50 & mentions demisexuality or being demisexual & 0.765 \\
51 & discusses how their heterosexual orientation shapes their relationships or interactions & 0.809 \\
52 & explicitly uses the word `straight' without elaboration or complexity beyond identifying as straight & 0.864 \\
53 & uses the phrase `identify as' to describe their orientation & 0.869 \\
54 & mentions requiring an emotional connection before feeling sexual attraction & 0.732 \\
55 & mentions attraction to both males and females & 0.778 \\
56 & mentions attraction to women without mentioning attraction to other genders & 0.798 \\
57 & uses a brief, single-sentence response to describe sexual orientation & 0.790 \\
58 & mentions being a woman in relation to their sexual orientation & 0.762 \\
59 & uses the term `queer' to describe their orientation & 1.000 \\
60 & uses the phrase `always been' to describe their sexual orientation or attraction & 0.775 \\
61 & uses the exact word `Heterosexual' without additional context or elaboration & 0.469 \\
62 & mentions personal fulfillment or life improvement due to sexual orientation & 0.180 \\
63 & explicitly mentions lack of gender preference in attraction & 0.840 \\
64 & uses the phrase `I am straight' or `I am heterosexual' explicitly & 0.752 \\
\bottomrule
\end{tabular}}
\end{table}

\FloatBarrier 

\bmhead{Acknowledgements}
The authors thank Serina Chang, Evan Dong, Frauke Kreuter, Hima Lakkaraju, Rajiv Movva, Eni Mustafaraj, Katie Piner, Divya Shanmugam, David Shin, and participants at the Pew Research Center research methods seminar for helpful comments. This work was supported by an NSF Graduate Research Fellowship, a Google Research Scholar award, an AI2050 Early Career Fellowship, NSF CAREER \#2142419, a CIFAR Azrieli Global scholarship, a gift to the LinkedIn-Cornell Bowers CIS Strategic Partnership, OpenAI API credits, the Survival and Flourishing Fund, Coefficient Giving, and the Zhang Family Endowed professorship.

\bmhead{Data availability}
The dataset is available upon request through our project website: \href{\projecturl}{\projecturl}. To mitigate any privacy risks, interested researchers must agree to a data usage agreement pledging not to re-identify individuals in the data, and to adhere to privacy-protecting measures when storing data and presenting results. 

\bmhead{Code availability}
Code to reproduce this analysis is available on GitHub: \href{\codeurl}{\codeurl}.
    
\bibliography{bibliography}

@article{carlson2025making,
  title={Making Interpretable Discoveries from Unstructured Data: A High-Dimensional Multiple Hypothesis Testing Approach},
  author={Carlson, Jacob},
  journal={arXiv preprint arXiv:2511.01680},
  year={2025}
}

@inproceedings{lieberum2024gemma,
  title={Gemma scope: Open sparse autoencoders everywhere all at once on gemma 2},
  author={Lieberum, Tom and Rajamanoharan, Senthooran and Conmy, Arthur and Smith, Lewis and Sonnerat, Nicolas and Varma, Vikrant and Kram{\'a}r, J{\'a}nos and Dragan, Anca and Shah, Rohin and Nanda, Neel},
  booktitle={Proceedings of the 7th BlackboxNLP Workshop: Analyzing and Interpreting Neural Networks for NLP},
  pages={278--300},
  year={2024}
}

@article{singer2017some,
  title={Some methodological uses of responses to open questions and other verbatim comments in quantitative surveys},
  author={Singer, Eleanor and Couper, Mick P},
  journal={Methods, data, analyses: a journal for quantitative methods and survey methodology (mda)},
  volume={11},
  number={2},
  pages={115--134},
  year={2017},
  publisher={DEU}
}

@article{nelson2020computational,
  title={Computational grounded theory: A methodological framework},
  author={Nelson, Laura K},
  journal={Sociological methods \& research},
  volume={49},
  number={1},
  pages={3--42},
  year={2020},
  publisher={Sage Publications Sage CA: Los Angeles, CA}
}

@article{than2025updating,
  title={Updating “the future of coding”: Qualitative coding with generative large language models},
  author={Than, Nga and Fan, Leanne and Law, Tina and Nelson, Laura K and McCall, Leslie},
  journal={Sociological Methods \& Research},
  volume={54},
  number={3},
  pages={849--888},
  year={2025},
  publisher={SAGE Publications Sage CA: Los Angeles, CA}
}

@article{wong2025other,
  title={What to do with “other, describe”},
  author={Wong, Jaclyn S and Valentino, Lauren and Pao, Christina and Donnelly Moran, Katie and Compton, D’Lane and Kaufman, Gayle},
  journal={Sociological Methodology},
  volume={55},
  number={2},
  pages={244--268},
  year={2025},
  publisher={SAGE Publications Sage CA: Los Angeles, CA}
}

@article{pao2025case,
  title={The Case for “Other”: Measuring Gender and Sexual Identity in Survey Research},
  author={Pao, Christina and Donnelly Moran, Katie and Compton, D′ Lane and Kaufman, Gayle and Dowling, Julie A},
  journal={Sociology Compass},
  volume={19},
  number={1},
  pages={e70031},
  year={2025},
  publisher={Wiley Online Library}
}

@techreport{ennis2024examining,
	title        = {Examining racial identity responses among people with middle eastern and north African ancestry in the American community survey},
	author       = {Ennis, Sharon and Tiv, Mehrgol and Fernandez, Leticia and Bhaskar, Renuka and Porter, Sonya},
	year         = 2024,
	institution  = {U.S. Census Bureau, Center for Economic Studies}

}

@article{shiller2017narrative,
  author = {Shiller, Robert J.},
  title = {Narrative Economics},
  journal = {American Economic Review},
  year = {2017},
  volume = {107},
  number = {4},
  pages = {967--1004}
}

@article{dunivin2025scaling,
  title={Scaling hermeneutics: a guide to qualitative coding with LLMs for reflexive content analysis},
  author={Dunivin, Zackary Okun},
  journal={EPJ Data Science},
  volume={14},
  number={1},
  pages={28},
  year={2025},
  publisher={Springer}
}

@inproceedings{ferrario2022eliciting,
  title={Eliciting people’s first-order concerns: Text analysis of open-ended survey questions},
  author={Ferrario, Beatrice and Stantcheva, Stefanie},
  booktitle={AEA papers and proceedings},
  volume={112},
  pages={163--169},
  year={2022},
  organization={American Economic Association 2014 Broadway, Suite 305, Nashville, TN 37203}
}

@article{walton2011brief,
  title={A brief social-belonging intervention improves academic and health outcomes of minority students},
  author={Walton, Gregory M and Cohen, Geoffrey L},
  journal={Science},
  volume={331},
  number={6023},
  pages={1447--1451},
  year={2011},
  publisher={American Association for the Advancement of Science}
}

@article{garland2018legislating,
	title        = {Legislating intersex equality: Building the resilience of intersex people through law},
	author       = {Garland, Fae and Travis, Mitchell},
	year         = 2018,
	journal      = {Legal Studies},
	publisher    = {Cambridge University Press},
	volume       = 38,
	number       = 4,
	pages        = {587--606}
}

@article{weideman2025research,
	title        = {Research Funded by National Institutes of Health Concerning Sexual and Gender Minoritized Populations: A Tracking Update for 2012 to 2022},
	author       = {Weideman, Ben CD and Ecklund, Alexandra M and Alley, Rhea and Rosser, BR Simon and Rider, G Nic},
	year         = 2025,
	journal      = {American Journal of Public Health},
	publisher    = {American Public Health Association},
	number       = {0},
	pages        = {e1--e13}
}

@article{morgan2022equal,
	title        = {Equal Is Not Good Enough: An Analysis of School Funding Equity across the US and within Each State.},
	author       = {Morgan, Ivy},
	year         = 2022,
	journal      = {Education Trust},
	publisher    = {ERIC}
}

@article{rothschild2024opportunities,
	title        = {Opportunities and risks of LLMs in survey research},
	author       = {Rothschild, David M and Brand, James and Schroeder, Hope and Wang, Jenny},
	year         = 2024,
	journal      = {Available at SSRN}
}

@article{mediabiasdetector,
	title        = {Media Bias Detector: Designing and Implementing a Tool for Real-Time Selection and Framing Bias Analysis in News Coverage},
	author       = {Wang, Jenny S and Haider, Samar and Tohidi, Amir and Gupta, Anushkaa and Zhang, Yuxuan and Callison-Burch, Chris and Rothschild, David and Watts, Duncan J},
	year         = 2025,
	journal    = {Proceedings of the 2025 CHI Conference on Human Factors in Computing Systems},
	location     = {},
	publisher    = {Association for Computing Machinery},
	url          = {https://doi.org/10.1145/3706598.3713716},
	articleno    = 790,
	numpages     = 27,
}

@article{bartholomay2023doing,
  title={Doing sexuality: How married bisexual, queer, and pansexual people navigate passing and erasure},
  author={Bartholomay, Daniel J and Pendleton, Meagan},
  journal={The Sociological Quarterly},
  volume={64},
  number={3},
  pages={520--539},
  year={2023},
  publisher={Taylor \& Francis}
}

@article{hughes2022guidance,
	title        = {Guidance for researchers when using inclusive demographic questions for surveys: Improved and updated questions},
	author       = {Hughes, Jennifer L and Camden, Abigail A and Yangchen, Tenzin and Smith, Gabrielle PA and Domenech Rodr{\'\i}guez, Melanie M and Rouse, Steven V and McDonald, C Peeper and Lopez, Stella},
	year         = 2022,
	journal      = {Psi Chi Journal of Psychological Research},
	volume       = 27,
	number       = 4,
	pages        = {232--255}
}

@book{national2022measuring,
  author = {{National Academies of Sciences, Engineering, and Medicine}},
  editor    = "Nancy Bates and Marshall Chin and Tara Becker",
  title     = "Measuring Sex, Gender Identity, and Sexual Orientation",
  isbn      = "978-0-309-27510-1",
  year      = 2022,
  publisher = "The National Academies Press",
  address   = "Washington, DC"
}

@techreport{amaya2020adapting,
  title={Adapting how we ask about the gender of our survey respondents},
  author={Amaya, Ashley and Vogels, Emily A and Brown, Anna},
  institution={Pew Research Center},
  url={https://www.pewresearch.org/decoded/2020/09/11/adapting-how-we-ask-about-the-gender-of-our-survey-respondents/},
  year={2020}
}

@article{revesz2024revisions,
  title={Revisions to OMB’s statistical policy directive no. 15: standards for maintaining, collecting, and presenting federal data on race and ethnicity},
  author={Revesz, RL},
  journal={Federal Register. Published March},
  volume={29},
  year={2024}
}

@misc{gonzalez2024kff,
  title={KFF Survey on Racism, Discrimination and Health: Views on Racism and Trust in Key US Institutions},
  author={Gonzalez-Barrera, A and Hamel, L and Artiga, S and Presiado, M},
  year={2024},
  publisher={KFF},
  url = {https://files.kff.org/attachment/Topline-KFF-Survey-on-Racism-Discrimination-and-Health.pdf}
}

@article{saperstein2025recognizing,
  title={Recognizing Identity Fluidity in Demographic Research},
  author={Saperstein, Aliya},
  journal={Population and Development Review},
  volume={51},
  number={1},
  pages={519--538},
  year={2025},
  publisher={Wiley Online Library}
}

@techreport{pew2015multiracial_appendix,
  author       = {{Pew Research Center}},
  title        = {Appendix C: Multiracial Survey Topline},
  booktitle    = {Multiracial in America: Proud, Diverse and Growing in Numbers},
  institution  = {Pew Research Center},
  address      = {Washington, DC},
  year         = {2015},
  url          = {https://www.pewresearch.org/social-trends/wp-content/uploads/sites/3/2015/06/2015-06-11_multiracial-in-america_final-updated.pdf},
}

@article{kilpatrick1960self,
  title={Self-anchoring scaling: A measure of individuals' unique reality worlds},
  author={Kilpatrick, Franklin Pierce and Cantril, Hadley},
  journal={Journal of Individual psychology},
  volume={16},
  number={2},
  pages={158},
  year={1960},
  publisher={North American Society of Adlerian Psychology.}
}

@article{anderssen2017oversampling,
  title={Oversampling as a methodological strategy for the study of self-reported health among lesbian, gay and bisexual populations},
  author={Anderssen, Norman and Malterud, Kirsti},
  journal={Scandinavian Journal of Public Health},
  volume={45},
  number={6},
  pages={637--646},
  year={2017},
  publisher={SAGE Publications Sage UK: London, England}
}

@misc{vaughan2017oversampling,
  title={Oversampling in health surveys: Why, when, and how?},
  author={Vaughan, Roger},
  journal={American Journal of Public Health},
  volume={107},
  number={8},
  pages={1214--1215},
  year={2017},
  publisher={American Public Health Association}
}

@inproceedings{movva2023coarse,
	title        = {Coarse race data conceals disparities in clinical risk score performance},
	author       = {Movva, Rajiv and Shanmugam, Divya and Hou, Kaihua and Pathak, Priya and Guttag, John and Garg, Nikhil and Pierson, Emma},
	year         = 2023,
	booktitle    = {Machine Learning for Healthcare Conference},
	pages        = {443--472},
	organization = {PMLR}
}

@misc{TrevorProject2021Gay,
	title        = {{Understanding Gay and Lesbian Identities}},
	author       = {{The Trevor Project}},
	year         = 2021,
	month        = aug,
	howpublished = {\url{https://www.thetrevorproject.org/resources/article/understanding-gay-lesbian-identities/}}
}

@misc{TrevorProjectBisexuality,
	title        = {{Understanding Bisexuality}},
	author       = {{The Trevor Project}},
	year         = 2021,
	month        = aug,
	howpublished = {\url{https://www.thetrevorproject.org/resources/article/understanding-bisexuality/}}
}

@misc{HRC2023Queer,
	title        = {{Glossary of Terms}},
	author       = {{Human Rights Campaign}},
	year         = 2023,
	month        = may,
	url = {https://www.hrc.org/resources/glossary-of-terms}
}

@article{petukhova2025text,
  title={Text clustering with large language model embeddings},
  author={Petukhova, Alina and Matos-Carvalho, Joao P and Fachada, Nuno},
  journal={International Journal of Cognitive Computing in Engineering},
  volume={6},
  pages={100--108},
  year={2025},
  publisher={Elsevier}
}

@article{arseniev2024theoretical,
  title={Theoretical foundations and limits of word embeddings: what types of meaning can they capture?},
  author={Arseniev-Koehler, Alina},
  journal={Sociological Methods \& Research},
  volume={53},
  number={4},
  pages={1753--1793},
  year={2024},
  publisher={Sage Publications Sage CA: Los Angeles, CA}
}

@article{cunningham2023sparse,
  title={Sparse autoencoders find highly interpretable features in language models},
  author={Cunningham, Hoagy and Ewart, Aidan and Riggs, Logan and Huben, Robert and Sharkey, Lee},
  journal={arXiv preprint arXiv:2309.08600},
  year={2023}
}

@article{bricken2023towards,
  title={Towards monosemanticity: Decomposing language models with dictionary learning},
  author={Bricken, Trenton and Templeton, Adly and Batson, Joshua and Chen, Brian and Jermyn, Adam and Conerly, Tom and Turner, Nick and Anil, Cem and Denison, Carson and Askell, Amanda and others},
  journal={Transformer Circuits Thread},
  volume={2},
  year={2023}
}

@article{o2024disentangling,
  title={Disentangling dense embeddings with sparse autoencoders},
  author={O'Neill, Charles and Ye, Christine and Iyer, Kartheik and Wu, John F},
  journal={arXiv preprint arXiv:2408.00657},
  year={2024}
}

@misc{templeton2024scaling,
  title={Scaling monosemanticity: Extracting interpretable features from claude 3 sonnet},
  author={Templeton, Adly},
  year={2024},
  publisher={Anthropic}
}

@article{gao2024scalingevaluatingsparseautoencoders,
	title        = {Scaling and evaluating sparse autoencoders},
	author       = {Leo Gao and Tom Dupré la Tour and Henk Tillman and Gabriel Goh and Rajan Troll and Alec Radford and Ilya Sutskever and Jan Leike and Jeffrey Wu},
	year         = 2024,
	url          = {https://arxiv.org/abs/2406.04093},
	journal = {arXiv preprint 2406.04093},
}

@article{movva2025s,
  title={What's In My Human Feedback? Learning Interpretable Descriptions of Preference Data},
  author={Movva, Rajiv and Milli, Smitha and Min, Sewon and Pierson, Emma},
  journal={ICLR},
  year={2026}
}

@article{hypothesaes,
  title = 	 {Sparse Autoencoders for Hypothesis Generation},
  author =       {Movva, Rajiv and Peng, Kenny and Garg, Nikhil and Kleinberg, Jon and Pierson, Emma},
  journal = {Proceedings of the 42nd International Conference on Machine Learning},
  pages = 	 {44997--45023},
  year = 	 {2025},
  volume = 	 {267},
  month = 	 {13--19 Jul},
  publisher =    {PMLR},
  url = 	 {https://proceedings.mlr.press/v267/movva25a.html},
}

@article{williams1997racial,
  title={Racial differences in physical and mental health: Socio-economic status, stress and discrimination},
  author={Williams, David R and Yu, Yan and Jackson, James S and Anderson, Norman B},
  journal={Journal of health psychology},
  volume={2},
  number={3},
  pages={335--351},
  year={1997},
  publisher={Sage Publications Sage CA: Thousand Oaks, CA}
}

@article{phinney1990ethnic,
	title        = {Ethnic identity in college students from four ethnic groups},
	author       = {Phinney, Jean S and Alipuria, Linda Line},
	year         = 1990,
	journal      = {Journal of adolescence},
	publisher    = {Wiley Online Library},
	volume       = 13,
	number       = 2,
	pages        = {171--183}
}

@article{porter1993minority,
	title        = {Minority identity and self-esteem},
	author       = {Porter, Judith R and Washington, Robert E},
	year         = 1993,
	journal      = {Annual review of sociology},
	publisher    = {Annual Reviews 4139 El Camino Way, PO Box 10139, Palo Alto, CA 94303-0139, USA},
	volume       = 19,
	number       = 1,
	pages        = {139--161}
}

@techreport{pew2019race,
	title        = {The Role of Race and Ethnicity in Americans' Personal Lives},
	author       = {Menasce Horowitz, Juliana and Brown, Anna and Cox, Kiana},
	year         = 2019,
	month        = {April},
	address      = {Washington, D.C.},
	url          = {https://www.pewresearch.org/social-trends/2019/04/09/the-role-of-race-and-ethnicity-in-americans-personal-lives/},
	institution  = {Pew Research Center}
}

@article{gulgoz2019similarity,
	title        = {Similarity in transgender and cisgender children’s gender development},
	author       = {G{\"u}lg{\"o}z, Selin and Glazier, Jessica J and Enright, Elizabeth A and Alonso, Daniel J and Durwood, Lily J and Fast, Anne A and Lowe, Riley and Ji, Chonghui and Heer, Jeffrey and Martin, Carol Lynn and others},
	year         = 2019,
	journal      = {Proceedings of the National Academy of Sciences},
	publisher    = {National Academy of Sciences},
	volume       = 116,
	number       = 49,
	pages        = {24480--24485}
}

@article{landis1977measurement,
  title={The measurement of observer agreement for categorical data},
  author={Landis, J Richard and Koch, Gary G},
  journal={biometrics},
  pages={159--174},
  year={1977},
  publisher={JSTOR}
}

@article{watkins2000interobserver,
  title={Interobserver agreement in behavioral research: Importance and calculation},
  author={Watkins, Marley W and Pacheco, Miriam},
  journal={Journal of Behavioral Education},
  volume={10},
  number={4},
  pages={205--212},
  year={2000},
  publisher={Springer}
}

@article{galupo2014conceptualization,
	title        = {Conceptualization of sexual orientation identity among sexual minorities: Patterns across sexual and gender identity},
	author       = {Galupo, M Paz and Davis, Kyle S and Grynkiewicz, Ashley L and Mitchell, Renae C},
	year         = 2014,
	journal      = {Journal of Bisexuality},
	publisher    = {Taylor \& Francis},
	volume       = 14,
	number       = {3-4},
	pages        = {433--456}
}

@article{blei2003latent,
	title        = {Latent dirichlet allocation},
	author       = {Blei, David M and Ng, Andrew Y and Jordan, Michael I},
	year         = 2003,
	journal      = {Journal of Machine Learning research},
	volume       = 3,
	number       = {Jan},
	pages        = {993--1022}
}

@article{grootendorst2022bertopic,
	title        = {BERTopic: Neural topic modeling with a class-based TF-IDF procedure},
	author       = {Grootendorst, Maarten},
	year         = 2022,
	journal      = {arXiv preprint arXiv:2203.05794}
}

@article{pham-etal-2024-topicgpt,
	title        = {{T}opic{GPT}: A Prompt-based Topic Modeling Framework},
	author       = {Pham, Chau Minh  and Hoyle, Alexander  and Sun, Simeng  and Resnik, Philip  and Iyyer, Mohit},
	year         = 2024,
	month        = jun,
	journal    = {Proceedings of the 2024 Conference of the North American Chapter of the Association for Computational Linguistics},
	publisher    = {Association for Computational Linguistics},
	address      = {Mexico City, Mexico},
	pages        = {2956--2984},
	url          = {https://aclanthology.org/2024.naacl-long.164/},
	editor       = {Duh, Kevin  and Gomez, Helena  and Bethard, Steven}
}

@article{michalos2001ethnicity,
	title        = {Ethnicity, modern prejudice and the quality of life},
	author       = {Michalos, Alex C and Zumbo, Bruno D},
	year         = 2001,
	journal      = {Social Indicators Research},
	publisher    = {Springer},
	volume       = 53,
	pages        = {189--222}
}

@book{crenshaw2013demarginalizing,
  title={Demarginalizing the intersection of race and sex: A black feminist critique of antidiscrimination doctrine, feminist theory and antiracist politics},
  author={Crenshaw, Kimberl{\'e}},
  booktitle={Feminist legal theories},
  pages={23--51},
  year={1997},
  publisher={Routledge},
  address={New York, NY}
}

@techreport{asirvatham2026gpt,
  title={GPT as a Measurement Tool},
  author={Asirvatham, Hemanth and Mokski, Elliott and Shleifer, Andrei},
  year={2026},
  institution={National Bureau of Economic Research}
}

@misc{prolific2025,
  author       = {{Prolific}},
  title        = {Prolific},
  year         = {2025},
  note         = {Online platform. Version: Dec 2024-Jan 2025. London, UK},
  howpublished = {\url{https://www.prolific.com}}
}

@article{jiangsun2025interp_embed,
    title={Interpretable Embeddings with Sparse Autoencoders: A Data Analysis Toolkit},
    author={Nick Jiang and Xiaoqing Sun and Lisa Dunlap and Lewis Smith and Neel Nanda},
    journal={arXiv preprint 2512.10092},
    url={https://arxiv.org/abs/2512.10092}, 
    year={2025}
}

@article{verkuyten1989happiness,
	title        = {Happiness among adolescents in the Netherlands: Ethnic and sex differences},
	author       = {Verkuyten, Maykel},
	year         = 1989,
	journal      = {Psychological Reports},
	publisher    = {SAGE Publications Sage CA: Los Angeles, CA},
	volume       = 65,
	number       = 2,
	pages        = {577--578}
}

@article{conron2010population,
	title        = {A population-based study of sexual orientation identity and gender differences in adult health},
	author       = {Conron, Kerith J and Mimiaga, Matthew J and Landers, Stewart J},
	year         = 2010,
	journal      = {American journal of public health},
	publisher    = {American Public Health Association},
	volume       = 100,
	number       = 10,
	pages        = {1953--1960}
}

@article{yip2018ethnic,
	title        = {Ethnic/racial identity-A double-edged sword? Associations with discrimination and psychological outcomes},
	author       = {Yip, Tiffany},
	year         = 2018,
	journal      = {Current directions in psychological science},
	publisher    = {Sage Publications Sage CA: Los Angeles, CA},
	volume       = 27,
	number       = 3,
	pages        = {170--175}
}

@article{burrow2010racial,
	title        = {Racial identity as a moderator of daily exposure and reactivity to racial discrimination},
	author       = {Burrow, Anthony L and Ong, Anthony D},
	year         = 2010,
	journal      = {Self and Identity},
	publisher    = {Taylor \& Francis},
	volume       = 9,
	number       = 4,
	pages        = {383--402}
}

@book{goffman2009stigma,
  title={Stigma: Notes on the Management of Spoiled Identity},
  author={Goffman, Erving},
  year={1963},
  publisher={Prentice-Hall},
  address={Englewood Cliffs, NJ}
}

@article{quinn2013concealable,
  title={Concealable stigmatized identities and psychological well-being},
  author={Quinn, Diane M and Earnshaw, Valerie A},
  journal={Social and personality psychology compass},
  volume={7},
  number={1},
  pages={40--51},
  year={2013},
  publisher={Wiley Online Library}
}

@article{klein1978klein,
  title={Klein sexual orientation grid},
  author={Klein, Fritz and Sepekoff, Barry and Wolf, Timothy J},
  journal={Journal of Homosexuality},
  year={1978}
}

@misc{magliozzi2016scaling,
  title={Scaling up: Representing gender diversity in survey research. Socius, 2, 1--11},
  author={Magliozzi, Devon and Saperstein, Aliya and Westbrook, Laurel},
  year={2016}
}

@article{gottgens2022impact,
	title        = {The impact of multiple gender dimensions on health-related quality of life in persons with Parkinson’s disease: an exploratory study},
	author       = {G{\"o}ttgens, Irene and Darweesh, Sirwan KL and Bloem, Bastiaan R and Oertelt-Prigione, Sabine},
	year         = 2022,
	journal      = {Journal of Neurology},
	publisher    = {Springer},
	volume       = 269,
	number       = 11,
	pages        = {5963--5972}
}

@article{bostwick2010dimensions,
	title        = {Dimensions of sexual orientation and the prevalence of mood and anxiety disorders in the United States},
	author       = {Bostwick, Wendy B and Boyd, Carol J and Hughes, Tonda L and McCabe, Sean Esteban},
	year         = 2010,
	journal      = {American journal of public health},
	publisher    = {American Public Health Association},
	volume       = 100,
	number       = 3,
	pages        = {468--475}
}

@misc{openai2024embeddings,
  author = {OpenAI},
  title = {OpenAI text embeddings},
  howpublished = {\url{https://platform.openai.com/docs/guides/embeddings}},
  note = {Accessed August 2025}
}

@article{allen1997testing,
  title={Testing hypotheses in nested regression models},
  author={Allen, Michael Patrick},
  journal={Understanding regression analysis},
  pages={113--117},
  year={1997},
  publisher={Springer US}
}

@book{greene2018econometric,
  title={Econometric Analysis},
  author={Greene, William H.},
  year={2018},
  publisher={Pearson Education},
  address = {Upper Saddle River,
NJ}
}

@article{benjamini1995controlling,
  title={Controlling the false discovery rate: a practical and powerful approach to multiple testing},
  author={Benjamini, Yoav and Hochberg, Yosef},
  journal={Journal of the Royal statistical society: series B (Methodological)},
  volume={57},
  number={1},
  pages={289--300},
  year={1995},
  publisher={Wiley Online Library}
}

@article{mays2003classification,
  title={Classification of race and ethnicity: implications for public health},
  author={Mays, Vickie M and Ponce, Ninez A and Washington, Donna L and Cochran, Susan D},
  journal={Annual review of public health},
  volume={24},
  number={1},
  pages={83--110},
  year={2003},
  publisher={Annual Reviews 4139 El Camino Way, PO Box 10139, Palo Alto, CA 94303-0139, USA}
}

@misc{ct2022classification,
  title={Connecticut Department of Public Health: Race and Ethnicity Classification and Data Collection},
  url={https://portal.ct.gov/dph/health-information-systems--reporting/race-ethn-classification/race-ethn-classification},
  year={2022},
  author={{Connecticut Department of Public Health}}
}

@article{antecol2006unhealthy,
  title={Unhealthy assimilation: why do immigrants converge to American health status levels?},
  author={Antecol, Heather and Bedard, Kelly},
  journal={Demography},
  volume={43},
  number={2},
  pages={337--360},
  year={2006},
  publisher={Springer}
}

@article{berkman1979social,
  title={Social networks, host resistance, and mortality: a nine-year follow-up study of Alameda County residents},
  author={Berkman, Lisa F and Syme, S Leonard},
  journal={American journal of Epidemiology},
  volume={109},
  number={2},
  pages={186--204},
  year={1979},
  publisher={Oxford University Press}
}

@article{house1988social,
  title={Social relationships and health},
  author={House, James S and Landis, Karl R and Umberson, Debra},
  journal={Science},
  volume={241},
  number={4865},
  pages={540--545},
  year={1988},
  publisher={American Association for the Advancement of Science}
}

@article{pickett2008people,
  title={People like us: ethnic group density effects on health},
  author={Pickett, Kate E and Wilkinson, Richard G},
  journal={Ethnicity \& health},
  volume={13},
  number={4},
  pages={321--334},
  year={2008},
  publisher={Taylor \& Francis}
}

@article{garcia2016converging,
  title={Converging to American: Healthy immigrant effect in children of immigrants},
  author={Garc{\'\i}a-P{\'e}rez, M{\'o}nica},
  journal={American Economic Review},
  volume={106},
  number={5},
  pages={461--466},
  year={2016},
  publisher={American Economic Association 2014 Broadway, Suite 305, Nashville, TN 37203}
}

@article{udry2003health,
  title={Health and behavior risks of adolescents with mixed-race identity},
  author={Udry, J Richard and Li, Rose Maria and Hendrickson-Smith, Janet},
  journal={American journal of public health},
  volume={93},
  number={11},
  pages={1865--1870},
  year={2003},
  publisher={American Public Health Association}
}

@article{vora2024influence,
  title={The influence of identity on Multiracial emerging adults’ health and experiences seeking Healthcare in the United States: a qualitative study},
  author={Vora, Anjali S and Grilo, Stephanie A},
  journal={Journal of Racial and Ethnic Health Disparities},
  volume={11},
  number={6},
  pages={3313--3325},
  year={2024},
  publisher={Springer}
}

@article{elhage2022toymodelssuperposition,
      title={Toy Models of Superposition}, 
      author={Nelson Elhage and Tristan Hume and Catherine Olsson and Nicholas Schiefer and Tom Henighan and Shauna Kravec and Zac Hatfield-Dodds and Robert Lasenby and Dawn Drain and Carol Chen and Roger Grosse and Sam McCandlish and Jared Kaplan and Dario Amodei and Martin Wattenberg and Christopher Olah},
      year={2022},
      journal={arXiv preprint 2209.10652},
      url={https://arxiv.org/abs/2209.10652}, 
}

@article{kang2015multiple,
  title={Multiple identities in social perception and interaction: Challenges and opportunities},
  author={Kang, Sonia K and Bodenhausen, Galen V},
  journal={Annual review of psychology},
  volume={66},
  number={1},
  pages={547--574},
  year={2015},
  publisher={Annual Reviews}
}

@article{aspinall2012answer,
  title={Answer formats in British census and survey ethnicity questions: does open response better capture ‘superdiversity’?},
  author={Aspinall, Peter J},
  journal={Sociology},
  volume={46},
  number={2},
  pages={354--364},
  year={2012},
  publisher={Sage Publications Sage UK: London, England}
}

@article{prewitt2005racial,
  title={Racial classification in America: where do we go from here?},
  author={Prewitt, Kenneth},
  journal={Daedalus},
  volume={134},
  number={1},
  pages={5--17},
  year={2005},
  publisher={MIT Press One Rogers Street, Cambridge, MA 02142-1209, USA journals-info~…}
}

@article{jones2008using,
  title={Using “socially assigned race” to probe white advantages in health status},
  author={Jones, Camara Phyllis and Truman, Benedict I and Elam-Evans, Laurie D and Jones, Camille A and Jones, Clara Y and Jiles, Ruth and Rumisha, Susan F and Perry, Geraldine S},
  journal={Ethnicity \& disease},
  volume={18},
  number={4},
  pages={496--504},
  year={2008},
  publisher={JSTOR}
}

@article{kim1999racial,
  title={The racial triangulation of Asian Americans},
  author={Kim, Claire Jean},
  journal={Politics \& society},
  volume={27},
  number={1},
  pages={105--138},
  year={1999},
  publisher={SAGE Publications, Inc.}
}

@article{schachter2021intersecting,
  title={Intersecting Boundaries: Comparing Stereotypes of Native-and Foreign-Born Members of Ethnoracial Groups},
  author={Schachter, Ariela},
  journal={Social Forces},
  volume={100},
  number={2},
  pages={506--539},
  year={2021},
  publisher={Oxford University Press}
}

@article{bem1974measurement,
  title={The measurement of psychological androgyny.},
  author={Bem, Sandra L},
  journal={Journal of consulting and clinical psychology},
  volume={42},
  number={2},
  pages={155},
  year={1974},
  publisher={American Psychological Association}
}

@article{croll2019race,
  title={Race as an open field: Exploring identity beyond fixed choices},
  author={Croll, Paul R and Gerteis, Joseph},
  journal={Sociology of Race and Ethnicity},
  volume={5},
  number={1},
  pages={55--69},
  year={2019},
  publisher={SAGE Publications Sage CA: Los Angeles, CA}
}

@article{garbarski2025improving,
  title={Improving the Measurement of Gender in Surveys: Effects of Categorical Versus Open-Ended Response Formats on Measurement and Data Quality Among College Students},
  author={Garbarski, Dana and Dykema, Jennifer and Yonker, James A and Bae, Rosie Eungyuhl and Rosenfeld, Rachel A},
  journal={Journal of Survey Statistics and Methodology},
  volume={13},
  number={1},
  pages={18--38},
  year={2025},
  publisher={Oxford University Press}
}

@article{wylie2010socially,
  title={Socially assigned gender nonconformity: A brief measure for use in surveillance and investigation of health disparities},
  author={Wylie, Sarah A and Corliss, Heather L and Boulanger, Vanessa and Prokop, Lisa A and Austin, S Bryn},
  journal={Sex roles},
  volume={63},
  number={3},
  pages={264--276},
  year={2010},
  publisher={Springer}
}

@article{mishel2019intersections,
  title={Intersections between sexual identity, sexual attraction, and sexual behavior among a nationally representative sample of American men and women},
  author={Mishel, Emma},
  journal={Journal of Official Statistics},
  volume={35},
  number={4},
  pages={859--884},
  year={2019},
  publisher={SAGE Publications Sage UK: London, England}
}

@incollection{gentile2014generational,
  title     = {Generational cultures},
  author    = {Gentile, Brittany and Campbell, W. Keith and Twenge, Jean M.},
  booktitle = {Culture Reexamined: Broadening Our Understanding of Social and Evolutionary Influences},
  editor    = {Cohen, Adam B.},
  pages     = {31--48},
  year      = {2014},
  publisher = {American Psychological Association},
  address   = {Washington, DC}
}

@article{garbarski2023measurement,
  title={The measurement of gender expression in survey research},
  author={Garbarski, Dana},
  journal={Social Science Research},
  volume={110},
  pages={102845},
  year={2023},
  publisher={Elsevier}
}

@article{suen2020,
  author  = {Suen, L. W. and Lunn, M. R. and Katuzny, K. and Finn, S. and Duncan, L. and Sevelius, J. and Obedin-Maliver, J. and others},
  title   = {What sexual and gender minority people want researchers to know about sexual orientation and gender identity questions: A qualitative study},
  journal = {Archives of sexual behavior},
  year    = {2020},
  volume  = {49},
  number  = {7},
  pages   = {2301--2318},
  publisher = {Springer}
}

@article{lee2024ibelong,
  title={The iBelong Scale: Construction and validation of a measure of racial--ethnic--cultural belonging.},
  author={Lee, B Andi and Neville, Helen A},
  journal={Journal of Counseling Psychology},
  volume={71},
  number={3},
  pages={139},
  year={2024},
  publisher={American Psychological Association}
}

@article{russell2009teens,
  title={Are teens “post-gay”? Contemporary adolescents’ sexual identity labels},
  author={Russell, Stephen T and Clarke, Thomas J and Clary, Justin},
  journal={Journal of Youth and Adolescence},
  volume={38},
  number={7},
  pages={884--890},
  year={2009},
  publisher={Springer}
}

@article{lopez2018s,
  title={What’s your “street race”? Leveraging multidimensional measures of race and intersectionality for examining physical and mental health status among Latinxs},
  author={L{\'o}pez, Nancy and Vargas, Edward and Juarez, Melina and Cacari-Stone, Lisa and Bettez, Sonia},
  journal={Sociology of Race and Ethnicity},
  volume={4},
  number={1},
  pages={49--66},
  year={2018},
  publisher={SAGE Publications Sage CA: Los Angeles, CA}
}

@article{douglas2023data,
  title={Data quality in online human-subjects research: Comparisons between MTurk, Prolific, CloudResearch, Qualtrics, and SONA},
  author={Douglas, Benjamin D and Ewell, Patrick J and Brauer, Markus},
  journal={Plos one},
  volume={18},
  number={3},
  pages={e0279720},
  year={2023},
  publisher={Public Library of Science San Francisco, CA USA}
}

@article{smyth2009open,
  title={Open-ended questions in web surveys: Can increasing the size of answer boxes and providing extra verbal instructions improve response quality?},
  author={Smyth, Jolene D and Dillman, Don A and Christian, Leah Melani and McBride, Mallory},
  journal={Public Opinion Quarterly},
  volume={73},
  number={2},
  pages={325--337},
  year={2009},
  publisher={Oxford University Press}
}

@article{phinney1996we,
  title={When we talk about American ethnic groups, what do we mean?},
  author={Phinney, Jean S},
  journal={American psychologist},
  volume={51},
  number={9},
  pages={918},
  year={1996},
  publisher={American Psychological Association}
}

@article{mcadams2001psychology,
  title={The psychology of life stories},
  author={McAdams, Dan P},
  journal={Review of general psychology},
  volume={5},
  number={2},
  pages={100--122},
  year={2001},
  publisher={SAGE Publications Sage CA: Los Angeles, CA}
}

@article{o2004any,
  title={``Any other comments?" Open questions on questionnaires--a bane or a bonus to research?},
  author={O'Cathain, Alicia and Thomas, Kate J},
  journal={BMC medical research methodology},
  volume={4},
  number={1},
  pages={25},
  year={2004},
  publisher={Springer}
}

@article{riiskjaer2012value,
  title={The value of open-ended questions in surveys on patient experience: number of comments and perceived usefulness from a hospital perspective},
  author={Riiskj{\ae}r, Erik and Ammentorp, Jette and Kofoed, Poul-Erik},
  journal={International Journal for Quality in Health Care},
  volume={24},
  number={5},
  pages={509--516},
  year={2012},
  publisher={Oxford University Press}
}

@article{read2021disaggregating,
  title={Disaggregating heterogeneity among non-Hispanic Whites: evidence and implications for US racial/ethnic health disparities},
  author={Read, Jen’nan Ghazal and Lynch, Scott M and West, Jessica S},
  journal={Population research and policy review},
  volume={40},
  number={1},
  pages={9--31},
  year={2021},
  publisher={Springer}
}

@article{tortora2010gallup,
  title={The Gallup world poll},
  author={Tortora, Robert D and Srinivasan, Rajesh and Esipova, Neli},
  journal={Survey methods in multinational, multiregional, and multicultural contexts},
  pages={535--543},
  year={2010},
  publisher={Wiley Online Library}
}

@misc{pew2025_discrimination_questionnaire,
  title        = {2025 Pew Research Center’s American Trends Panel Wave 167 Politics Survey: Final Questionnaire, April 7-13, 2025},
  year={2025},
  author       = {{Pew Research Center}},
  url = {https://www.pewresearch.org/wp-content/uploads/sites/20/2025/05/PP_2025.5.20_discrimination_questionnaire.pdf},
}

@article{powers2011evaluation,
  author = {Powers, D. M. W.},
  journal = {Journal of Machine Learning Technologies},
  number = 1,
  pages = {37--63},
  title = {{Evaluation: From precision, recall and f-measure to roc., informedness, markedness \& correlation}},
  volume = 2,
  year = 2011
}

@misc{kff2023_consumer_experiences_health_insurance,
  title        = {KFF Survey of Consumer Experiences with Health Insurance},
  author       = {Karen Pollitz and Kaye Pestaina and Alex Montero and Lunna Lopes and Isabelle Valdes and Ashley Kirzinger and Mollyann Brodie},
  year         = {2023},
  month        = jun,
  url = {https://www.kff.org/affordable-care-act/kff-survey-of-consumer-experiences-with-health-insurance},
}

@book{angrist2009mostly,
  title={Mostly Harmless Econometrics: An Empiricist's Companion},
  author={Angrist, Joshua D and Pischke, J{\"o}rn-Steffen},
  year={2009},
  publisher={Princeton University Press},
  address={Princeton, NJ},
  isbn={9780691120355}
}

@inproceedings{zhong2025hicode,
  title={HICode: Hierarchical inductive coding with LLMs},
  author={Zhong, Mian and Wang, Pristina and Field, Anjalie},
  booktitle={Proceedings of the 2025 Conference on Empirical Methods in Natural Language Processing},
  pages={31048--31066},
  year={2025}
}

@article{beresford2022coding,
  title={Coding qualitative data at scale: Guidance for large coder teams based on 18 studies},
  author={Beresford, Melissa and Wutich, Amber and du Bray, Margaret V and Ruth, Alissa and Stotts, Rhian and SturtzSreetharan, Cindi and Brewis, Alexandra},
  journal={International Journal of Qualitative Methods},
  volume={21},
  year={2022},
  publisher={SAGE Publications Sage CA: Los Angeles, CA}
}

@book{glaser2017discovery,
  title={Discovery of grounded theory: Strategies for qualitative research},
  author={Glaser, Barney and Strauss, Anselm},
  year={2017},
  publisher={Routledge},
  address   = {New York, NY}
}

@article{venkatesh2013bridging,
  title={Bridging the qualitative-quantitative divide: Guidelines for conducting mixed methods research in information systems},
  author={Venkatesh, Viswanath and Brown, Susan A and Bala, Hillol},
  journal={MIS quarterly},
  pages={21--54},
  year={2013},
  publisher={JSTOR}
}

@inproceedings{schofield2017pulling,
  title={Pulling out the stops: Rethinking stopword removal for topic models},
  author={Schofield, Alexandra and Magnusson, M{\aa}ns and Mimno, David},
  booktitle={Proceedings of the 15th Conference of the European Chapter of the Association for Computational Linguistics},
  pages={432--436},
  year={2017}
}

@misc{antoniak2022topic,
  author = {Antoniak, Maria},
  title = {Topic Modeling for the People},
  howpublished = {Blog by M. Antoniak},
  year = {2022},
  url = {https://maria-antoniak.github.io//2022/07/27/topic-modeling-for-the-people.html},
}

\end{document}